# Dynamically Optimal Treatment Allocation

Karun Adusumilli, Friedrich Geiecke, and Claudio Schilter *

October 31, 2024

**Abstract**

Dynamic decisions are pivotal to economic policy making. We show how existing evidence from randomized control trials can be utilized to guide personalized decisions in challenging dynamic environments with budget and capacity constraints. Recent advances in reinforcement learning now enable the solution of many complex, real-world problems for the first time. We allow for restricted classes of policy functions and prove that their regret decays at rate $n^{-1/2}$, the same as in the static case. Applying our methods to job training, we find that by exploiting the problem's dynamic structure, we achieve significantly higher welfare compared to static approaches.

*Adusumilli: Department of Economics, University of Pennsylvania, akarun@sas.upenn.edu.
Geiecke: Department of Methodology, London School of Economics and Politicial Science, f.c.geiecke@lse.ac.uk.
Schilter: Department of Economics, University of Bern, claudio.schilter@unibe.ch.
We would like to thank the co-editor and three anonymous referees for valuable comments that substantially improved the paper. Thanks also to Facundo Albornoz, Tim Armstrong, Debopam Bhattacharya, Xiaohong Chen, Wouter Den Haan, Frank Diebold, Johan Egebark, John Ham, Stephen Hansen, Toru Kitagawa, Anders Bredahl Kock, Damian Kozbur, Franck Portier, Marcia Schafgans, Frank Schorfheide, Aleksey Tetenov, Brendan Tracey, Alwyn Young and seminar participants at Stanford, Brown, University of Pennsylvania, Vanderbilt, Yale, Greater NY area Econometrics conference, Bristol Econometrics study group, the HSG causal machine learning workshop, AEA-ESEM meetings, and the CEMMAP/UCL workshop on Personalized Treatment for helpful comments.

# 1 Introduction

Many critical treatment assignment problems are inherently dynamic in nature. For example, consider the allocation of job-training programs to newly unemployed individuals. Such programs are often funded through a fixed yearly budget determined in advance by the legislature, while job-seekers arrive at job centers sequentially throughout the year. Policymakers must decide whether to allocate training to each individual, recognizing that each decision affects the remaining budget and the availability of training for future applicants who may benefit even more. In essence, policymakers face a dynamic optimization problem, balancing immediate needs against long-term budgetary constraints. Other examples of such decision problems include a small ICU unit deciding whether to admit a patient at full capacity or send them on a longer, riskier transport, and a non-perpetual foundation determining how to best allocate its funds before excessive amounts are consumed by operating costs. In all these examples, improving allocations can lead to substantial welfare gains.

While identifying optimal decision rules in dynamic environments is critical, the complexity of dynamic optimization has historically constrained real-world applications, particularly when the decision-making process involves many state variables. For instance, individualized job training decisions should consider factors such as applicants' socioeconomic backgrounds and work histories. Recent advancements in computational methods, notably reinforcement learning (RL), can enable solutions to many of these complex dynamic problems for the first time. Building on these developments, we aim to construct policy rules that are interpretable—essential for ethical and legal compliance—guaranteed by theory to converge efficiently to the optimal welfare (abstracting from computational error), and practically feasible.

In this paper, we develop a framework for deriving such policy rules in dynamic settings, using data from randomized controlled trials (RCTs) or observational studies. Specifically, we estimate the optimal treatment rule by optimizing a sample analogue of the ex-ante welfare function within a pre-specified policy class. This approach extends the static Empirical Welfare Maximization (EWM) framework introduced by Kitagawa and Tetenov (2018) to dynamic decision-making.

Existing simultaneous allocation methods, such as EWM, are often inadequate or suboptimal in dynamic settings for several reasons. First, our approach accommodates scenarios with a fixed budget but no time constraint—a context where simultaneous allocation methods generally fall short. These methods focus solely on solving the allocation problem, without addressing how the budget should be managed over time or how long it should last. Specifically, they lack the inherently dynamic concept of discounting.

Second, even in settings without discounting and where both budget and time constraints are present, solving the associated simultaneous allocation problem fails to yield a dynamically optimal policy. For instance, running down the budget before the time constraint can often be optimal. When the class of policy functions is constrained, nominally relaxing the budget constraint—by allowing a larger fraction of the population to be treated in a simultaneous setting—can result in policies that leave no budget unused and can even yield a higher average treatment effect among the treated, without violating the original budget constraint on the number of treated individuals. This leads to higher welfare. In our empirical application, we find a welfare gain of 8% compared to the traditional simultaneous allocation approach, just from employing such 'softer' budget constraints.[1]

[1]This gain varies with the initial budget. In our empirical setting, we show it can be as high as 17%. Simultaneous methods could deal with this specific issue by conducting a cumbersome grid search over all possible budget constraints to find the highest total welfare.

Third, in settings where the distribution of outcomes varies over time, simultaneous allocation rules fail to account for the timing of candidate arrivals. In contrast, dynamically optimal policies, even when stationary (i.e., not allowed to vary with budget or time), recognize that by strategically selecting the point at which the budget is exhausted, it is possible to effectively influence the population distribution of outcomes. For example, if individuals with higher potential outcomes tend to arrive earlier, it may be optimal to deplete the budget more quickly. In addition, dynamic policies can identify the optimal policy conditional on this relevant population distribution—allowing for improved solutions by concentrating on the subset of the population that is practically relevant. In contrast, simultaneous allocation methods are unable to account for this endogenous relationship between the chosen policy and the resulting distribution of outcomes. Additionally, if non-stationary policies—those that adapt over time and with changes in budget—are employed, the welfare gains can be even more significant. In our empirical application, we observe welfare improvements of 27% and 133% for stationary and non-stationary policies compared to a simultaneous allocation policy.[2]

We provide theoretical guarantees that the welfare under our estimated decision rule converges rapidly to the optimal welfare as the dataset sample size increases. To obtain this result, we study a suitable limiting version of the dynamic optimization problem by following recent literature and taking the rate of arrivals to infinity (Wager and Xu, 2024; Fan and Glynn, 2021; Adusumilli, 2021), resulting in a continuous time approximation of the problem. Exploiting Partial Differential Equation (PDE) methods in a novel manner, we are able to prove that regret decays at a $n^{-1/2}$ rate, where $n$ is the sample size of the RCT.[3] This matches the rate in the static setting, though it is not immediately obvious that it should extend to the dynamic context, given the substantial dependence involved. Specifically, any individual's covariates influence the actions taken for all subsequently arriving individuals, e.g., through budget depletion.[4]

[2]Even if we add the gridsearch over budget constraints to the simultaneous allocation approach, the improvements are still high at 17% and 116% respectively.

[3]Regret is defined as the difference in welfare from using the estimated policy rule, relative to the optimal policy within a candidate class of policy rules.

In our empirical illustration, we apply our methods to derive optimal job-training allocations in a setting that is calibrated to the institutional details of Sweden's job-assistance program, KROM. As we detail in Section 2, the program inherently operates under both time and budget constraints. Additionally, the arrival of candidates is time-varying, and the costs of providing treatment differ across individuals. We compute optimal policy functions that maximize a utilitarian welfare criterion, chosen from a flexible class of policy functions that can be freely selected by the policymaker. These policy classes can involve stationary or non-stationary policies (i.e., those that can vary with budget and/or time). The latter naturally yield weakly higher welfare outcomes. However, a policymaker might favor stationary policies for their simplicity or because they are less vulnerable to manipulation through strategically timed arrivals—especially when the welfare gain from a non-stationary policy over a stationary one is minimal. We estimate both types of policies and find that this welfare gain ranges from 0% to 84% depending on the specific setup.

The RL algorithm we use to solve our setup, Proximal Policy Optimization (PPO) by Schulman et al. (2017), is a well-established policy gradient method for obtaining optimal policies in dynamic decision problems with high-dimensional state spaces. PPO's scalability to problems with numerous states is essential, as our non-stationary policy includes 15 inputs plus an intercept, rendering traditional dynamic programming approaches impractical. Notably, only a single hyperpa-rameter required tuning in our setup. PPO can readily apply to future research in this domain with small modifications, even for applications involving large RCT datasets with many covariates. With its neural network-based value function, PPO also handles complex timing and seasonal variations effectively.

[4]Our primary interest in this paper is on estimation error and we do not present results on the effect of computational errors on regret. Theoretical results on the accuracy of PPO under realistic assumptions is still an active area of research. Also, all computational errors could in principle be reduced to 0 with enough computational time.

At a broader scale, we demonstrate that causal inference enables the construction of novel dynamic simulations directly from empirical data. We estimate the causal effects of policy actions on individuals by leveraging RCT data, which allows us to simulate realistic economic environments. Such simulations can support the training and development of RL methods in previously uncharted domains. To date, RL advances have often occurred in contexts like games or physical systems where the effects of actions on the environment are precisely known, e.g. through theoretical models. By constructing training environments based on empirical (economic) data, we can broaden the application of RL to more diverse real-world settings only because causal inference provides the essential framework to estimate the effect of the policy's actions on the simulated world.

## 1.1 Related Literature

If the dynamic aspect can be ignored, there exist a number of methods for estimating an optimal policy function that maximizes social welfare, starting from the seminal contribution of Manski (2004), and extended by Hirano and Porter (2009), Stoye (2009, 2012), Chamberlain (2011), Bhattacharya and Dupas (2012), Tetenov (2012), and others. More recently, Kitagawa and Tetenov (2018), Athey and Wager (2021), and Mbakop and Tabord-Meehan (2021) proposed and extended Empirical Welfare Maximization in this context. Our contribution here is to build on this literature and extend it to a dynamic setting.

Bandits with knapsacks (Slivkins, 2019) are a variant of bandit algorithms with resource constraints. Since they are devised for online learning, it is not straightfor-ward to compare them to our approach. Also, we solve the full dynamic optimization problem, whereas knapsack bandits are heuristic in nature.[5]

Dynamic Treatment Regimes (DTRs) are another related literature, see e.g., Laber et al. (2014) for an overview.[6] DTRs consist of a sequence of individualized treatment decisions. The data in these studies is a sequence of dynamic paths for each individual. By contrast, the dynamics in our setting are faced by the policy maker, not the individual. Another difference is that the number of decision points in DTRs is quite small (often in the single digits), whereas, in our setting, this corresponds to the number of arrivals, which is very high.

There is a vast literature on computing policy functions with RL. Unlike value function-based methods (e.g., Q-learning), policy gradient methods directly optimize policy parameters, allowing us to constrain the policy function space, e.g., to the class of logistic functions or neural networks. While other policy gradient methods like A3C (Mnih et al., 2016) and SAC (Haarnoja et al., 2018) could also be used, PPO has become a benchmark due to its balance of performance and stability. Recently, PPO has, e.g., been prominently used to align several widely discussed large language models (LLMs), including the GPT series (Ouyang et al., 2022).

In economics, dynamic optimization problems have traditionally been solved using Monte Carlo or grid-based methods like generalized policy iteration (e.g., Benitez-Silva et al., 2000). PPO provides significant advantages through its clipping mechanism in the objective function, which curtails overly aggressive policy updates. Moreover, its use of advantage estimation with bootstrapping and its ease of parallelization enhance both stability and scalability, making it well-suited to

[5] Additionally, we optimize within a constrained policy class, which is (currently) not possible with knapsack bandits.

[6] See Sakaguchi (2024) for an extension of the EWM approach to DTRs. In the biostatistics literature, RL algorithms are also commonly employed to identify optimal policies for DTRs; however, unlike our approach, these methods typically involve ”off-policy” RL algorithms.

handling high-dimensional state spaces effectively.[7]

In making use of the knowledge of dynamics, our methods are arguably closest to the domain of model-based RL. Recent studies (Yu et al., 2020; Kidambi et al., 2020) have argued that model based methods greatly expand the sample efficiency and applicability of RL. Compared to these works, our methods are narrowly tailored to the specific setup studied here, but for this same reason are more efficient. For instance, in a case with limited budget, our algorithm does not need to visit all budget values to learn the optimal policy. In contrast, generic model-based RL methods behave conservatively when not all the states are reached, and typically minimize only a lower bound on regret.

# 2 Setup

Sweden recently implemented a job-search assistance initiative known as KROM. Under this program, the Swedish Public Employment Service contracts private providers to assist job seekers in securing employment, with an emphasis on facilitating a swift transition back to work. Upon becoming unemployed, individuals meet with a caseworker who enters their information into a system to assess their eligibility for KROM. If eligible, they receive vouchers for training services provided by third-party organizations.[8] KROM operates with two distinct and fixed annual budgets: one allocated for paying training vouchers and another for unemployment insurance payments, with the latter serving as a binding constraint.[9] Consequently, eligibility thresholds for KROM can be adjusted based on the availability of funds in the unemployment insurance budget. In this section, we describe how our methods can be used to determine optimal eligibility rules under such constraints.

[7] For example, obtaining a policy function for LLMs, where each of thousands of token represents a state, has been achived with algorithms such as PPO.

[8] The caseworker may alternatively decide that the unemployed person should enter a more intense training program (or none at all). We discuss imperfect compliance in the supplementary material. In practice, if the system recommends KROM, compliance is around 50 to 55%; a no from the system means the candidate is not eligible for KROM, which cannot be overturned.

[9] Those not receiving KROM continue to receive unemployment insurance through another government agency.

We suppose that the decision maker faces a series of individuals who arrive as they get unemployed. The state variables are denoted by $s := (x, z, t)$, where $x$ denotes the vector of individual covariates, $z$ the current budget, and $t$ is time. For the job-training example $t = 1$ corresponds to 1 year.

The arrival rates are modeled as a time-varying Poisson process with parameter $\lambda(t)N$, where $N$ is a scaling factor representing the expected number of individuals arriving within a unit period (e.g., one year). We interpret $N$ as the effective population size. The function $\lambda(t)$ is normalized so that $\int_0^1 \lambda(t)\,dt = 1$. Given the strong seasonal fluctuations in unemployment—higher in winter than in summer—it is essential to allow for time varying $\lambda(t)$. Following recent literature (Wager and Xu, 2024; Fan and Glynn, 2021; Adusumilli, 2021), we let $N \to \infty$ to derive a PDE characterization of welfare. In practice, $N$ is already large, as it is the number of unemployed individuals in a year.

Any approach to determine policy rules must use one or more forecasts (or estimates) of the arrival function $\lambda(t)$ and use this as optimally as possible. Therefore, for the most part of this paper, we take $\lambda(t)$ as given (density forecasts are also possible, and discussed in Appendix F.4.1).

Based on the state $s$, the policy maker makes a decision on whether to provide the treatment, i.e., job-training ($a = 1$) or not ($a = 0$). The decision is governed by the policy rule, $\pi(a|s) \in [0, 1]$, that specifies the probability of choosing action $a$ given state $s$. The set of all policy classes under consideration is denoted by $\Pi$. The policies are indexed by some (possibly infinite dimensional) parameter $\theta \in \Theta$. We discuss policy classes in Section 2.3.

Once an action $a$ is chosen, the policy maker receives an instantaneous utility $Y(a, z, t)/N$. We typically have $Y(a, z, t) = Y(a) + f(z, t)$, where $Y(\cdot)$ is the potential outcome of an individual under treatment $a$, and the function $f(\cdot)$ allows for the fact that the treatment costs may vary with $z,t$. The utilities are divided by $N$ to rescale them to a 'per-person' quantity, which will be convenient when we take $N \to \infty$.

We focus on utilitarian social welfare criteria: the welfare from choosing actions $\{a_i\}_{i=1}^{\infty}$ to the sequence of individuals $i = 1, 2, \ldots$, arriving at times $\{t_i\}_{i=1}^{\infty}$, is given by $\frac{1}{N}\sum_{i=1}^{\infty} e^{-\beta t_i}(Y_i(a_i, z_i, t_i) - Y_i(0, z_i, t_i))$. Note that we always define welfare relative to not treating anyone. This is a convenient normalization that ensures welfare is 0 if the budget is 0.

***Modeling assumptions.*** We make two key modeling assumptions. The first is that the chosen policy $\pi$ does not affect the arrival rates $\lambda(t)$. We argue that this assumption is justified in our setup for at least three reasons. First, individuals typically have limited control over the timing of their unemployment, as it is predominantly shaped by external factors such as macroeconomic conditions, industry-wide trends, or firm-specific conditions. Additionally, it is unlikely that individuals would deliberately delay receiving unemployment benefits to increase their chances of accessing job training, since the benefits of such training are uncertain and comparatively modest, whereas the immediate financial loss from delaying unemployment support is significant.[10] Second, unemployed individuals typically lack detailed knowledge of the policy function, as is currently the case with KROM. Finally, one can design policy rules, such as stationary policies, that are robust to strategic behavior.

Our second key modeling assumption is that the covariates and potential outcomes for each individual are drawn from a joint distribution $F$ that is independent of $\pi, z$ (however, we allow $F$ to vary by month in our empirical illustration in Section 5). This supposes that the chosen policy does not impact the covariate distribution of the unemployed. In the context of job training, it is possible that the provision of training could affect the length of subsequent employment spells, influencing how soon individuals return to being unemployed. Arguably, however, this effect is small, as it is unlikely for individuals to become unemployed twice within a calendar year (which is the duration of our period). Moreover, it seems plausible that the covariate distribution of the unemployed is primarily influenced by broad economic factors, such as the unemployment rate, rather than by the chosen policy.

[10]For instance, the JTPA study estimates that the average treatment effect of job training results in an increase of $1,500 in cumulative earnings over 30 months, which is often outweighed by the earnings lost from deferring unemployment benefits for just one month.

Apart from the above, our setup also implicitly assumes that a treatment decision is made immediately after arrival, but this is not strictly necessary: as outlined in Section 3, we can allow the policy maker to pool candidates into small batches and then decide on a policy rule simultaneously for each batch. Compared to the actual implementation of KROM, we do make a slight simplification by supposing that rejected candidates are not re-evaluated. Under KROM, individuals rejected for job training can be reconsidered if they remain unemployed after 4 months. Summary data obtained from KROM shows that this is the case for 35% of those initially evaluated. While we abstract away from this issue for simplicity, it is possible to extend our setup to accommodate this scenario. This can be done by modeling the situation as a queue, where the policy maker has three available actions: accept, reject outright, or place in a queue. Individuals in the queue are reconsidered for eligibility if they remain unemployed after a specific period (in this case, 4 months).

***Evolution of budget.*** Let $z_0$ denote the budget at the start of the program, measured in units of $N$ (i.e., the actual budget is $z_0 N$). In practice, we would determine $z_0$ by taking the total budget and dividing it by $N$; it can then be interpreted as the amount of money available 'per-person' in the population. Conditional on $(a,s)$, the change to the budget $z$ is determined by the law of motion $z' - z = \frac{G_a(s)}{N}$, where $G_a(\cdot); a \in \{0,1\}$ may be a known function or a random variable with a known dis-tribution conditional on $s$. The division by $N$ ensures that we measure changes to budget in a relative manner, in units of $N$. Intuitively, it captures the idea that the relative change to budget from treating a single individual is quite small. If the cost of the treatment, $c$, were uniform across individuals, we would have

$$G_a(s) = -c, \text{ if } a = 1 \text{ and } z > 0. \quad (1)$$

In the Swedish context, $zN$ represents the total budget available for unemployment insurance, while $G_a(s)$ denotes the payment to each new claimant. This payment depends on the individual's prior earnings-up to a maximum of 80% of previous earnings or the minimum wage, whichever is higher-and the length of their unemployment spell. These payments vary based on individual covariates $x$. Since the unemployment duration is unknown at the time of treatment assignment, the policy rule cannot depend on it. As our framework accommodates restricted policy classes, optimal policies can still be estimated as long as we have access to observational data on unemployment durations.

***Terminal condition.*** To complete the model, we need to specify a terminal condition. In the context of KROM, the program ends either upon depletion of the budget or at the year's close. Specifically, there is no borrowing if the budget is insufficient; the program can become stricter if the budget depletes faster than expected. Indeed, there have been instances where no newly unemployed individuals were able to enter KROM because the budget was exhausted. Additionally, the budget operates under a ”use-it-or-lose-it” requirement, meaning any unused funds are returned to the government. While this might appear sub-optimal initially, Khalil et al. (2019) document that such rules are common, likely because their ex-ante positive incentive effects can outweigh ex-post inefficiencies. As discussed in Section 3.1, our methodology also allows other types of terminal conditions.

## 2.1 Other Examples

While we use job-training in Sweden as a running example, our framework applies broadly to other settings as well. One possible application is to 'sunsetting' foundations, discussed in detail in Appendix D. Another example involves a hospital with a small intensive care unit. If a life-threatening accident occurs nearby, the hospital can either offer one of its limited beds or refer the patient to a larger hospital. Referring the patient involves a risk of injury or death due to the delay. Conversely, if the ICU is full, the next patient in urgent need would need to be sent to another hospital, potentially facing a higher risk of death. This scenario meets all our modeling assumptions: patients arrive sequentially, decisions are made immediately, and re-application is not an option as the patients require immediate treatment. Also, ICUs have strict capacity constraints due to their use of specialized equipment, and hospitals can adjust their emergency admission rules based on their current capacity. This environment is thus well-suited for our model.

## 2.2 Representing Welfare using Value Functions

Define $r(s,a) = E[Y(a,z,t)|s]$, where the expectation is taken under the distribution $F$, as the expected utility for the policy maker from choosing action $a$ for an individual when the state variable is $s$. Given our relative welfare criterion, it will be convenient to normalize $r(s,0) = 0$, and set $r(s,1) := E[Y(1,z,t) - Y(0,z,t)|s]$.

Let $V_{\theta,N}(s)$ denote the value function at state $s$ under policy $\pi_\theta(\cdot)$, defined as the expected social welfare from implementing this policy when the initial state is $s$ (recall that $\theta$ indexes the policy rules). We can characterize $V_{\theta,N}(s)$ recursively as

$$V_{\theta,N}(s) = \begin{cases} \frac{r_\theta(s,1)\pi_\theta(1|s)}{N} + E_\theta\left[e^{-\beta(t'-t)}V_{\theta,N}(s')\,|s\right] & \text{if } z > 0, t < 1 \\ 0 & \text{otherwise} \end{cases}, \tag{2}$$

where $s' \equiv (x', z', t')$ denotes the state variable for the next arrival immediately fol-lowing $s$, and $E_\theta[\cdot|s]$ is the expectation jointly over (i) the new covariate $x' \sim F$, (ii) the waiting time $t'-t \sim \text{Exponential}(N\lambda(t))$, and (iii) the randomized policy $\pi_\theta(\cdot)$, which influences the value of $z'$ via $z' = z + G_A(s)/N$; $A \sim \text{Bernoulli}(\pi_\theta(s))$. Note that we have imposed the terminal condition that $V_{\theta,N}(s) = 0$ if budget runs out, or we reach the end of the year.

Unfortunately, the dependence of $V_{\theta,N}(s)$ on $x$ makes it hard to determine its properties (for one, $V_{\theta,N}(\cdot)$ does not have a stable limit as $N \to \infty$). Furthermore, the iid nature of arrivals implies $x'$ is independent of $s$, so it is convenient to integrate this out and achieve a dimension reduction for the value function. This motivates the integrated value function $h_{\theta,N}(z,t) := E_{x\sim F}[V_{\theta,N}(x,z,t)]$. By taking expectations with respect to $E[\cdot|z,t]$ on both sides of (2), we can characterize $h_{\theta,N}(\cdot)$ recursively as

$$h_{\theta,N}(z,t) = \begin{cases} \frac{\bar{r}_\theta(z,t)}{N} + E_\theta\left[e^{-\beta(t'-t)} h_{\theta,N}(z',t') \,|z,t\right] & \text{if } z > 0, t < 1 \\ 0 & \text{otherwise} \end{cases}, \tag{3}$$

where $\bar{r}_\theta(z,t) := E_{x\sim F}[r(s,1)\pi_\theta(1|s)|z,t]$.

## 2.3 Policy Classes and Optimal Policies

Our theory is quite general and allows for a restricted class $\Pi$ of policies indexed by some (possibly infinite dimensional) parameter $\theta \in \Theta$. For computation using RL methods, however, we require $\pi_\theta(.)$ to be differentiable in $\theta$. This still allows for rich spaces of policy functions. The class of logistic functions is convenient and commonly used in RL (see Sutton and Barto, 2018 ). Let $f(s)$ denote a vector of functions of dimension $k$. The logistic policy rule takes the form

$$\pi_\theta(1|s) = \frac{\exp(\theta^\intercal f(s))}{1+\exp(\theta^\intercal f(s))}. \tag{4}$$

If we take $\|\theta\| \to \infty$, the class of logistic policies converges to the class of Generalized Eligibility Scores (Kitagawa and Tetenov, 2018), which are of the form $\mathbb{I}\{\tilde{\theta}^\intercal f(s) > 0\}$, where $\tilde{\theta} := \theta/\|\theta\|$. More broadly, this class can approximate any deterministic policy - including the first-best policy rule (i.e., the policy that maximizes $h_\pi(z_0, t_0)$ across all possible $\pi$) - with arbitrary precision, provided the dimension $k$ is sufficiently large. The logistic policy class offers distinct advantages in terms of interpretability and mitigating the risk of overfitting the data. But our framework can also accommodate more flexible policy classes like multi-layer neural networks, and we present results on using these in Appendix C.

When using RL methods, we initialize $\theta$ randomly and let the algorithm optimize over $\theta$. We found that the algorithms automatically take $\|\theta\| \to \infty$ over the course of training, meaning that the optimal policy eventually becomes deterministic (e.g., Figure A4 in Appendix C shows that the relative coefficient values stays relatively constant, but the scale increases slowly).

Note that for each policy $\pi_\theta$, we can define the integrated value function, $h_\theta(z,t)$, as in (3). The policy maker's objective is to choose $\theta^*$ that maximizes the welfare at the initial values, $(z_0, t_0)$, of $(z,t)$:

$$\theta^* = \arg\max_{\theta \in \Theta} h_{\theta,N}(z_0, t_0). \tag{5}$$

# 3 Continuous Time Approximation

The integrated value function $h_{\theta,N}(z,t)$ describes the expected welfare that could be achieved starting from state $z,t$ onwards, when employing the policy $\pi_\theta$. However, two factors make the analysis of its properties, and the subsequent derivation of regret bounds, challenging: First, the transition $z' - z$ is random as it depends on the realized covariates $x$. This implies that the second and higher order moments of $z' - z$ could affect $h_{\theta,N}(\cdot)$. Second, the high correlation between $z'$ and $z$ implies that the rewards are also serially correlated over time.

For these reasons, we take $N \to \infty$ to obtain a more tractable characterization of the integrated value function in continuous time. To this end, denote

$$\bar{G}_\theta(z,t) := E_{x\sim F}\left[G_1(s)\pi_\theta(1|s) + G_0(s)\pi_\theta(0|s)|z,t\right].$$

We can interpret $\lambda(t)\bar{G}_\theta(z,t)$ as the average change to $z$ under $\pi_\theta$ when the state variable is $(z,t)$. As $N\to\infty$, $h_{\theta,N}(z,t)$ converges to the solution of the PDE:

$$\beta h_\theta(z,t) - \lambda(t)\bar{G}_\theta(z,t)\partial_z h_\theta(z,t) - \partial_t h_\theta(z,t) - \lambda(t)\bar{r}_\theta(z,t) = 0 \text{ on } \mathscr{U}, \quad (6)$$

where $\mathscr{U}$ is the domain of the PDE (more on this below).

PDE (6) is a key contribution of this paper. The terms, $\lambda(t)\bar{r}_\theta(z,t)$ and $\lambda(t)\bar{G}_\theta(z,t)$, are the average utilities and average change to $z$ under $\pi_\theta$ when the state variable is $(z,t)$. Now, $\lambda(t)\bar{r}_\theta(z,t)$ can be interpreted as expected flow utility, while $\lambda(t)\bar{G}_\theta(z,t)\partial_z h_\theta(z,t) + \partial_t h_\theta(z,t)$ can be interpreted as the total rate of change of $h_\theta(\cdot)$ with respect to time. PDE (6) then equates the flow value of welfare, $\beta h_\theta(z,t)$, to the sum of the flow utilities and rate of change of $h_\theta(z,t)$. Thus the PDE has an asset pricing interpretation if we think of $\beta$ as the natural interest rate.

Remarkably, PDE (6) characterizes $h_\theta(\cdot)$ in terms of only the the expected (given $z,t$) rate of change, $\lambda(t)\bar{G}_\theta(z,t)$, of $z$, so e.g., the variance of $\delta(z) := z' - z$ does not matter for welfare. The reason for this is that as $N\to\infty$, the variance of $z' - z$ declines at the rate $N^{-2}$, while the expected change is of the order $N^{-1}$, so the mean dominates. Note also that the PDE depends only on the current state $s$, thus enabling us to abstract away from the time series dependence of $z$. These facts show how PDE (6) ameliorates the issues with the discrete-time formulation. We remark on a few additional properties of the PDE below:

***Large N approximation.*** Theorem 1, stated later in this section, shows that

$$\sup_{(z,t),\theta\in\Theta}\left|h_{\theta,N}(z,t) - h_\theta(z,t)\right| \le K\sqrt{\frac{1}{N}}. \quad (7)$$

for some $K<\infty$. In our running example, $N$ corresponds to the number of unemployed individuals per year, which is a large number. Hence, we expect PDE (6) to provide a very good approximation in this setting.

***General arrival processes.*** While it is natural to model arrivals as an exponential process, this is not necessary. In fact, the randomness of the arrivals becomes irrelevant for the properties of $h_\theta(z,t)$ as $N \to \infty$. Let $\Delta_t$ denote the waiting time between arrivals when the last arrival was at time $t$. We require $\lim_{N\to\infty} \mathbb{E}[N\Delta_t] = \lambda(t)$ and $\lim_{N\to\infty} \text{Var}[N\Delta_t] = 0$ at each $t$ (note that these are satisfied by the exponential distribution with parameter $\lambda(t)N$). Any arrival process that satisfies these conditions will lead to PDE (6).

***Batching.*** In practice, the policy maker may want to group arrivals into small batches and apply the same policy rule on each batch. Let $B$ denote the batch size. Then, PDE (6) continues to apply as long as $B/N \to 0$, so applying our proposed policy rules with such small batches delivers the same welfare as not using batching. We can also account for larger batch sizes, $B/N \to c \in (0,1)$, by using piece-wise constant policy rules (see Section 5). These are policy rules that remain unchanged for some specific time intervals.

## 3.1 Boundary Conditions

To complete our description of the PDE, we need to specify a boundary condition that determines its domain $\mathscr{U}$. Different boundary conditions correspond to different institutional constraints and terminal conditions.

Suppose that the policy maker faces only a budget constraint (the program ends when budget runs out), but no time constraint (the program continues indefinitely until the budget runs out). This corresponds to a boundary condition of the form $h_\theta(0,t) = 0 \; \forall\, t$. In our job training example, there is actually also a time constraint, so the boundary condition is specified as: $h_\theta(0,t) = 0 \; \forall\, t$ and $h_\theta(z,T) = 0 \; \forall\, z$.

Generalizing things a bit, we allow for Dirichlet boundary conditions of the form

$$h_\theta(z,t) = 0 \text{ on } \Gamma \equiv \{\{\underline{z}\} \times [t_0, T]\} \cup \{(\underline{z}, \infty) \times \{T\}\}. \tag{8}$$

Here, $t_0$ is initial time, $\underline{z}$ is the lower bound on budget (0 in our empirical example) and $T$ is the time constraint (1 year in our example). The corresponding domain of PDE (6) is $\mathscr{U} \equiv (\underline{z}, \infty) \times [t_0, T)$.

The boundary condition does not impose any restrictions on $\pi_\theta(\cdot)$. It is possible to have $\lim_{(z,t)\to\Gamma} \pi_\theta(1|x,z,t) \neq 0$. In fact, it is generally optimal to let $\lim_{(z,t)\to\Gamma} \pi_\theta(1|x,z,t) = 1$ to treat as many individuals as possible before the constraints are hit. The policy rule $\pi_\theta(a|\cdot)$ is only meant to be applied in the interior of $\mathscr{U}$. On the boundary, the terminal conditions take precedence. As the program terminates when the boundary is reached, there may be an effective discontinuity in the policy makers' actions before and after the boundary. However, this does not affect the continuity of $h_\theta(z,t)$.

Appendix E describes two other types of boundary conditions: the periodic and Neumann boundary conditions. The first is designed to accommodate seasonality. The second accommodates scenarios where dynamics at the boundary are different from those in the interior. This is the case, e.g., under a borrowing constraint, where hitting the borrowing limit implies a different transition law for $z$ as compared to when the borrowing limit is not binding.

## 3.2 Informal Derivation of PDE (6) in a simplified setting

Because of the importance of PDE (6), we provide an intuitive derivation of it in a simplified setting with a budget constraint $h_\theta(0,t) = 0 \ \forall\, t$, and where $\lambda(t) = 1$ for all $t$, $r(s,a)$ is independent of $z,t$ and the budget evolves as $G_a(s) = -\mathbb{I}(a=1)$. In short, this corresponds to a setting where the policy maker starts with an initial allocation of money that can be distributed among incoming arrivals, the treatment costs are constant and the program ends when the budget runs out. In this setting, $t$ is no longer a state variable. We can represent $V_{\theta,N}(x,z)$ in recursive form as

$$V_{\theta,N}(x,z)=\frac{r(x,1)}{N}\pi_\theta(1|x,z)+\left(1-\frac{\beta}{N}\right)E_{x'\sim F}\left[V_{\theta,N}\left(x',z-\frac{1}{N}\right)\pi_\theta(1|x,z)+V_{\theta,N}(x',z)\pi_\theta(0|x,z)\right]$$
$$\text{for } z\geq 1/N, \quad \text{and} \quad V_{\theta,N}(x,z)=0, \quad \text{otherwise.}$$

In deriving the above, we used $E[e^{-\beta\Delta t}]=1-\frac{\tilde{\beta}}{N}$, where $\tilde{\beta}=\beta+O(N^{-1})$, and replaced $\tilde{\beta}$ with $\beta$ to simplify notation. Define $\bar{\pi}_\theta(a|z)=E_{x\sim F}[\pi_\theta(a|x,z)]$ and $\bar{r}_\theta(z)=E_{x\sim F}[r(x,1)\pi_\theta(1|x,z)]$. Then, taking expectations with respect to $x\sim F$ on both sides of the recursion for $V_{\theta,N}(\cdot)$ gives

$$h_{\theta,N}(z)=\frac{\bar{r}_\theta(z)}{N}+\left(1-\frac{\beta}{N}\right)\left\{h_{\theta,N}\left(z-\frac{1}{N}\right)\bar{\pi}_\theta(1|z)+h_{\theta,N}(z)\bar{\pi}_\theta(0|z)\right\} \tag{9}$$
$$\text{for } z\geq 1/N, \quad \text{and} \quad h_{\theta,N}(z)=0, \quad \text{otherwise.}$$

Now subtract $h_{\theta,N}(z)$ from both sides, take a second order Taylor expansion of $h_{\theta,N}(z-1/N)$, and multiply the entire expression by $N$ (the derivation is only for intuition, in reality, $h_\theta(\cdot)$ need not be twice differentiable). Then, in the limit as $N\to\infty$, we end up with the following Ordinary Differential Equation (ODE) for the evolution of $h_\theta(.)$:[11]

$$\beta h_\theta(z)=\bar{r}_\theta(z)-\bar{\pi}_\theta(1|z)\partial_z h_\theta(z), \qquad h_\theta(0)=0. \tag{10}$$

This is just a special case of PDE (6).

Suppose we restrict attention to stationary policies that do not change with $z$. Then, $\bar{\pi}_\theta(1|z)\equiv\bar{\pi}_\theta, \bar{r}_\theta(z)\equiv\bar{r}_\theta$ and we can obtain an explicit solution to (10) as $h_\theta(z)=\beta^{-1}\bar{r}_\theta\left(1-e^{-\beta z/\bar{\pi}_\theta}\right)$. The optimal policy $\theta^*$ solves $\max_\theta h_\theta(z_0)$, where $z_0$ is the initial budget. On the other hand, the EWM rule with budget constraints corresponds to the population problem: $\max_\theta \bar{r}_\theta$ s.t. $\bar{\pi}_\theta\leq\alpha$ for some given $\alpha\in[0,1]$. Clearly the two optimization problems are not equivalent. One can of course solve the EWM problem for each $\alpha$ and choose the one that results in the best value for $h_\theta(z_0)$, but this requires knowledge of $h_\theta(\cdot)$ provided in this paper (and this derivation is only applicable to stationary policies).

[11] Sufficient conditions for a unique solution to (10) are provided in Appendix F.5.

## 3.3 Formal Properties of PDE (6)

We now provide formal results on the existence and uniqueness of PDE (6), along with rates on the quality of the continuous time approximation.

For semi-linear PDEs of the form (6), it is well known that a classical solution (i.e., a solution $h_\theta(z,t)$ that is continuously differentiable) does not exist. We instead employ the weak solution concept of a viscosity solution (Crandall and Lions, 1983). This is a common solution concept for equations of the HJB form; see Crandall *et al* (1992) for a user's guide, and Achdou *et al* (2017) for a useful discussion. The following ensures existence of a unique, continuous viscosity solution to (6):

**Assumption 1.** *(i)* $\bar{G}_\theta(z,t)$, $\bar{r}_\theta(z,t)$ *are Lipschitz continuous uniformly over* $\theta \in \Theta$.

*(ii)* $\lambda(t)$ *is bounded, Lipschitz continuous, and bounded away from* 0.

*(iii) There exists* $M < \infty$ *such that* $|\bar{r}_\theta(z,t)|, |\bar{G}_\theta(z,t)| \leq M$ *for all* $\theta, z, t$.

*(iv) There exists* $\delta > 0$ *s.t* $\bar{G}_\theta(z,t) < -\delta$. *This is only required for the case of the boundary condition with* $\underline{z} > -\infty$.

The sole purpose of Assumption 1(i) is to ensure $h_\theta(z,t)$ exists and is uniform Lipschitz continuous. However, in certain contexts, Assumption 1(i) can be relaxed while still maintaining these properties. For instance, depending on the boundary condition, we can allow $\bar{G}_\theta(z,t), \bar{r}_\theta(z,t)$ to be discontinuous in one of the arguments, see Appendix F.5. For ODE (10), just integrability of $\bar{r}_\theta(z), \bar{\pi}_\theta(1|z)$ is sufficient. We do not address the question of minimal sufficient conditions, but make do with Assumption 1(i) for simplicity. Appendix F.5 provides primitive conditions for verifying Assumption 1(i) under the soft-max policy class (4). Briefly, we require either $||\theta||$ be bounded away from $\infty$, or at least one of the covariates be continuous. With purely discrete covariates and $||\theta|| \to \infty$, $\bar{G}_\theta(z,t)$ and $\bar{r}_\theta(z,t)$ will typically be discontinuous, unless the policies are stationary (i.e., they only depend on $x$).[12]

[12]Note that $G_a(s)$ is discontinuous in $a$. But this does not affect the continuity of $\bar{G}_\theta(z,t)$ since

Assumption 1(ii) implies the arrival rates vary smoothly with $t$ and are bounded away from 0. Assumption 1(iii) is a mild requirement ensuring the expected rewards and changes to $z$ are bounded. In fact, we later impose a stronger assumption (Assumption 2(i)) that nests this assumption.

Assumption 1(iv) is required in situations where the boundary condition is determined partly by $z$. In these settings, $z$ plays the role of time, so we need the coefficient, $\bar{G}_\theta(z,t)$, multiplying $\partial_z h_\theta$ in PDE (6) to be non-singular. Typically, we already have $\bar{G}_\theta(z,t) < 0$ in such settings (e.g., the budget can only be depleted), so Assumption 2(ii) additionally requires $\bar{G}_\theta(z,t)$ to be strictly bounded away from 0. This is a mild restriction: if there exist some people that benefit from treatment and $\beta > 0$, it is a dominant strategy to always treat some fraction of the population. We also note that this assumption is not required for showing existence and uniqueness of a viscosity solution to PDE (6), but only for obtaining a bound on the quality of the continuous time approximation.

**Lemma 1.** *Suppose Assumptions 1(i)-(iii) hold. Then for each $\theta$, there exists a unique viscosity solution $h_\theta(z,t)$ to (6) under the boundary conditions (8).*

We further obtain the following result on the quality of the continuous time approximation as a function of the population size:

**Theorem 1.** *Under Assumption 1, there exists $C < \infty$ independent of $(\theta, z, t)$ such that, under the Dirichlet boundary condition (8),*

$$\sup_{(z,t)\in\bar{\mathscr{U}},\theta\in\Theta} \left| h_{\theta,N}(z,t) - h_\theta(z,t) \right| \leq C\sqrt{\frac{1}{N}}.$$

Theorem 1 is a special case of the more general Theorem 2 described in Section 4, so we omit its proof (specifically, $N$ plays the same role as $b_n$ in Theorem 2). The

$G_1(s)$ and $G_0(s)$ enter separately in the definition of $\bar{G}_\theta(z,t)$.

precise value of $C$ depends on the Lipschitz constants and the quantities $M, \delta$ from Assumption 1.

# 4 Sample Data and Regret Bounds

## 4.1 Estimation Procedure

The unknown quantities in the policy maker's problem are the rewards, $r(s,a)$, from treating candidates and the distribution of arrivals, i.e., the population distribution $F$ introduced in Section 2. We suppose that the policy maker has access to RCT or observational data that allows one to obtain estimates $F_n$ and $\hat{r}(s,a)$ of $F$ and $r(s,a)$. In the context of our empirical example, this corresponds to a pre-existing RCT on the effect of job-training on wages. For the most part, we consider the setting where $r(s,a) \equiv r(x,a)$ i.e., the rewards only depend on the candidate's characteristics, $x$, but not on $z,t$.[13] The dataset is then a random sample of size $n$ in the format $\{Y_i, W_i, X_i\}_{i=1}^n$. Here, $W_i$ is a dummy indicating treatment, $X_i$ are the covariates, and $Y_i = W_i Y_i(1) + (1 - W_i) Y_i(0)$, where $Y_i(1), Y_i(0)$ denote the potential outcomes for treatment and no treatment. We assume there was no selection into the RCT. Formally, we suppose that our sample is a representative draw from a joint population distribution over $(Y_i(1), Y_i(0), W_i, X_i)$, i.e., each $(Y_i(1), Y_i(0), X_i)$ is a random draw from $F$. Moreover, we assume ignorability, i.e., $(Y_i(0), Y_i(1)) \perp\!\!\!\perp W_i | X_i$, which is trivially satisfied for an RCT. For ease of notation we denote the population distribution of $(Y_i(1), Y_i(0), W_i, X_i)$ by $F$ as well. The empirical distribution, $F_n$, of these observations is thus a good proxy for $F$.

In practice, the population distribution of covariates may differ from the observed data distribution. For example, we could have time-varying covariate shifts. If the population distribution is known, which is conceivable in the case of unem-ployment data, it is possible to to use time-varying importance weights to correct for this difference in covariate distributions. Our empirical illustration in Section 5 suggests one way to do this using different arrival distributions for each month.

[13] An extension to the scenario where $r(s,a) = r(x,a) + f(z,t)$, with $f(\cdot)$ denoting a known cost of treatment, is straightforward.

We recommend a doubly robust method to estimate the rewards. Let $\mu(x,w) := E[Y(w)|X = x]$ denote the conditional expectations for $w \in \{0,1\}$, and $p(x) = E[W|X = x]$, the propensity score. We can then estimate $r(x,1)$ over $x \in \text{support}(F_n)$:

$$\hat{r}(X_i,1) = \hat{\mu}(X_i,1) - \hat{\mu}(X_i,0) + (2W_i - 1)\frac{Y_i - \hat{\mu}(X_i,W_i)}{W_i\hat{p}(X_i) + (1-W_i)(1-\hat{p}(X_i))}, \quad (11)$$

where $\hat{\mu}(x,w)$ and $\hat{p}(x)$ are non-parametric estimates of $\mu(x,w)$ and $p(x)$.

Using the estimates $\hat{r}$ and $F_n$, we can set up a sample version of the dynamic optimization problem for determining the optimal policy. First, we estimate the quantities $\bar{r}_\theta(z,t), \bar{G}_\theta(z,t)$ as:

$$\hat{r}_\theta(z,t) := E_{x\sim F_n}[\hat{r}(s,1)\pi_\theta(1|x,z,t)], \text{ and}$$

$$\hat{G}_\theta(z,t) := E_{x\sim F_n}\left[G_1(x,z,t)\pi_\theta(1|x,z,t) + G_0(x,z,t)\pi_\theta(0|x,z,t)\right].$$

Based on the above we can construct a sample version of PDE (6) as

$$\beta\hat{h}_\theta(z,t) - \lambda(t)\hat{G}_\theta(z,t)\partial_z\hat{h}_\theta(z,t) - \partial_t\hat{h}_\theta(z,t) - \lambda(t)\hat{r}_\theta(z,t) = 0 \text{ on } \mathscr{U}, \quad (12)$$

with the boundary condition (8). However, due to technical reasons, the existence of a unique solution to PDE (12) is not guaranteed without additional assumptions. As a result, it is helpful to view PDE (12) as a heuristic tool. In practice, we rely on a sample-based approximation of the (discrete) integrated-value function (3), detailed below, which circumvents these existence issues, but nevertheless approximates PDE (6) as $n \to \infty$.

The approximation discretizes arrivals so that the difference between arrival times is specified as (the 'prime' notation denotes one-step ahead quantities)

$$t' - t \sim \min\{\text{Exponential}(\lambda(t)b_n), T - t\}, \quad (13)$$

with the censoring at $T$ used as a device to impose a finite horizon boundary condi-tion. Here, $b_n$ plays a similar role as the population size $N$ in Section 2, except that it is now a computational choice. A large value of $b_n$ provides a closer approximation to the continuous time limit at the expense of higher computational costs. In practice, following our theoretical results, we recommend choosing $b_n = O(n)$.

Continuing with the approximation scheme, the law of motion for $z$ is given by

$$z' = \max\left\{z + b_n^{-1} G_a(s), \underline{z}\right\}. \tag{14}$$

The rest of environment is the same as in Section 2. For this discretized setup, define $\tilde{h}_\theta(z,t)$ as the integrated value function at the state $(z,t)$, when an individual *happens* to arrive at that state. This can be obtained as the fixed point to the following dynamic optimization problem:

$$\tilde{h}_\theta(z,t) = \begin{cases} \frac{\hat{r}_\theta(z,t)}{b_n} + E_{n,\theta}\left[e^{-\beta(t'-t)}\tilde{h}_\theta(z',t')\,|z,t\right] & \\ 0 & \text{for } (z,t) \in \Gamma \end{cases}, \text{ where} \tag{15}$$

$$E_{n,\theta}\left[e^{-\beta(t'-t)} f(z',t')\,|z,t\right] := \int e^{-\beta\frac{\omega}{b_n}} E_{x\sim F_n}\left[f\left(\max\left\{z + \frac{G_1(x,t,z)}{b_n}, \underline{z}\right\}, t + \frac{\omega}{b_n}\right)\pi_\theta(1|x,z,t)\right.$$
$$\left. + f\left(\max\left\{z + \frac{G_0(x,t,z)}{b_n}, \underline{z}\right\}, t + \frac{\omega}{b_n}\right)\pi_\theta(0|x,z,t)\right] g_{\lambda(t)}(\omega)d\omega$$

for any function $f$, and $g_{\lambda(t)}(\omega)$ denotes the right censored exponential distribution with parameter $\lambda(t)$ and censoring at $\omega = b_n(T-t)$. The usual contraction mapping argument ensures that $\tilde{h}_\theta$ always exists as long as $T < \infty$ or $\beta < 1$.

To obtain estimates for $\theta$, we can use $\tilde{h}_\theta$ as the feasible sample counterpart of $h_\theta$, and solve the sample version of the policy maker's problem:

$$\tilde{\theta} = \arg\max_{\theta \in \Theta} \tilde{h}_\theta(z_0, t_0). \tag{16}$$

In simple instances, see Section 5.2, equation (16) can be solved exactly using PDE methods. For all other cases, we use RL.

## 4.2 Assumptions for Regret Bounds

We are interested in obtaining a probabilistic bound on the regret, $h_{\theta^*}(z_0,t_0) - h_{\tilde{\theta}}(z_0,t_0)$, from employing $\pi_{\tilde{\theta}}$ as the policy rule. To this end, in addition to Assumption 1, we impose:

**Assumption 2.** *(i) There exists $M < \infty$ such that $|Y(a,z,t)|, |G_a(s)| \leq M \; \forall \; (a,s)$.*

*(ii) (Complexity of the policy function space) The collection of functions*[14]

$$\mathscr{I} \equiv \left\{\pi_\theta(1|\cdot,z,t) : (z,t) \in \bar{\mathscr{U}}, \theta \in \Theta\right\}$$

*indexed by $z,t$ and $\theta$, is a VC-subgraph class with finite VC index $v_1$. Furthermore, for each $a = 0,1$, the collection of functions*

$$\mathscr{G}_a \equiv \left\{\pi_\theta(a|\cdot,z,t)G_a(\cdot,z,t) : (z,t) \in \bar{\mathscr{U}}, \theta \in \Theta\right\}$$

*is also a VC-subgraph class with finite VC index $v_2$. Let $v := \max\{v_1, v_2\}$.*

Assumption 2(i) is a boundedness assumption, imposed mainly for ease of deriving the theoretical results (see, e.g., Kitagawa and Tetenov, 2018). In practice, budget changes are almost certainly bounded by a specific finite value.

The term $v$ in Assumption 2(ii) is a measure of the complexity of the policy class. Since the policy class is chosen by the policy maker, this only puts some mild restrictions on that choice. Often, $G_a(s)$ is independent of $x$, as in equation (1), in which case $v_1 = v_2$. In calculating the VC dimension, we treat $z,t$ as an index to the functions $\pi_\theta(1|\cdot,z,t)$, similarly to $\theta$. This is intuitive, since how rapidly the policy rules change with budget and time is also a measure of their complexity.

The next set of assumptions relate to the properties of the observational data from which we estimate $\hat{r}(s,a)$; see Section 4 for the terminology. Here, we con-sider the case where $(z,t)$ does not impact the potential outcomes, with extensions discussed in Appendix F.2.

[14]Except for the Neumann boundary conditions, the domain of $(z,t)$ in the definitions of $\mathscr{I}$ and $\mathscr{G}_a$ can be taken to be $\mathscr{U}$ instead of $\bar{\mathscr{U}}$. For the Neumann boundary conditions, we require $\mathscr{I}$ and $\mathscr{G}_a$ to be defined by continuously extending the 'interior' values of $\pi_\theta(1|\cdot)$ and $G_a(\cdot)$ to the boundary, even though the actual policy and law of motion at the boundary may be quite different.

**Assumption 3.** *(i)* $Y(a,z,t) \equiv Y(a)$.

*(ii)* $\{Y_i(1), Y_i(0), W_i, X_i\}_{i=1}^n$ *are an iid draw from the distribution* $F$.

*(iii) (Selection on observables)* $(Y_i(1), Y_i(0)) \perp\!\!\!\perp W_i | X_i$.

*(iv) There exists* $\kappa > 0$ *such that* $p(x) \in [\kappa, 1-\kappa]$ *for all* $x \in$ *support*$(F)$.

Assumption 3(ii) assumes the observed data is representative of the population. Assumption 3(iii) requires the observational data to satisfy ignorability. Assumption 3(iv) ensures the propensity scores are bounded away from 0 and 1. All three are trivially satisfied for RCT data.

Under Assumptions 2 and 3, there exist many different estimates of the rewards, $\hat{r}(x,1)$, that are consistent for $r(x,1)$. In this paper, we use the doubly robust estimates given in (11). We assume that the estimates $\hat{\mu}(X_i,w), \hat{p}(X_i)$ of $\mu(X_i,w), p(X_i)$ are obtained through cross-fitting (see Chernozhukov *et al*, 2018, or Athey and Wager, 2021 for a description). Specifically, we choose some non-parametric procedures, $\tilde{\mu}(x,w), \tilde{p}(x)$ for estimating $\mu(x,w), p(x)$, and apply cross-fitting to weaken the assumptions required and reduce bias. We impose the following high-level conditions on $\tilde{\mu}(x,w), \tilde{p}(x)$:

**Assumption 4.** *(i) (Sup convergence) There exists* $c > 0$ *such that, for* $w \in \{0,1\}$,

$$\sup_x |\tilde{\mu}(x,w) - \mu(x,w)| = O_p(n^{-c}), \quad \sup_x |\tilde{p}(x) - p(x)| = O_p(n^{-c}).$$

*(ii) ($L^2$ convergence) There exists some* $\xi > 1/2$ *such that*

$$E\left[|\tilde{\mu}(x,w) - \mu(x,w)|^2\right] \lesssim n^{-\xi}, \qquad E\left[|\tilde{p}(x) - p(x)|^2\right] \lesssim n^{-\xi}.$$

Assumption 4 is taken from Athey and Wager (2021). The requirements imposed are weak and satisfied by almost all non-parametric procedures such as series estimation, LASSO, random forests etc (using appropriate tuning parameters).

## 4.3 Regret Bounds

We bound the regret, $h_{\theta^*}(z_0,t_0) - h_{\tilde{\theta}}(z_0,t_0)$, by bounding the maximal difference between the integrated value functions, i.e., $\sup_{(z,t)\in\bar{\mathscr{U}},\theta\in\Theta} |\tilde{h}_\theta(z,t) - h_\theta(z,t)|$. The latter suffices since (see, e.g., Kitagawa and Tetenov, 2018)

$$h_{\theta^*}(z_0,t_0) - h_{\tilde{\theta}}(z_0,t_0) \leq 2 \sup_{(z,t)\in\bar{\mathscr{U}},\theta\in\Theta} |\tilde{h}_\theta(z,t) - h_\theta(z,t)|.$$

As computing $\tilde{\theta}$ requires choosing a 'approximation' factor $b_n$, we characterize the numerical error resulting from any sequence $b_n \to \infty$.

**Theorem 2.** *Suppose that Assumptions 1-4 hold and $\beta > 0$. Then, with probability approaching one, there exists $K < \infty$ independent of $(\theta, z, t)$ such that*

$$\sup_{(z,t)\in\bar{\mathscr{U}},\theta\in\Theta} \left|\tilde{h}_\theta(z,t) - h_\theta(z,t)\right| \leq K\left(\sqrt{\frac{v}{n}} + \sqrt{\frac{1}{b_n}}\right),$$

*under the Dirichlet boundary condition (8).*

The term $b_n^{-1/2}$ in Theorem 2 represents the approximation error. By setting $b_n \propto n$, this error matches the order of statistical regret. However, the theorem does not account for computational error from using RL algorithms, although PDE methods can provide exact solutions in simple setups as we illustrate in Section 5.1. Even if the global optimum is not found, Theorem 2 remains useful for policy comparisons: if the empirical welfare of the estimated policy $\pi_{\tilde{\theta}}$ exceeds that of another policy $\pi_\theta$, the regret from using $\pi_{\tilde{\theta}}$ over $\pi_\theta$ is bounded by $1/\sqrt{n}$.

The continuous time approximation provides some intuition as to why the regret rate of $n^{-1/2}$ is the same as in the static case: when the rate of arrivals goes to infinity, the flow utility and the rate of change of budget becomes predictable as the stochastic fluctuations are averaged out in every (arbitrarily small) time interval. Consequently, only $\bar{G}_\theta(z,t)$ and $\bar{r}_\theta(z,t)$, i.e., the expected flow utility and the expected rate of change of budget matter for welfare. By Kitagawa and Tetenov (2018) and Athey and Wager (2021), $\bar{G}_\theta(z,t), \bar{r}_\theta(z,t)$ and $\hat{G}_\theta(z,t), \hat{r}_\theta(z,t)$, are uniformly close. This suggests that the welfares in the population and sample settings, determined as the solutions to PDEs (6) and (12), should be uniformly close as well, which we verify using the theory of viscosity solutions.[15] The $n^{-1/2}$ rate for the regret cannot be improved, since Kitagawa and Tetenov (2018) show that this rate is optimal in the static case.

Theorem E.1 in Appendix E describes extensions to periodic and Neumann boundary conditions. It also covers the case of $\beta = 0$.

# 5 Empirical Application

In this section, we empirically examine two dynamic decision problems related to the provision of job training. Additional applications can be found in Appendix D.

We use data from the JTPA study which randomized unemployed job center applicants into receiving job training or not.[16] This dataset allows us to compare our findings with Kitagawa and Tetenov (2018), who examine the simultaneous allocation problem using this data, while also enabling us to build a simulated environment closely mirroring the KROM setting in Sweden—a useful approach given the lack of RCT data in the Swedish context. We follow the sample selection strategy of Kitagawa and Tetenov (2018), resulting in 9223 observations. The data contains baseline information about the participants and their subsequent earnings over 30 months.

We consider two different settings for the provision of job-training. In Setting 1, we impose a finite budget without a time constraint to demonstrate three advantages of our approach. First, we can reach a solution even if there is no inherently predictable number of candidates. The algorithm trades off the need for selectivity against discounting future rewards. Methods devised for simultaneous allocation are not applicable as they rely on a fixed number of candidates. Second, we can obtain both the optimal stationary and non-stationary policy and analyze the welfare cost of choosing the former. Third, we illustrate how to compute solutions analytically in such simple settings. The more extensive Setting 2 is closely tailored to the institutional details of Sweden's KROM system. Both budget and time are constrained. We compare our policy rules with a budget-constrained (static) EWM benchmark to demonstrate the welfare cost of ignoring dynamics. We also use this computationally demanding second setting to introduce our RL approach.

[15] A technical complication here is that a solution to the empirical PDE (12) may not exist. For this reason, we work with the solution to the approximation scheme (15).

[16] Local centers could choose to provide training, job-search assistance, or other support. Following Kitagawa and Tetenov (2018), we consolidate all support forms as 'job training'.

## 5.1 Setting 1: Constrained Budget, Unconstrained Time

As mentioned earlier, we assume constant arrival rates and no time constraints for this setting. In Setting 2, we extend this by incorporating time-varying arrival rates and the time constraint of a fiscal year, which are both more representative of real-world job training programs. However, the current environment does mirror some real-world situations, such as non-perpetual foundations. In Appendix D, we provide an empirical application involving such a foundation.[17]

***Setup.*** The policymaker begins with an initial budget $z_0$, normalized to represent the fraction of expected annual arrivals that can be treated. For example, if $z_0 = 0.25$ and 5000 candidates are expected, the budget allows treatment for up to 1250 individuals, with the cost of training $c$ set to $c = 1/5000$. For each arrival, the policymaker decides whether to offer job training based on the remaining budget and the individual's characteristics. Training is free for individuals but costly to the policymaker. The reward, i.e., the outcome from treatment assignment, is the wage

[17] The lack of a time constraint means no pre-defined duration, not an infinite one. In our application in Appendix D, the non-stationary policy's budget lasts just over eight years (using a budget for 25% of arrivals, 5% administrative costs, and a discount factor of -ln(0.9)).

gain from treatment, with wage gains under no training normalized to 0.

***Solving the Problem.*** We obtain reward estimates $\hat{r}(x,1)$ from a cross-fitted doubly robust procedure, where we use OLS to estimate the conditional means, $\mu(x,a)$, and the propensity score is 2/3, as set by the RCT.

We consider two policy classes: (A) a stationary one of the form $\pi_\theta^A(1|s) = \mathbb{I}\{\theta^\intercal x \geq 0\}$, and (B) a non-stationary one $\pi_\theta^B(1|s) = \mathbb{I}\{f(z)^\intercal x \geq 0\}$, where $x = (1, \text{age, education, previous earnings})$ and the functions $f$ are allowed to vary unrestrictedly with budget ($z$). We compute the optimal stationary and non-stationary policies using piecewise constant policy approximations (see Krylov, 1999). Intuitively, we solve a sequence of allocation problems, starting with the case where the budget allows for the treatment of only a single individual. We then work backward, incrementally adding budget, until we solve the problem for the full initial budget. To achieve this, we approximate $f(\cdot)$ with functions that stay constant within a set of grid points $\{0, \Delta z, 2\Delta z, \dots, z_0\}$ for budget. The values of $f(z)$ within each grid are computed recursively, starting from $z = 0$. By letting $\Delta z \to 0$, we can approximate the optimal $f$ arbitrarily well. We use $\Delta z = 1/5000$ (halving $\Delta z$ barely affects the results, see Appendix C). These ODE techniques are feasible due to the single state variable, $z$, in $h_\theta(\cdot)$. With more state variables, this method quickly becomes intractable and we recommend employing RL algorithms.

***Results and Discussion.*** The policy rules select a group of people for treatment given each value of (initial) budget. Figure A3 in Appendix C illustrates how the coefficients of the optimal policy change with budget. For instance, for a budget of 0.5, the coefficients for age and education are negative, implying that the non-stationary policy selects young, low-educated individuals. The selected individuals comprise only about 12% of the sample (see Figure 1), but their rewards are roughly three times higher than the sample average. As the policy is restricted to be linear, it excludes some high reward individuals. The current group of treated people, for example, excludes the ten individuals with the highest-rewards. Including any of them would either lower the group's average reward or reduce its size, leading to longer waiting times and, consequently, lower welfare due to discounting.

Treated groups vary across budget levels and may not overlap. With a higher budget, it can be advantageous to target a larger group with lower average rewards to realize rewards sooner. Under a large budget, a more selective policy would delay rewards further into the future, resulting in welfare loss due to discounting. Consequently, a higher budget tends to support a less restrictive policy.

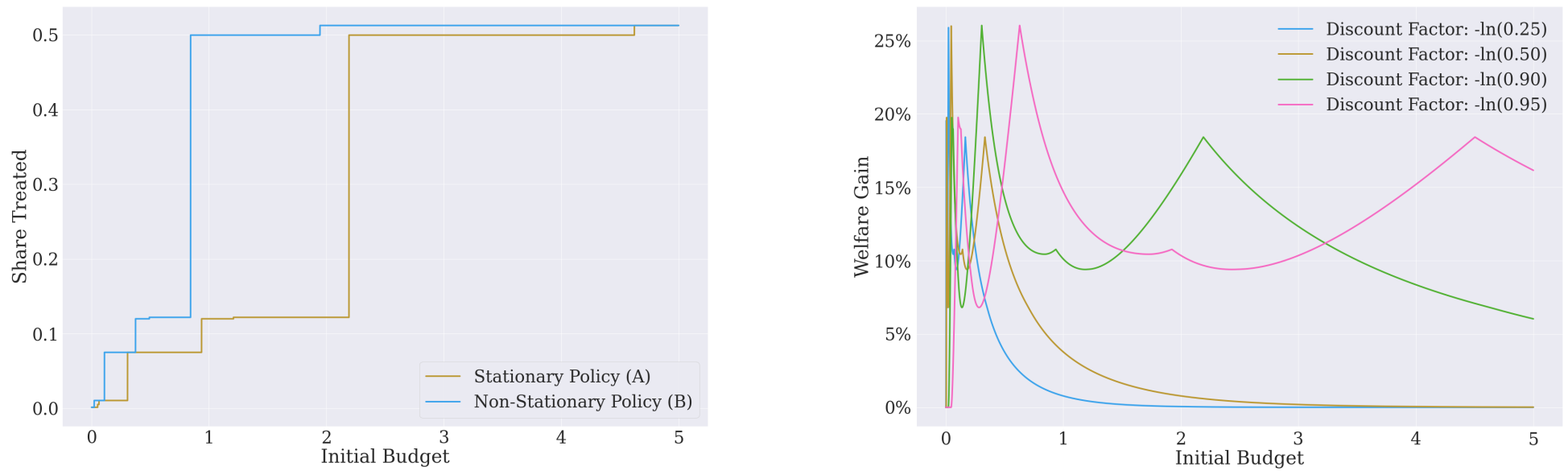

A budget of 1 corresponds to the expected cost of treating all arrivals of one year. The share treated remains constant between an initial budget of 5 and 10. If not indicated otherwise, we use a discount factor of $-\log(0.9)$.

Figure 1: Effects of the Initial Budget

Figure 1 compares stationary and non-stationary policies.[18] The optimal stationary policy varies with the initial budget, $z_0$, but always treats a weakly smaller share of arrivals when $z = z_0$. This is because the non-stationary policy can be more generous initially and more selective as the budget is depleted. The stationary policy must remain constant even as the budget is used up and is inferior due to this restriction (Figure A6 in Appendix D shows how policies respond to current budget). The welfare gain from the non-stationary policy varies with the initial budget level. For very large budgets, the gain approaches zero because the differences mainly occur in the distant future, which is heavily discounted. This is easily noticeable in Figure 1 under the lower discount factors -ln(0.25) and -ln(0.5), but it is generally true for all discount factors with sufficiently high initial budgets. For a tiny budget (say one that allows treating only 1 person) the gain is also 0 as both policies solve the same problem. In between, the welfare gain from using the non-stationary policy is non-monotonic in budget due to the discrete nature of the covariates. The latter causes the stationary policy to abruptly change its treated group based on the initial budget, thereby shifting the reference point for the welfare gain. Overall, the welfare gains range between 0% and 26%, depending on the initial budgets and discount factors.

[18] We emphasize that the welfare differences presented are based solely on in-sample evaluations. Due to potential biases introduced by in-sample optimization and the considerable variability in welfare outcomes, we caution that these differences may not reach statistical significance.

## 5.2 Setting 2: Constrained Budget and Time

***Setup.*** Setting 2 is closely tailored to the institutional details of Sweden's KROM system (see Section 2). Contrary to Setting 1, costs are individual-specific. To mirror the KROM program, treatment costs are set to 80% of the job applicant's previous earnings over the last three months, bounded by the minimum wage on the lower end and the average wage rate on the upper end (we calculate the equivalents for the US during the time of JTPA). These costs are then normalized so that a budget of $z_0 = 1$ would allow us to treat exactly 25% of expected yearly arrivals. This broadly resembles the fraction of applicants that receive KROM. We also allow both the arrival rates and the distribution of arrivals to be time-varying to reflect the seasonal nature of unemployment. To make comparisons to EWM simpler, and because time is bounded by only one administrative year in this setting, we do not discount; however, our methods can easily incorporate discounting.

***Solving the Problem.*** Due to the complexity of this setting, we construct a dynamic environment aligned with our setup and solve it using an RL algorithm. We begin by using the JTPA dataset to estimate month-specific distribution of arrivals and arrival rates. These estimates are then applied to build a simulation similar to the KROM setup. The dynamics of this simulation are illustrated with the following exemplary episode (an *episode* starts at the beginning of the year and ends if budget or time constraints are hit):

First, we draw the time of the first arrival from an exponential distribution utilizing a January-specific estimate of monthly arrivals. Once we know the time of the first arrival, we sample an observation from those individuals in the dataset who were originally considered in January. Based on a state $s$ consisting of the individual's age, education, and previous earnings (as well as possibly budget and the month of the year), we sample the treatment decision from the policy function $\pi_\theta(1|s)$. In case of treatment, the budget decreases by the cost associated with the specific individual $c_i$ and the policy maker receives the reward $\hat{r}(x_i, 1)$—we use the same doubly robust reward estimates as in Setting 1. Afterwards, we draw a time increment for the next arrival, sample an individual for that arrival, and so on. This process repeats until the budget is exhausted or the end of the year is reached.

We let an RL algorithm (PPO) iterate through such simulated episodes thousands of times and let it find the optimal parameters of the policy function. We again consider two classes of policies: (A) a non-stationary policy class: $\log(\pi_\theta^A(1|s)/(1-\pi_\theta^A(1|s)) = \theta_1^\intercal x + \theta_2 z + \theta_3^\intercal \cdot \mathbf{D_t}$, and (B) a stationary one: $\log(\pi_\theta^B(1|s)/(1-\pi_\theta^B(1|s)) = \theta^\intercal x$. The covariate vector $x$ is the same as in Setting 1 and $\mathbf{D_t}$ is a vector with 11 monthly dummies. While the policy functions are logistic, the value function—used only to determine policy function updates during RL training—is approximated by a flexible neural network. This value function neural network has approximately 4,500 parameters and always includes individual covariates, budget, and continuous time in its state. Among the many hyperparameters of the PPO algorithm, the only one that needs tuning in our setup is the learning rate of the policy function. Other hyper-parameters can be kept at the default values suggested by the Spinning-Up implementation (Achiam, 2018). See Appendix C for more details.

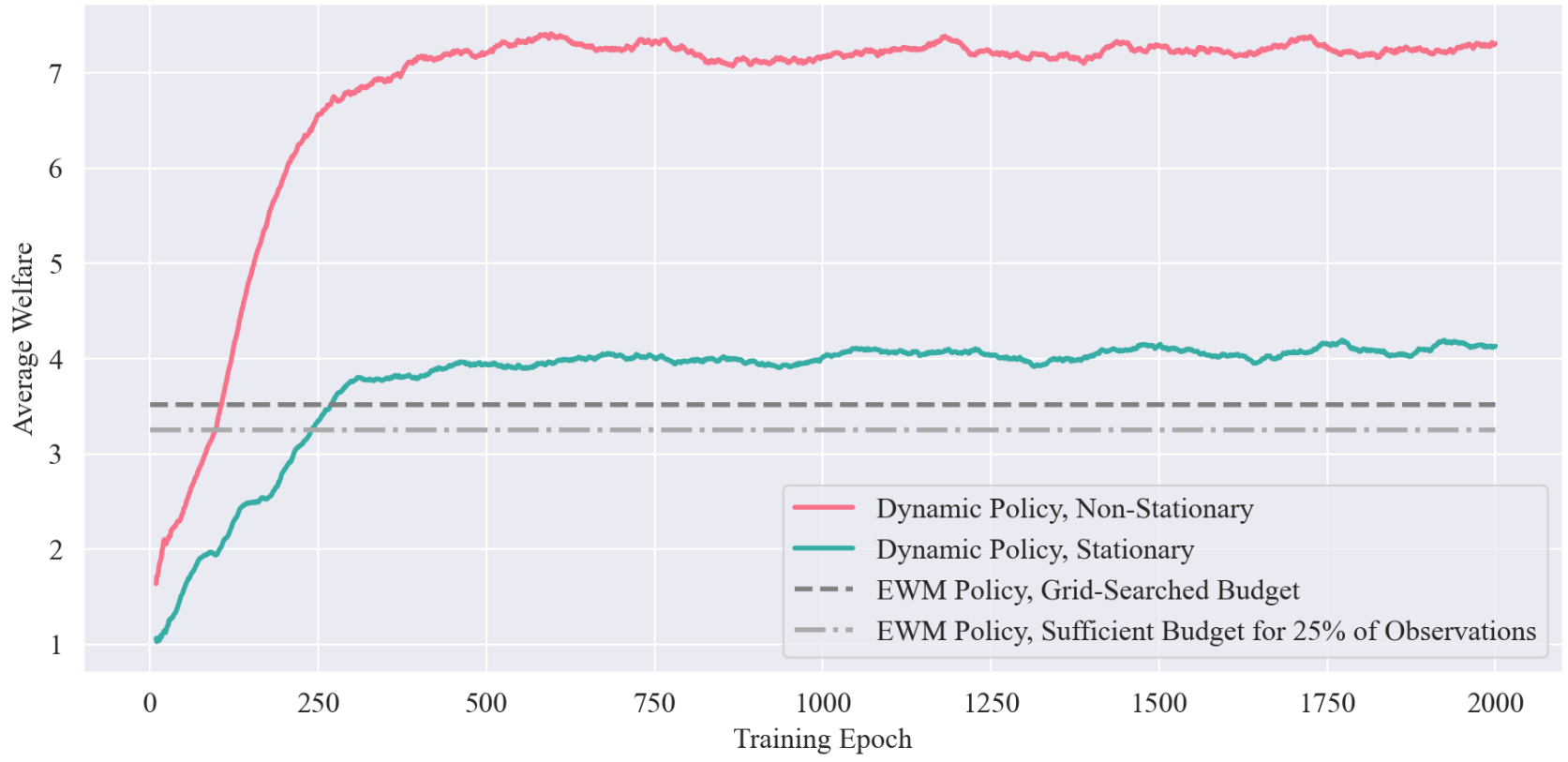

Learning trajectory for the two dynamic policies start at epoch 10 and are moving averages over the last (where available) 100 training epochs. The welfare under a purely random policy (50% treatment probability) is normalized to 1.

Figure 2: Welfare Achieved on the Dynamic Problem

Training was performed over 50 seeds per policy function class, each with 2000 epochs and 10 parallel processes.[19] We select the best-performing seed for each class, though, especially for the stationary policy, the top seeds were often identical. Appendix C provides further details on the training, along with additional results for neural network policy classes.

***Results and Discussion.*** Figure 2 shows how the welfare under our policy rules evolves over the course of their training. We also compare our optimal dynamic policies to those obtained from EWM (Kitagawa and Tetenov, 2018). The lower dashed line depicts the welfare under an EWM policy obtained by solving the simultaneous allocation problem with a nominal budget of 25%.[20] The upper dashed line depicts an improved EWM policy, where the nominal budget was chosen using grid-search to maximize welfare, as described in the next section. We discuss the welfare differences between these four policies in Section 5.2.1 below.

The stationary policy is the most straightforward to interpret. We find that the probability of receiving treatment decreases with age and, to a lesser extent, with education, while it increases with previous earnings.

[19] An epoch is essentially a batch of episodes - see Appendix C for more details.

[20] We used a weighted version of EWM to account for the fact that the yearly distribution of arrivals is slightly different from the RCT distribution.

### 5.2.1 Why Dynamics Matter

There are number of reasons why a policy rule based on simultaneous allocation such as EWM is not optimal in a dynamic setting.

First, the budget constraint in the dynamic setting is inherently 'softer' when the class of policies is restricted. Consider a simplified version of our example with constant arrival rates (i.e., no variation across months) and a constant cost of treatment, with a budget that enables treating at most 25% of the arrivals. Even in this simplified framework, the EWM policy based on a population budget constraint of 25% need not be optimal. This is because we can alternatively employ a policy with a looser 'nominal' budget constraint, e.g., one that treats 30% of the population in a simultaneous setting. Under sequential arrivals, this policy would still only treat 25% of arrivals before the money runs out. But because the class of policies is restricted, it can be the case that this looser policy is associated with a higher average reward per treated person, and therefore lead to greater welfare overall.[21] Or, in cases where the policy under the tighter nominal budget does not exhaust the full budget, loosening the nominal budget could generate higher total welfare by allowing more individuals to be treated.[22] Simultaneous allocations, constrained by strict budget rules, do not not naturally account for these nuances. In fact, one should search over all possible budget values above the given constraint to determine the EWM policy that leads to the highest welfare. A detailed analysis of this

[21]The restriction on the policy class is crucial. If we allow unrestricted policy rules the average reward per treated person is monotonically decreasing in the budget constraint.

[22]The main reason for not exhausting the budget is again the restriction on the policy class, but it can also occur as a consequence of several candidates having the exact same covariates.

effect is provided in Appendix B.[23]

Second, in settings with time-varying outcome distributions, adjusting the nominal budget can alter the effective population distribution faced by the policy, as the program would then terminate earlier. For instance, if the program ends in mid-June (as with our stationary policy in Figure 2), the relevant population distribution is based on arrivals from January through mid-June, rather than the full year. While EWM can be trained on an empirical distribution approximating yearly arrivals, it cannot adapt to distributional shifts resulting from varying budgets, as it assumes each arrival is an independent draw from a fixed yearly distribution. In contrast, our approach simultaneously optimizes the nominal budget and accounts for the timing of arrivals, allowing it to effectively choose the optimal population distribution and also maximize welfare within that distribution.[24] This is particularly important when covariate-reward relationships evolve over time or when earlier arrivals yield higher rewards in ways not captured by covariates or the policy design.

Of course, a non-stationary policy that employs time as a covariate in a flexible manner can also exploit such a dependence between rewards and time. However, incorporating time as a covariate in the EWM setting can be problematic. If time affects rewards, it would be natural to also suppose that it affects the distributions of the covariates. But the latter is not straightforward to handle in EWM.[25]

Finally, simultaneous methods cannot accommodate discounting, and the extent of this limitation depends on the discount rate. We discus discounting extensively in Setting 1, though we omit it from Setting 2 to provide a cleaner comparison with EWM. However, we do believe that discounting is appropriate in the job-training setting as, ceteris paribus, a policymaker would prefer to stop paying unemployment benefits and start receiving taxes earlier.

[23] Additionally, varying candidate arrivals mean the share of treated individuals will vary under a fixed budget. Thus, treating exactly 25% of candidates will result in leftover budget approximately half the time. However, this disadvantage of the simultaneous solution diminishes as the number of candidates increases and vanishes in the limit.

[24] It is able to achieve this because it maximizes a value function that is a nonlinear function of $t$. Consequently, even the choice of optimal stationary policies - where $t$ is absent from the policy function - is indirectly influenced by the presence of $t$ in the value function.

[25] One option, as in our empirical example, could be to split the data into segments based on monthly or seasonal dummies, but this creates even more problems for determining the right nominal budgets within each segment.

In the empirical setting illustrated in Figure 2, the first explanation accounts for an 8% improvement in total welfare.[26] This improvement arises from comparing the 'classic' EWM policy, which operates with a nominal budget set at 25%, against an adjusted EWM policy where the nominal budget is optimized through grid-search to maximize welfare in a context where each arrival is assumed to be an independent draw from the fixed yearly distribution of arrivals. The remaining welfare gap between our dynamically optimal policies and the EWM policies stems from the second explanation. The optimal stationary policy achieves 27% higher welfare than the 'classic' EWM policy and 17% higher welfare than the EWM policy with the optimal nominal budget. The non-stationary policy even achieves a welfare improvement of 133% and 116% over these two EWM policies. The non-stationary policy is able to achieve substantially higher welfare than the stationary one as it is able to adapt to the time-varying outcome distribution.

## 5.3 Limitations and Possible Extensions

Every RCT aims to inform future policy using past experimental data, assuming treatment effects remain stable. This assumption is more realistic if the time between RCT and application is short and the respective populations are similar. Our paper is no exception in this regard. In case RCT data is trusted to inform policy, we aim to provide guidance how this can be done optimally. Our algorithm could incorporate online learning (see Appendix F.3), which is relevant for settings where the episodes are particularly long or treatment effects suspected to be unstable, but this is not the current paper's main focus.

[26] This magnitude varies depending on the initial budget (and the dataset). In Appendix B, we use a simplified setting with only two covariates and a large grid search to find optimal coefficient values for each budget. We find that this effect varies, but it can reduce total welfare by as much as 17%. With more covariates, such a grid search becomes infeasible, but our RL algorithm computes the optimal policy directly (and thereby also automatically determines the right nominal budget level).

# 6 Conclusion

We consider a dynamic setting where candidates arrive sequentially, and a policy-maker with limited budget or capacity must decide whether to provide treatment. Our approach involves estimating a policy function from a pre-specified functional class that maximizes a given utilitarian welfare criterion. We demonstrate that reinforcement learning (RL) can effectively solve this optimization problem. The resulting policy function is particularly appealing because its regret decreases at a rate of $n^{-1/2}$, consistent with the performance seen in static settings.

In an empirical illustration calibrated to the institutional details of Sweden's job training program, we find that our proposed methods increase welfare by up to 133% compared to existing approaches that solve a simultaneous allocation problem and thus ignore dynamics.

*For Online Publication*

## SUPPLEMENTAL APPENDIX TO:

# DYNAMICALLY OPTIMAL TREATMENT ALLOCATION

KARUN ADUSUMILLI[⋆], FRIEDRICH GEIECKE[†] & CLAUDIO SCHILTER[‡]

[⋆]Department of Economics, University of Pennsylvania; akarun@sas.upenn.edu

[†]Department of Methodology, London School of Economics; f.c.geiecke@lse.ac.uk

[‡]Department of Economics, University of Bern; claudio.schilter@unibe.ch

Supplementary material for this paper (not intended for publication) can be accessed here.

# A Proofs

We start by recalling the definition of a viscosity solution. We focus here on the Dirichlet boundary condition. The other boundary conditions are discussed in Appendix E. Consider the Dirichlet problem

$$F\left(z,t,u(z,t),Du(z,t)\right)=0 \text{ on } \mathscr{U}; \quad u=0 \text{ on } \Gamma, \tag{17}$$

where $Du$ denotes the derivative with respect to $(z,t)$, $\mathscr{U}$ is the domain of the PDE, and $\Gamma\subseteq\partial\mathscr{U}$ is the set on which the boundary condition is specified. In what follows, let $y:=(z,t)$. Also, $\mathscr{C}^2(\mathscr{U})$ denotes the space of all twice continuously differentiable functions on $\mathscr{U}$.

**Definition 1.** *A bounded continuous $u(\cdot)$ is a viscosity sub-solution to (17) if:*

*(i) $u\leq 0$ on $\Gamma$, and*

*(ii) for each $\phi\in\mathscr{C}^2(\mathscr{U})$, if $u-\phi$ has a local maximum at $y\in\mathscr{U}$, then*

$$F\left(y,u(y),D\phi(y)\right)\leq 0.$$

*Similarly, a bounded continuous $u(\cdot)$ is a viscosity super-solution to (17) if:*

*(i) $u\geq 0$ on $\Gamma$, and*

*(ii) for each $\phi\in\mathscr{C}^2(\mathscr{U})$, if $u-\phi$ has a local minimum at $y\in\mathscr{U}$, then*

$$F\left(y,u(y),D\phi(y)\right)\geq 0.$$

*Finally, $u$ is a viscosity solution to (17) if it is both a sub- and a super-solution.*

We say that $u$ is a viscosity sub-solution to (17) *on* $\mathscr{U}$ if only condition (ii) holds, i.e., it need not be the case that $u\leq 0$ on $\Gamma$. Similarly, $u$ is a viscosity super-solution to (17) *on* $\mathscr{U}$ if only condition (ii) holds. Henceforth, whenever we refer to a viscosity super- or sub-solution, we will implicitly assume that it is bounded and continuous.

We say that a PDE is in Hamiltonian form if

$$F\left(y,u(y),Du(y)\right)=\partial_t u(y)+H(y,u(y),\partial_z u(y)),$$

for some Hamiltonian $H(\cdot)$. Suppose that the PDE (17) can be written in Hamiltonian form. Then there exists a unique viscosity solutions to (17) if the following regularity conditions are satisfied (see, e.g., Crandall *et al*, 1992):

(R1) $H(y,u,p)$ is uniformly continuous in all its arguments.

(R2) There exists a modulus of continuity $\omega(\cdot)$ such that $\forall\ (y_1,y_2)\in\mathscr{U}$,

$$\begin{aligned}&|H(y_1,u,p_1)-H(y_2,u,p_2)|\leq\omega\left(\|y_1-y_2\|+\|p_1-p_2\|\right), \quad \text{and}\\ &|H(y_1,u,p_1)-H(y_2,u,p_1)|\leq\omega\left(\|y_1-y_2\|\,|1+\|p_1\||\right).\end{aligned}$$

(R3) $H(y,u,p)$ is non-decreasing in $u$ for all $(y,p)$.

(R4) There exist some viscosity sub- and super-solutions to PDE (17).

### A.1 Proof of Lemma 1

Consider the PDE

$$\begin{aligned}\partial_\tau u_\theta+H_\theta(z,\tau,\partial_z u_\theta)&=0 \quad \text{on } \Upsilon\equiv(\underline{z},\infty)\times(0,T]; \qquad (18)\\ u_\theta&=0 \quad \text{on } \mathscr{B}\equiv\{\{\underline{z}\}\times[0,T]\}\cup\{(\underline{z},\infty)\times\{0\}\},\end{aligned}$$

where $H_\theta(\cdot)$ is defined as

$$H_\theta(z,\tau,p):=-e^{\beta\tau}\lambda(\tau)\bar{r}_\theta(z,\tau)-\lambda(\tau)\bar{G}_\theta(z,\tau)p. \qquad (19)$$

Let $u_\theta$ denote a viscosity solution to (18). By Lemma G.1 in Appendix G and the subsequent discussion, there is a one-to-one transformation between $u_\theta$ and any solution, $h_\theta$, for PDE (6) under the Dirichlet boundary condition (8); this transformation is given by $u_\theta(z,t) := e^{-\beta(T-t)} h_\theta(z, T-t)$. Hence, $h_\theta$ exists and is unique if and only if $u_\theta$ also exists and is unique.[1]

It thus suffices to show existence of a unique solution for (18). It is straightforward to verify that the function $H_\theta(\cdot)$ satisfies the regularity conditions (R1)-(R3) under Assumption 1. Furthermore, the set $\mathscr{B}$ satisfies the uniform exterior sphere condition.[2] When these properties are satisfied, Crandall (1997, Section 9) shows that a unique viscosity solution exists for (18), as long as we are able to exhibit continuous sub- and super-solutions to (18). Under Assumption 1, it can be verified that one such set is $-L\tau$ and $L\tau$, where $L < \infty$ is chosen to satisfy $|\lambda(\tau)\bar{r}_\theta(z,\tau)| \le M \sup_\tau \lambda(\tau) < L$.

## A.2 Proof of Theorem 2

The following proof is based on an argument sketched by Souganidis (2009). We focus here on the Dirichlet boundary condition with $T < \infty$ (but $\underline{z}$ could be $-\infty$). The argument for the setting with $T = \infty$ is similar, so we omit it.

All the statements in this section should be understood to be holding with probability approaching 1. In what follows, we drop this qualification for ease of notation and hold this to be implicit. We also employ the following notation: For any function $f$ over $(z,t)$, $Df$ denotes its Jacobean. Additionally, $\|\partial_z f\|, \|\partial_z f\|$ and $\|Df\|$ denote the Lipschitz constants for $f(\cdot,t), f(z,\cdot)$ and $f(\cdot,\cdot)$.

To begin with we note that by Kitagawa and Tetenov (2018) and Athey and

[1]The utility of this transformation is that we can now handle any $\beta \in \mathbb{R}$.

[2]A set $\mathscr{U}$ is said to satisfy the uniform exterior sphere condition if there exists $r_0 > 0$ such that every point $y \in \partial\mathscr{U}$ is on the boundary of a ball of radius $r_0$ that otherwise does not intersect $\bar{\mathscr{U}}$.

Wager (2021), we have (see, also, the supplementary material for the details)

$$\sup_{(z,\tau)\in\Upsilon,\theta\in\Theta} |\hat{r}_\theta(z,\tau) - \bar{r}_\theta(z,\tau)| \le C_0\sqrt{\frac{v_1}{n}}, \text{ and} \tag{20}$$
$$\sup_{(z,\tau)\in\Upsilon,\theta\in\Theta} \left|\hat{G}_\theta(z,\tau) - \bar{G}_\theta(z,\tau)\right| \le C_0\sqrt{\frac{v_2}{n}},$$

with probability approaching one (henceforth wpa1), for some $C_0 < \infty$.

We represent PDE (6) by

$$F_\theta(z,t,f,\partial_z f,\partial_t f) = 0 \text{ on } \mathscr{U}; \quad f = 0 \text{ on } \Gamma, \tag{21}$$

with $f$ denoting a function, and where

$$F_\theta(z,t,l,p,q) := -\lambda(t)\bar{G}_\theta(z,t)l - p + \beta q - \lambda(t)\bar{r}_\theta(z,t).$$

Additionally, denote our approximation scheme (15) by

$$S_\theta([f], f(z,t), z, t) = 0 \text{ on } \mathscr{U}; \quad f = 0 \text{ on } \Gamma, \tag{22}$$

where for any function $f_1$ and $u \in \mathbb{R}$,

$$S_\theta([f_1], u, z, t, b_n) := b_n\lambda(t)\left(u - E_{n,\theta}\left[e^{-\beta(t'-t)} f_1(z',t')|z,t\right]\right) - \lambda(t)\hat{r}_\theta(z,t). \tag{23}$$

Here $[f]$ refers to the fact that it is a functional argument. Note that $h_\theta$ and $\tilde{h}_\theta$ are the functional solutions to (21) and (22) respectively. We make use of the following two properties of $S_\theta(\cdot)$: First, $S_\theta(\cdot)$ is monotone in its first argument, i.e.,

$$S_\theta([f_1], f(z,t), z, t, b_n) \ge S_\theta([f_2], f(z,t), z, t, b_n)\ \forall\ f_2 \ge f_1. \tag{24}$$

Second, for any $r \in \mathbb{R}$ and $m \in \mathbb{R}^+$, it holds for all $t \leq T - b_n^{-1/2}$ that

$$S_\theta([f+m], r+m, z, t, b_n) \geq S_\theta([f], r, z, t) + \chi m, \tag{25}$$

where $\chi = \beta + O(b_n^{-1}) > 0$. The first property is trivial to show. As for the second, under Assumption 1 and $t \leq T - b_n^{-1/2}$, it is straightforward to show

$$\begin{aligned} S_\theta([f+m], r+m, z, t, b_n) - S_\theta([f], r, z, t) &= m b_n \lambda(t) \left(1 - E_{n,\theta}\left[e^{-\beta(t'-t)} | z, t\right]\right) \\ &= m(\beta + O(b_n^{-1})). \end{aligned}$$

Our proof also makes use of Lemmas G.2 and G.3, which ensure existence of $K_1 < \infty$ such that

$$\|h_\theta\| < K_1, \quad \text{and} \quad \sup_\theta \|Dh_\theta\| < K_1. \tag{26}$$

We now provide an upper bound for

$$m_\theta := \sup_{(z,t) \in \bar{\mathscr{U}}} \left(h_\theta(z,t) - \tilde{h}_\theta(z,t)\right). \tag{27}$$

A lower bound for $h_\theta - \tilde{h}_\theta$ can be obtained analogously. Clearly, we may assume $m_\theta > 0$, as otherwise we are done. Denote $(z_\theta^*, t_\theta^*)$ as the point at which the supremum is attained in (27), or, if such a point does not exist, where the right hand side of (27) is arbitrarily close to $m_\theta$. We consider three (not necessarily mutually exclusive) cases: (i) $|t_\theta^* - T| \leq 2K_1\varepsilon$, (ii) $|z_\theta^* - \underline{z}| \leq 2K_1\varepsilon$, and (iii) $|z_\theta^* - \underline{z}| > 2K_1\varepsilon$ and $|t_\theta^* - T| > 2K_1\varepsilon$. Here, $\varepsilon$ is any positive number satisfying $\varepsilon \geq \sqrt{b_n}$.

We start with Case (i). In view of (26), and the fact $|t_\theta^* - T| \leq 2K_1\varepsilon$, we have

$$|h_\theta(z_\theta^*, t_\theta^*)| \leq 4K_1^2\varepsilon. \tag{28}$$

Now, we claim $\tilde{h}_\theta(z,t) \le L\{(T-t)+b_n^{-1}\}$, for some $L<\infty$ independent of $\theta,z,t$. Let $N[t,T]$ be a random variable denoting the number of arrivals between $t$ and the end point $T$. Then $N[t,T]$ is first order stochastically dominated by $\bar{N}[t,T] \sim$ Poisson$(\bar{\lambda} b_n(T-t))$, where $\bar{\lambda} := \sup_t \lambda(t) < \infty$.[3] Hence, $E[N[t,T]] \le E[\bar{N}[t,T]] = \bar{\lambda} b_n(T-t)$. Furthermore, the reward from any given arrival is at most $\sup_{\theta,z,t} |\hat{r}_\theta(z,t)|/b_n \le 2M/b_n$ by Assumption 2(i) and (20). Consequently,

$$\tilde{h}_\theta(z,t) \le \frac{2M}{b_n} + E\left[N[t,T]\frac{2M}{b_n}\right] \le 2M\bar{\lambda}\left\{(T-t)+b_n^{-1}\right\} := L\left\{(T-t)+b_n^{-1}\right\}.$$

Considering that we are in the case $|t_\theta^* - T| \le 2K_1\varepsilon$, the previous statement implies

$$|\tilde{h}_\theta(z_\theta^*, t_\theta^*)| \le L\left(2K_1\varepsilon + b_n^{-1}\right). \tag{29}$$

In view of (28) and (29), we thus obtain

$$m_\theta \le (4K_1^2 + 2LK_1)\varepsilon + Lb_n^{-1}. \tag{30}$$

This completes the treatment of the first case, when $|t_\theta^* - T| \le 2K_1\varepsilon$.

We next consider Case (ii). At the end of this proof, we show that when $\bar{G}_\theta(z,t) < -\delta$ (cf. Assumption 1(iv)), the expected number of arrivals subsequent to state $z$ is bounded above by $2\delta^{-1}\{b_n(z-\underline{z})+C_1\}$ for some $C_1<\infty$ independent of $\theta,z,t$. Hence, by a similar argument as that leading to (29), we have $|\tilde{h}_\theta(z_\theta^*, t_\theta^*)| \le L_2\{2K_1\varepsilon + b_n^{-1}\}$ for some $L_2<\infty$. Combined with the Lipschitz continuity of $h_\theta$, this implies the bound (30) also holds for Case (ii).

[3]Note that $\bar{N}[t,T]$ is the number of arrivals between $t$ and $T$ under a Poisson process with parameter $\bar{\lambda} b_n$; the rate of arrivals here is always faster than under the approximation scheme.

We now turn to Case (iii), i.e., $|z^*_\theta - \underline{z}| > 2K_1\varepsilon$ and $|t^*_\theta - T| > 2K_1\varepsilon$. Denote

$$\mathscr{A} \equiv \{(z,t) \in \bar{\mathscr{U}} : \ |z-\underline{z}| > 2K_1\varepsilon \cap |t-T| > 2K_1\varepsilon\}.$$

Let $h^\varepsilon_\theta(z,t)$ denote the sup-convolution of $h_\theta(z,t)$:[4]

$$h^\varepsilon_\theta(z,t) := \sup_{(r,w)\in\bar{\mathscr{U}}} \left\{ h_\theta(r,w) - \frac{1}{\varepsilon}\left(|r-z|^2 + |w-t|^2\right) \right\}.$$

We make use of the following properties of $h^\varepsilon_\theta$ : First, $h^\varepsilon_\theta$ is a semi-convex function with coefficient $1/\varepsilon$ (Lemma H.2).[5] Second, by (26) and Lemma H.2,

$$\sup_{(z,t)\in\bar{\mathscr{U}}} |h_\theta(z,t) - h^\varepsilon_\theta(z,t)| \le 4K_1^2\varepsilon, \text{ and} \tag{31}$$

$$\sup_\theta \|Dh^\varepsilon_\theta\| \le 4\sup_\theta \|Dh_\theta\| \le 4K_1. \tag{32}$$

Third, by Lemma H.3 (Assumption 1 ensures all relevant regularity conditions for $F_\theta(\cdot)$ are satisfied), there exists $c < \infty$ independent of $(\theta, z, t)$ such that, in a viscosity sense,

$$F_\theta(z,t,h^\varepsilon_\theta, \partial_z h^\varepsilon_\theta, \partial_t h^\varepsilon_\theta) \le c\varepsilon \quad \text{on } \mathscr{A}. \tag{33}$$

We now compare $S_\theta(\cdot)$ and $F_\theta(\cdot)$ at the function $h^\varepsilon_\theta$. Consider any $(z,t) \in \mathscr{A}$. If $h^\varepsilon_\theta$ is not differentiable at $(z,t)$, let $(\dot{z},\dot{t})$ denote a point arbitrarily close to $(z,t)$ where it is indeed differentiable (note that $h^\varepsilon_\theta$ is Lipschitz continuous, so this point

[4] We discuss sup and inf-convolutions and their properties in Appendix H.

[5] See Appendix H for the definition of semi-convex functions.

can be chosen as close to $(z,t)$ as one wants). We can then expand

$$
\begin{aligned}
S_\theta([h^\varepsilon_\theta], h^\varepsilon_\theta(z,t), z, t, b_n) &\cong S_\theta([h^\varepsilon_\theta], h^\varepsilon_\theta(\dot z,\dot t), z, t, b_n) \\
&= b_n\lambda(t) h^\varepsilon_\theta(\dot z,\dot t)\left(1 - E_{n,\theta}\left[e^{-\beta(t'-t)}|z,t\right]\right) \qquad (34)\\
&+ b_n\lambda(t)E_{n,\theta}\left[e^{-\beta(t'-t)}\left\{h^\varepsilon_\theta(\dot z,\dot t) - h^\varepsilon_\theta(z',t')\right\}|z,t\right] + (-1)\lambda(t)\hat r_\theta(z,t)\\
&:= A^{(1)}_\theta(z,t) + A^{(2)}_\theta(z,t) + A^{(3)}_\theta(z,t), \qquad (35)
\end{aligned}
$$

where the notation '$\cong$' in the first step indicates that we are performing the arbitrarily small approximation of replacing $h^\varepsilon_\theta(z,t)$ in the second argument of $S_\theta(\cdot)$ with $h^\varepsilon_\theta(\dot z,\dot t)$ - recall that $S_\theta(\cdot)$ is linear in its second argument and $h^\varepsilon_\theta$ is Lipschitz continuous. Using $\|h^\varepsilon_\theta\| \le \|h_\theta\| \le K_1$ and Assumptions 1-4, it is straightforward to show

$$
A^{(1)}_\theta(z,t) \le \beta h^\varepsilon_\theta(\dot z,\dot t) + \frac{C_1}{b_n}, \qquad (36)
$$

for some $C_1$ independent of $\theta, z, t$. Next, consider $A^{(2)}_\theta(z,t)$. By semi-convexity of $h^\varepsilon_\theta$ (see, Lemma H.1),

$$
\begin{aligned}
h^\varepsilon_\theta(z',t') &\ge h^\varepsilon_\theta(\dot z,\dot t) + \partial_z h^\varepsilon_\theta(\dot z,\dot t)(z'-\dot z) + \partial_t h^\varepsilon_\theta(\dot z,\dot t)(t'-\dot t) - \frac{1}{2\varepsilon}\left\{|z'-\dot z|^2 + |t'-\dot t|^2\right\}\\
&\cong h^\varepsilon_\theta(\dot z,\dot t) + \partial_z h^\varepsilon_\theta(\dot z,\dot t)(z'-z) + \partial_t h^\varepsilon_\theta(\dot z,\dot t)(t'-t) - \frac{1}{2\varepsilon}\left\{|z'-z|^2 + |t'-t|^2\right\},
\end{aligned}
$$

where we again employ the arbitrarily small approximation of replacing $\dot z, \dot t$ in the second step with $z,t$ (note that (32) ensures $\partial_z h^\varepsilon_\theta, \partial_t h^\varepsilon_\theta$ are uniformly bounded). Substituting the above into the expression for $A^{(2)}_\theta(z,t)$, and using Assumption 1, (20) and (32), some straightforward algebra enables us to show that when $\varepsilon \ge b_n^{-1/2}$,

[6]

$$A^{(2)}_\theta(z,t) \le -\lambda(t)\bar{G}_\theta(z,t)\partial_z h^\varepsilon_\theta(\dot z,\dot t) - \partial_t h^\varepsilon_\theta(\dot z,\dot t) + C_2\left(\frac{1}{\varepsilon b_n} + \sqrt{\frac{v}{n}}\right)$$
$$\cong -\lambda(\dot t)\bar{G}_\theta(\dot z,\dot t)\partial_z h^\varepsilon_\theta(\dot z,\dot t) - \partial_t h^\varepsilon_\theta(\dot z,\dot t) + C_2\left(\frac{1}{\varepsilon b_n} + \sqrt{\frac{v}{n}}\right), \tag{37}$$

where $C_2$ is again independent of $(\theta,z,t)$, and we have made use of the continuity of $\lambda(\cdot),\bar{G}_\theta(\cdot)$ in the second step. Finally, to bound $A^{(3)}_\theta(z,t)$, we make use of Assumption 1(ii) and (20), which together ensure there exists $C_3$ independent of $\theta,z,t$ such that

$$A^{(3)}_\theta(z,t) \le -\lambda(t)\bar{r}_\theta(z,t) + C_3\sqrt{\frac{v}{n}}$$
$$\cong -\lambda(\dot t)\bar{r}_\theta(\dot z,\dot t) + C_3\sqrt{\frac{v}{n}}. \tag{38}$$

Combining (35)-(38), and setting $C = \max(C_1,C_2,C_3)$, we thus find

$$S_\theta([h^\varepsilon_\theta],h^\varepsilon_\theta(z,t),z,t,b_n) \le F_\theta(\dot z,\dot t,h^\varepsilon_\theta,\partial_z h^\varepsilon_\theta,\partial_t h^\varepsilon_\theta) + C\left\{\frac{1}{b_n}\left(1+\frac{1}{\varepsilon}\right) + \sqrt{\frac{v}{n}}\right\}. \tag{39}$$

In view of (39) and (33),

$$S_\theta([h^\varepsilon_\theta],h^\varepsilon_\theta(z,t),z,t,b_n) \le c\varepsilon + C\left\{\frac{1}{b_n}\left(1+\frac{1}{\varepsilon}\right) + \sqrt{\frac{v}{n}}\right\}. \tag{40}$$

Let (here $f^+ := \max(f,0)$)

$$m^\varepsilon_\theta := \sup_{(z,t)\in\mathscr{A}} \left(h^\varepsilon_\theta(z,t) - \tilde{h}_\theta(z,t)\right)^+,$$

[6] To show this, we note that $E_{n,\theta}\left[b_n(t'-t)|z,t\right] = \lambda(t)^{-1} + O(b_n^{1/2}\exp\{-\lambda(t)b_n^{1/2}\})$ and $E_{n,\theta}\left[b_n(z'-z)|z,t\right] = \hat{G}_\theta(z,t) = \bar{G}_\theta(z,t) + O(\sqrt{v/n})$ when $\varepsilon \ge b_n^{-1/2}$. As for the quadratic terms, observe that $E_{n,\theta}\left[(t'-t)^2|z,t\right] \le (b_n \inf_t \lambda(t))^{-2}$, and $E_{n,\theta}\left[(z'-z)^2|z,t\right] \le Cb_n^{-2}$ since $G_a(s)$ is bounded. All statements here should only be understood as holding with probability approaching 1.

and denote $(\breve{z}_\theta, \breve{t}_\theta)$ as the point at which the supremum is attained (or where the right hand side of the above expression is arbitrarily close to $m^\varepsilon_\theta$). Now, by definition,

$$h^\varepsilon_\theta \le \tilde{h}_\theta + m^\varepsilon_\theta \text{ on } \mathscr{A}.$$

Then, in view of the properties (24), (25) of $S(\cdot)$,

$$\begin{aligned}
\chi m^\varepsilon_\theta &= S_\theta\left([\tilde{h}_\theta], \tilde{h}_\theta(\breve{z}_\theta, \breve{t}_\theta), \breve{z}_\theta, \breve{t}_\theta, b_n\right) + \chi m^\varepsilon_\theta \\
&\le S_\theta\left([\tilde{h}_\theta + m^\varepsilon_\theta], \tilde{h}_\theta(\breve{z}_\theta, \breve{t}_\theta) + m^\varepsilon_\theta, \breve{z}_\theta, \breve{t}_\theta, b_n\right) \\
&\le S_\theta\left([h^\varepsilon_\theta], h^\varepsilon_\theta(\breve{z}_\theta, \breve{t}_\theta), \breve{z}_\theta, \breve{t}_\theta, b_n\right).
\end{aligned} \tag{41}$$

Combining (41) and (40) gives

$$m^\varepsilon_\theta \le c_1\varepsilon + C\left\{\frac{1}{b_n}\left(1 + \frac{1}{\varepsilon}\right) + \sqrt{\frac{v}{n}}\right\}, \tag{42}$$

where $c_1 = \chi^{-1}c$ and $C_1 = \chi^{-1}C$ are independent of $\theta, z, t$. Hence, in view of (31) and (42),

$$m_\theta \le (4K_1^2 + c_1)\varepsilon + C_1\left\{\frac{1}{b_n}\left(1 + \frac{1}{\varepsilon}\right) + \sqrt{\frac{v}{n}}\right\}. \tag{43}$$

This completes the derivation of the upper bound for $m_\theta$ under Case (iii).

Finally, in view of (30) and (43), setting $\varepsilon = b_n^{-1/2}$ gives the desired rate.

*Bound on expected number of arrivals after z:* It remains to show that the expected number of arrivals subsequent to a state with institutional variable $z$ is bounded by $2\delta^{-1}\{b_n(z - \underline{z}) + C_1\}$, as was needed for the analysis of Case (ii). Denote by $\{\bar{s}_i \equiv (x_i, z_i, t_i, a_i) : i = 1, 2, \ldots\}$ the sequence of state-action variables following any particular state-action variable $\bar{s}_0 = (x,z,t,a)$, and let

$$M_l := \sum_{i=1}^{l} \left\{ G_{a_i}(x_i,z_i,t_i) - \hat{G}_\theta(z_i,t_i) \right\}.$$

Clearly, $M_l$ is a martingale with respect to the filtration $\mathscr{F}_l := \sigma(\bar{s}_{l-1},\dots,\bar{s}_0)$. Let $\mathscr{N}[\bar{s}_0]$ be the random variable denoting the number of arrivals following $\bar{s}_0$, until either $z$ goes below $\underline{z}$ or time runs out. Then $\mathscr{N}[\bar{s}_0] = \tau[\bar{s}_0] - 1$, where $\tau[\bar{s}_0]$ is the stopping time

$$\tau[\bar{s}_0] := \inf\left\{ l \in \{1,2,\dots\} : G_a(x,z,t) + \sum_{i=1}^{l-1} G_{a_i}(x_i,z_i,t_i) \le -b_n(z-\underline{z}) \text{ or } t_{l-1} \ge T \right\}.$$

Now, Assumption 2(i) implies the martingale differences of $M_l$ are bounded. Hence, we can apply the Optional Stopping Theorem to obtain

$$E_{n,\theta}\left[M_{\tau[\bar{s}_0]}\right] = E_{n,\theta}[M_1] = 0.$$

In other words,

$$E_{n,\theta}\left[ \sum_{i=1}^{\tau[\bar{s}_0]} G_{a_i}(x_i,z_i,t_i) - \sum_{i=1}^{\tau[\bar{s}_0]} \hat{G}_\theta(z_i,t_i) \right] = 0.$$

By Assumption 2(ii) and (20), $-\sum_{i=1}^{\tau[\bar{s}_0]} \hat{G}_\theta(z_i,t_i) \ge (\delta/2)\tau[\bar{s}_0]$. Furthermore, by the definition of $\tau[\bar{s}_0]$ and the assumption $|G_a(s)| \le M$, we have

$$\begin{aligned} E_{n,\theta}\left[ \sum_{i=1}^{\tau[\bar{s}_0]} G_{a_i}(x_i,z_i,t_i) \right] &\ge E_{n,\theta}\left[ \mathbb{I}(\tau[\bar{s}_0] \ge 2) \left\{ G_a(x,z,t) + \sum_{i=1}^{\tau[\bar{s}_0]-2} G_{a_i}(x_i,z_i,t_i) \right\} \right] - 3M \\ &> -b_n(z-\underline{z}) - 3M. \end{aligned}$$

The above implies $(\delta/2)E_{n,\theta}[\tau[\bar{s}_0]] < b_n(z-\underline{z}) + 3M$ or $E_{n,\theta}[\mathscr{N}[\bar{s}_0]] < 2\delta^{-1}\{b_n(z-$

$\underline{z}) + C_1\}$ where $C_1 := 3M$. Note that this bound is independent of $(x, t, a)$ in the definition of $\bar{s}_0$.

# B Additional Details for Section 5.2.1

In Section 5.2.1 we provided two major arguments as to why a policy rule based on simultaneous allocation such as EWM is not optimal in a dynamic setting.

Our first argument posited that the budget constraint in a dynamic setting is inherently 'softer' and this has welfare implications. To evaluate the significance of this claim, we examine a simplified version of Setting 2, limiting the covariates to education and prior earnings, and keeping both arrival distributions and rates constant. Aside from sequential arrivals, this setup closely resembles that of Kitagawa and Tetenov (2018). To maintain comparability, we compute the rewards $r(x, 1)$ using inverse probability weighting, as in Kitagawa and Tetenov (2018). This setting is theoretically equivalent to EWM *if* it uses the optimal nominal budget.

To leverage the simplicity of this setting—and to preempt concerns regarding whether the algorithms in Kitagawa and Tetenov (2018) achieve the global maximum—we use a grid search to identify the optimal policies across different budget levels. Specifically, we apply a grid of one billion points across the intercept and two covariates. For each possible budget level, we determine both the optimal simultaneous allocation policy (conditional on the granularity of the grid), which maximizes welfare by treating up to a set number of individuals, and the optimal dynamic policy. The latter either coincides with the simultaneous allocation policy or is the policy with the highest average treatment effect (ATT) that exceeds budget capacity, depending on which yields greater welfare.[7]

Figure A1 illustrates the performance of these policies in a dynamic setting. In

[7] We also trained PPO policies on this simulated setting and subsequently compared the best PPO policy's performance with that of the grid search policy. The results indicate that the welfare achieved by PPO is only approximately 2.1% below that of the grid search policy.

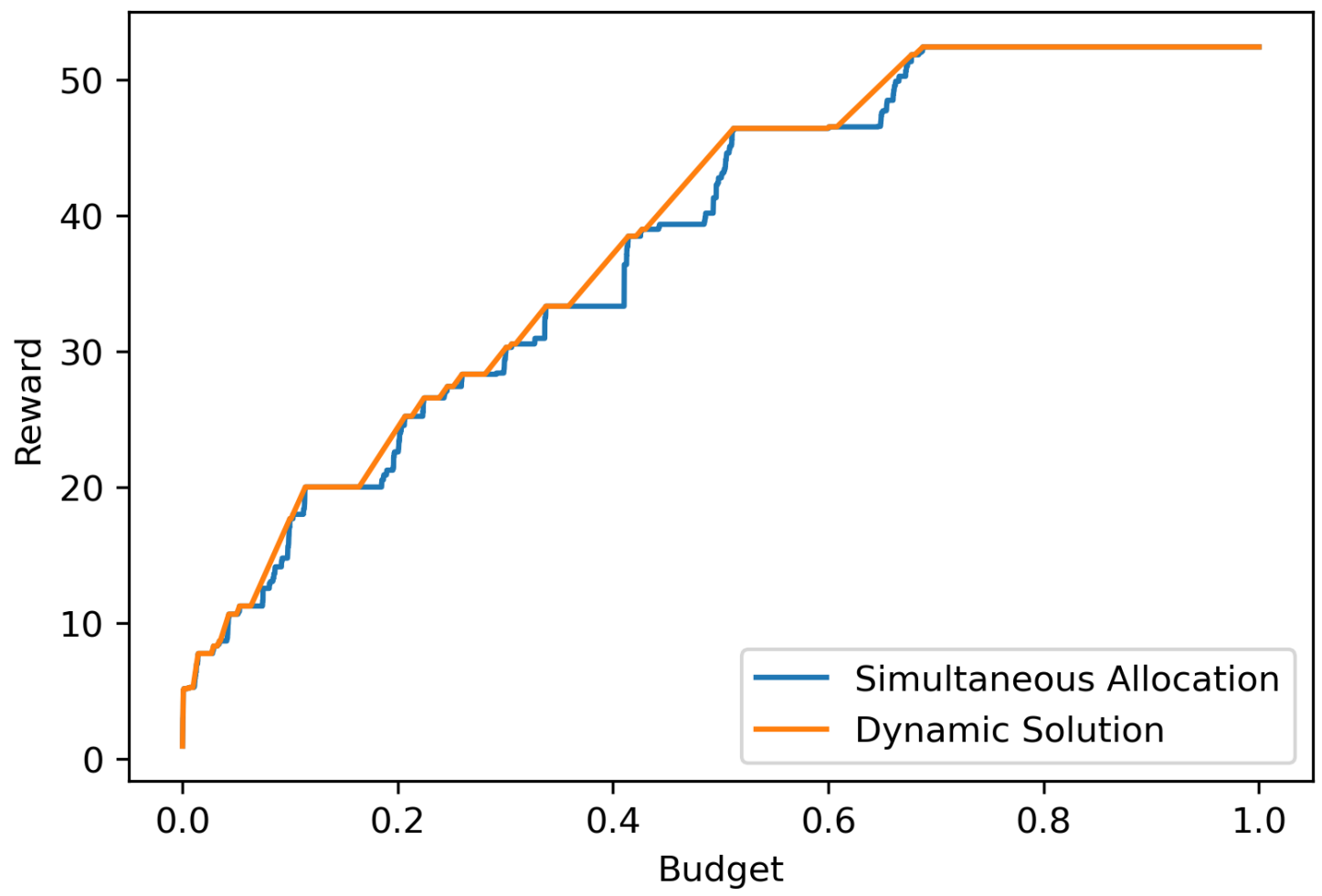

Figure A1: Welfare Consequences of Suboptimal Budget Constraints

The budget is normalized such that a budget of 1 is sufficient to treate the total sample of 9223 people.

40.9% of budget levels, the flexibility afforded by a 'softer' budget constraint results in an improved dynamic policy, with an average welfare gain of 5.3%, reaching up to 17.1% in some cases. Furthermore, in 49.7% of scenarios, this improvement is partially or fully due to a higher Average Treatment effect on the Treated (ATT) enabled by the softer budget constraint. All these considerations are apparent at the budget levels just below 0.2, where the welfare under both policies are initially identical. As the budget increases, more of it remains unused, ultimately making a policy with a lower ATT but complete budget utilization preferable. At this point, the dynamic policy diverges from the simultaneous allocation policy, providing higher welfare. With further budget increases, the welfare of the simultaneous allocation policy eventually begins to rise, indicating a shift in the policy rule so that it now fully exhausts the budget within the year. Nevertheless, a welfare gap persists between the two; this is due to the fact that the dynamic policy chooses a different nominal budget, enabling it to achieve a higher ATT.

Next, to assess the importance of the second argument provided in Section 5.2.1, we analyze the stationary dynamic policy from Section 5.2 in more detail. Simulations indicate that this policy considers 2,362 arrivals before reaching its budget limit, effectively resulting in a stopping point after 5–6 months based on average monthly arrivals. Figure A2(a) shows how the policy maximizes welfare per unit cost within these initial 5 months, disregarding the remainder of the year, which becomes inconsequential under this policy. Thus, the stationary dynamic policy not only establishes the optimal point at which the budget should be exhausted and the program concluded early, but also identifies the allocation rule that maximizes welfare for candidates arriving within this timeframe.

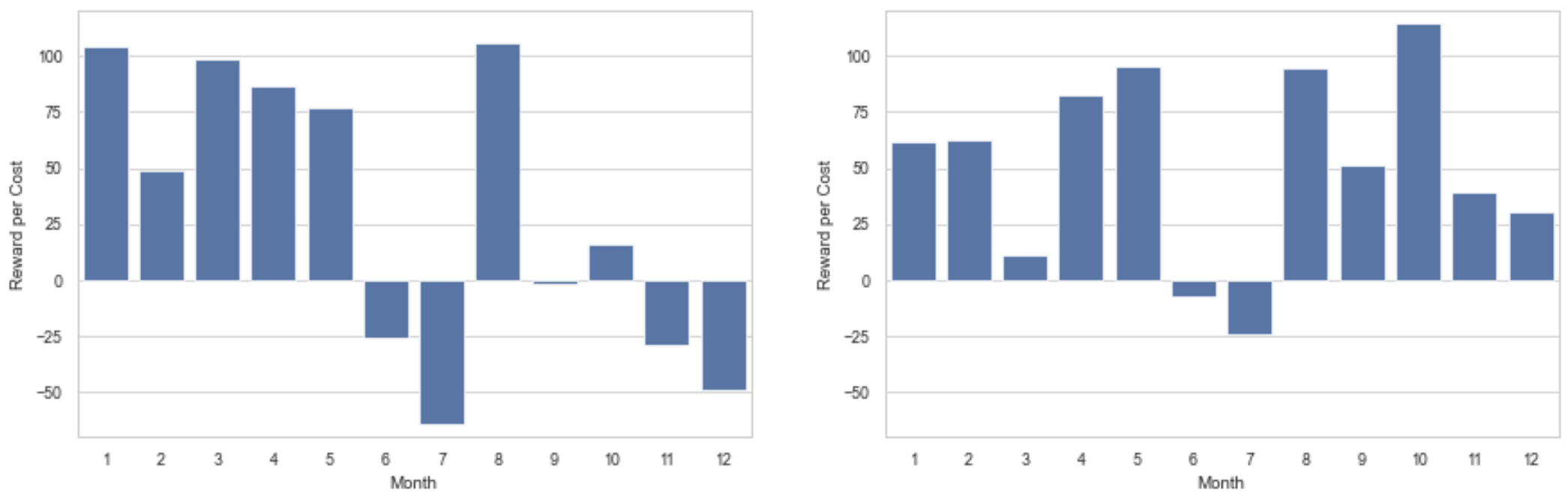

(a) Stationary Dynamic Policy (b) EWM Policy with ‘Dynamic Budget’

Figure A2: Monthly Analysis of Stationary and EWM Policy

Even if EWM is provided with the ‘optimal dynamic budget’—determined by applying the stationary dynamic policy to the simultaneous allocation problem to estimate the required budget—it remains suboptimal. This is because EWM fails to account for the fact that stopping early changes the effective population distribution of arrivals. It thus fails to strategically allocate this budget to maximize rewards under this changed population distribution (see Figure A2(b)). EWM exhausts this budget after approximately six months; however, unlike the stationary dynamic policy, it does not adjust to prioritize optimal allocation based on the pop-ulation distribution within this timeframe. As a result, while the EWM policy may be better suited for arrivals in the latter half of the year, this advantage is irrelevant since the budget is already depleted by that time.

In theory, providing EWM with this ‘optimal dynamic budget’ might improve its performance if early arrivals offered significantly higher rewards—although not to the extent that a dynamic policy would benefit, given its ability to optimize specifically for candidates arriving while the budget remains. However, in our setting, we found no evidence of high-reward individuals arriving disproportionately early in the year. Consequently, allocating EWM this optimal dynamic budget is actually detrimental, as it lacks the capacity to adjust accordingly.[8] Finally, even if this adjustment had improved EWM's performance, EWM could not have independently identified this ‘optimal dynamic budget’, even with a grid search over the various budget levels.

# C Additional Details for the Empirical Application

This section offers an in-depth overview of the empirical application, along with detailed guidance on implementing the PPO algorithm (for Setting 2).

***Setting 1.*** In Setting 1 and Appendix D, we normalize the initial budget $z_0$ to represent the fraction of expected annual arrivals that can be treated. Notably, due to the infinite horizon, this fraction may exceed one. Figure A3 illustrates how the coefficients of the optimal policy vary with budget. Given our assumption of constant cost, the cost of treating one individual is $1/N$, where $N$ is the expected annual arrival count. Ideally, we set $\Delta z = 1/N$ when using a piecewise constant policy approximation. If $N$ is large, however, this can be computationally intensive.

[8]The EWM policy could be improved by excluding candidates arriving in the second half of the year from the simultaneous allocation problem. However, it's important to note that this six-month cutoff is derived from the stationary dynamic policy—something that cannot be determined through simultaneous methods alone.

To address this, we experimented with different values in the denominator, adjusting $\Delta z$ by factors of up to four, and observed no significant differences in the results.

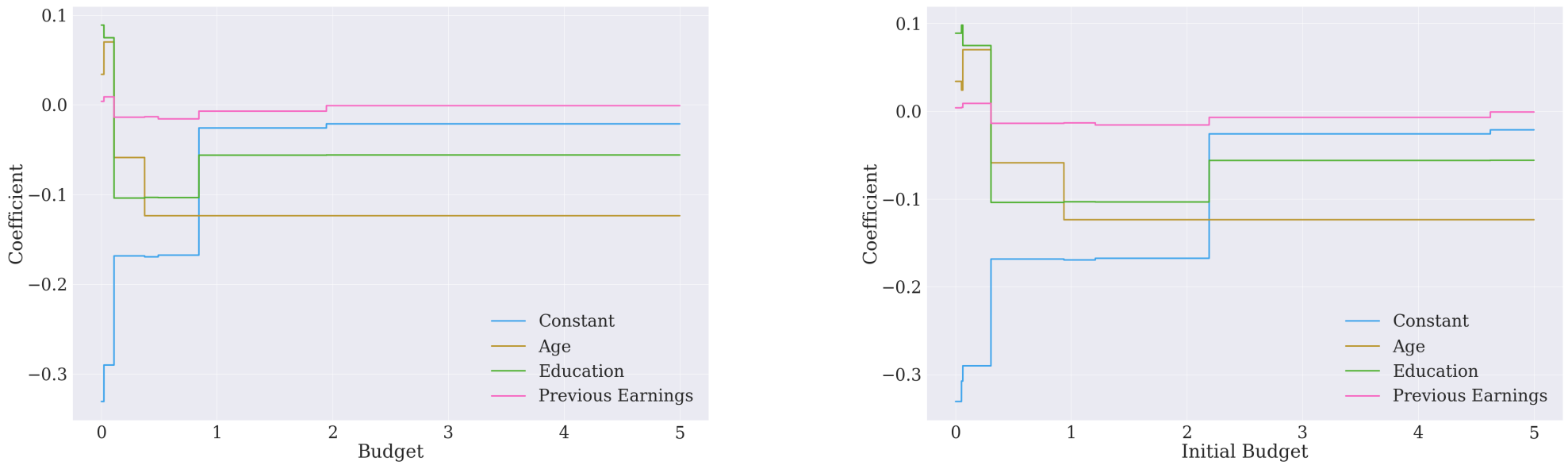

(a) Non-Stationary Policy

(b) Stationary Policy

Figure A3: Effect of Budget on Policies

A budget of 1 corresponds to the expected cost of treating all arrivals of one year. We use a discount factor of $\beta = -\ln(0.9)$, i.e., rewards accruing after 1 year are multiplied by 0.9. The stationary policy is only affected by initial budget as it is restricted to stay constant. The non-stationary policy responds to the remaining budget both at the initial value and over the course of its use.

We also use Setting 1 to assess the sensitivity of our approach to sample size by randomly splitting our data in two 50% subsamples. While we observe some sensitivity, it remains modest: averaging across 50,000 initial budgets from 0.0002 to 10, we find welfare outcomes from the 50% subsamples to be 12% and 15% higher for stationary and non-stationary policies, respectively, compared to the full sample.[9]

***Setting 2.*** Recall that in this setting, we normalize the budget so that $z_0 = 1$ corresponds to sufficient funds for treating 25% of expected arrivals in a given year if the treatment were assigned at random. Because of the variable costs, a strategic policy may therefore be able to treat slightly more than 25% eventually.

[9] Welfare outcomes from each 50% subsample were computed and averaged. The welfare from the first subsample is 10% and 12% higher for stationary and non-stationary policies, while the second subsample yields 14% and 18% higher welfare respectively.

All individual-level covariates $x$ are converted to their respective z-scores. Note that budget $z$ runs from 1 to 0 and time $t$ from 0 to 1. While we discretize time into monthly dummies for the non-stationary logistic policy functions, the neural network-based value functions always use continuous budget and time, along with individual-level covariates, as their states. We find that training results improve if we subtract 0.5 from budget and time, so that they become centered around zero just as the z-scores of the individual-level covariates.

The cost of treating an individual is determined by their approximate unemployment insurance payments. We follow the KROM setting and use the minimum wage as lower bound, the average wage as upper bound, and 80% of previous earnings in between. Since our data is from the late 1980s in the US, we use 419$ as lower and 1667$ as upper bound of monthly income as well as an unemployment duration of 3 months (see U.S. Department of Labor, 2024; Social Security Administration, 2024; St. Louis Fed, 2024).

To determine the optimal policy functions, we implement an adapted version of the PPO algorithm from Spinning Up (Achiam, 2018), tailored to our dynamic RCT environment. While most hyper-parameters can simply be kept at their default values, we observe sensitivity to the learning rate of the policy function when using logistic policies. A grid search over learning rates reveals that $\alpha_\pi = 0.1$ performs robustly for our logistic policies, which is notably higher than the default $\alpha_\pi = 0.0003$ typically recommended for neural networks. For logistic policies, this higher learning rate promotes more effective exploration before reaching relative convergence (where the convergence of coefficients is best evaluated relative to the intercept, as illustrated in Figure A4).[10] The learning rate for the value function

[10] The learning rate 0.1 performs robustly across stationary and non-stationary logistic policy functions and rewards. Small further setting-specific adjustments of this rate could yield a few residual percentage points in welfare in some instances.

($\alpha_v = 0.001$) appears to be less critical, so we retain the default setting. We also set the discount factor to 1 and adjust the steps per epoch and maximum steps per episode to align with the specifics of our environment.

The remaining hyper-parameters for PPO are kept at their default values: the clip ratio for the objective function (0.2), the learning rate for the value function (0.001), the structure of the value function's neural network (two hidden layers with 64 neurons each, using the tanh activation function), the number of training epochs per batch of collected data (80), the discount factor in the generalized advantage estimation (0.97), and the target Kullback-Leibler divergence between old and new policies (0.01). For full details on the algorithm, its pseudo-code, and its parameters, see the paper by Schulman et al. (2017) and the Spinning Up implementation at https://github.com/openai/spinningup/tree/master/spinup/algos/pytorch/ppo which incorporates some further heuristics such as normalization of advantage estimates.

In addition to tuning the learning rates for the policy function, we modify the Spinning Up PPO code to support logistic policy functions, accommodate differences in state representation across value functions, and distinguish between stationary and non-stationary policy functions in our application. We also create a dynamic environment that emulates the KROM setup, designed in a format compatible with standard reinforcement learning environments (Towers et al., 2024; Brockman et al., 2016) and included in the code accompanying this paper. For full project code, refer to the repository at https://github.com/friedrichgeiecke/dynamic_treatment. The broader idea of building dynamic environments—also referred to as *simulators*—based on real-world data and causal inference using RCT or observational data can be generalized. Future research could expand this approach to other datasets, enabling RL algorithm training and testing in new domains.

We train an optimal policy for our environment using the modified PPO algo-

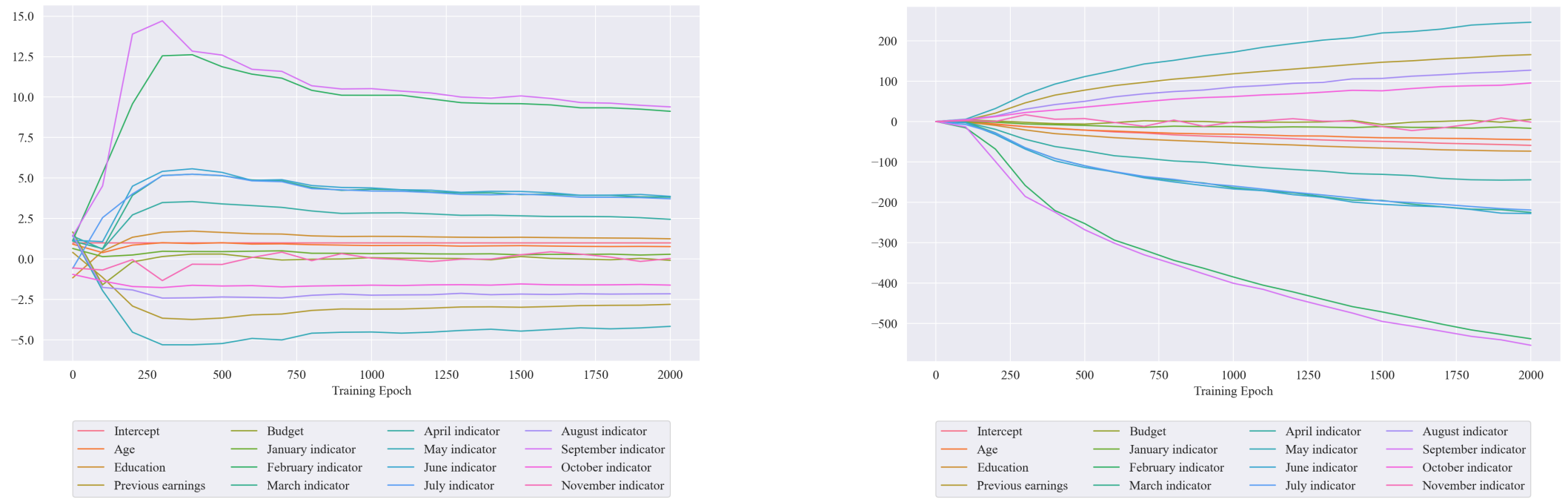

(a) Relative Coefficients over Training Epochs (Divided by $|intercept|$)

(b) Absolute Coefficients over the Training Epochs

Figure A4: Convergence of the Non-Stationary Policy Function Coefficients

rithm with 10 parallel processes (i.e., 10 simultaneous RL actors exploring the environment) over 2000 epochs. In each epoch, gradient-based parameter updates are performed using interaction data collected in buffers across the parallel processes, all of which update the same global parameters. We repeat this training process 50 times with different pseudo-random seeds to generate 50 sets of policy parameters. This approach is applied to both the stationary and non-stationary policies, ultimately selecting the highest-performing seed for each.[11]

To demonstrate that 2000 training epochs are sufficient for training the logistic policies in our problem, Figure A4 displays the convergence of the policy coefficients for the non-stationary policy. The relative values of the coefficients (relative to the intercept) converge rapidly, though their absolute values continue to increase gradually, making the policy more deterministic (i.e., action probabilities nearing 0 or 1). In practice, we can truncate training early and convert the logistic policy rule to a deterministic one for evaluation (i.e., treat if the treatment probability exceeds 50%).

***Neural Network Policies.*** To demonstrate the flexibility of this approach with

[11] Notably, for the stationary policy, most top seeds yield highly similar policy functions.

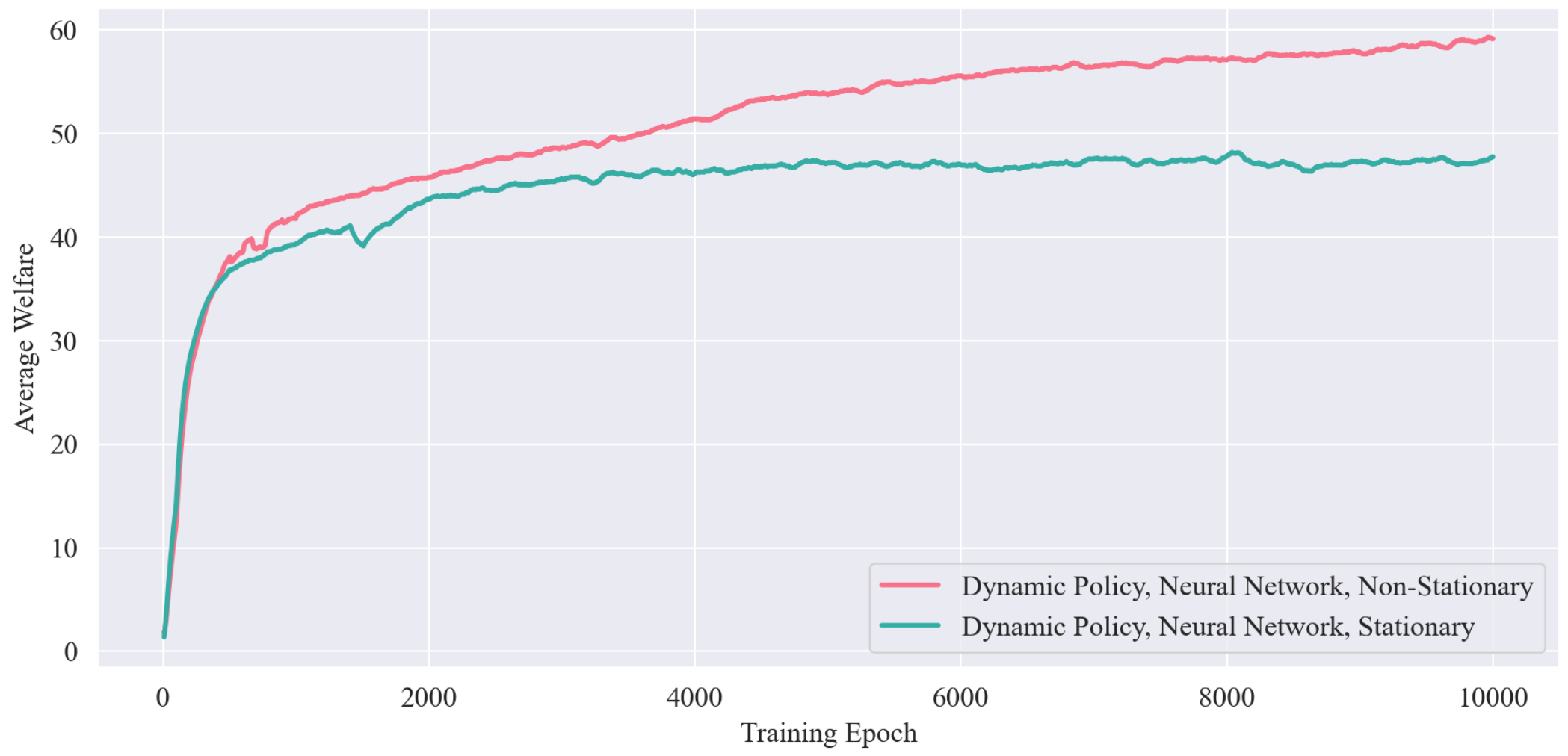

Figure A5: Welfare Achieved on the Dynamic Problem - Neural Network Policies

Learning trajectory for the two dynamic policies start at epoch 10 and are moving averages over the last (where available) 100 training epochs. The welfare under a purely random policy (50% treatment probability) is normalized to 1.

respect to the policy function class, we also train two example neural network policies for Setting 2 over 10,000 epochs. Figure A5 shows the resulting welfare trajectories over the course of training. The best-performing seed for each policy is displayed, though seed-to-seed welfare variation is relatively low for neural network policy functions in this context. The learning rate for these policies is set to 0.01, while all other hyperparameters remain unchanged, aside from those defining the neural network structure. Specifically, the policies follow Spinning Up's default settings with two hidden layers of 64 neurons each, resulting in approximately 4,500 parameters for our non-stationary policy. The policy networks employ tanh activations for the hidden layers and a softmax activation for the output layer. Similar to the value function, the non-stationary neural network policy takes as input individual-level covariates, budget, and continuous time, allowing it to leverage continuous variables without relying heavily on discretization, unlike the logistic policies.

Although these policies achieve substantially higher welfare, careful hyper-parameter tuning and out-of-sample evaluation would be necessary to prevent over-fitting. For decision-makers prioritizing interpretability, logistic policies may be preferable due to their straightforward structure. Compared to the logistic policies, we also observe that the welfare levels under stationary and non-stationary policies are closer to each other. This is likely because neural networks can effectively differentiate individuals based solely on their covariates, making them less reliant on temporal or seasonal information.

# D Additional Application: Non-Perpetual Foundations

Non-perpetual ("sunsetting") foundations are designed to exhaust their assets in pursuit of their mission and dissolve once their assets are exhausted. A prominent example is the Bill & Melinda Gates Foundation, but this structure is also appealing for smaller foundations, where the administrative demands of perpetual existence could detract from focusing resources effectively and timely on their mission (see Galligan and Milway, 2020). While these foundations do not evaluate each candidate project immediately, they do review them periodically. The foundation's lifespan remains unspecified, with its duration implicitly governed by the budget constraint.

Due to this structure, non-perpetual foundations that periodically review and fund projects generally meet the three key assumptions outlined in Setting 1 (see section 5.1). First, applications are commonly reviewed only once. Second, the budget constraint is binding, by the legal definition of a non-perpetual foundation. Third, the assumption that candidates appear sequentially is approximately satisfied: if candidates apply in sequence and decisions are made immediately, this holds exactly. More often, foundations review candidates in periodic batches. If the number of candidates per period is small relative to the total number of candidates, our results still hold approximately; they can even hold exactly under our PDE based methods (see the discussion on batching in Section 3). Whether or not strategic arrival is a concern depends on the specific setting. If it is, a stationary policy would arguably be more appropriate.

As mentioned above, non-perpetual foundations are often motivated by considerations of administrative costs. It is straightforward to incorporate such costs into our setting. We suppose, in addition to the cost of treatment, that the budget declines linearly with time as $\delta z = u\delta t$. Since we normalize the cost of treating everyone in a single year to 1, the parameter $u$ represents the fraction of the budget expended on administrative overhead. For a fixed $u$, the ODE for the evolution of $h_\theta(.)$ changes very slightly to:

$$\beta h_\theta(z) = \bar{r}_\theta(z) - (\bar{\pi}_\theta(1|z) + u)\partial_z h_\theta(z), \qquad h(0) = 0. \tag{44}$$

To illustrate how our methods can be applied in these settings, we use the RCT dataset from Allcott et al. (2020) on the welfare effects of social media. Consider a non-perpetual foundation dedicated to supporting individuals struggling with social media addiction by offering free social-media detox retreats. The foundation aims to maximize cumulative welfare and reviews applications on a weekly basis until its budget is depleted. For our analysis, we employ three covariates—age, education, and earnings—to construct doubly robust estimates of the treatment's effect on the subjective well-being index (SWB) from Allcott et al. (2020). We assume annual operating costs amounting to 5% of the initial budget. For simplicity, we adopt the assumptions of constant treatment costs and applicant arrival rates, as in Setting 1.

As before, the budget is normalized so that $z_0$ represents enough money to treat the expected number of arrivals in a given year. If we assume, for example, 10

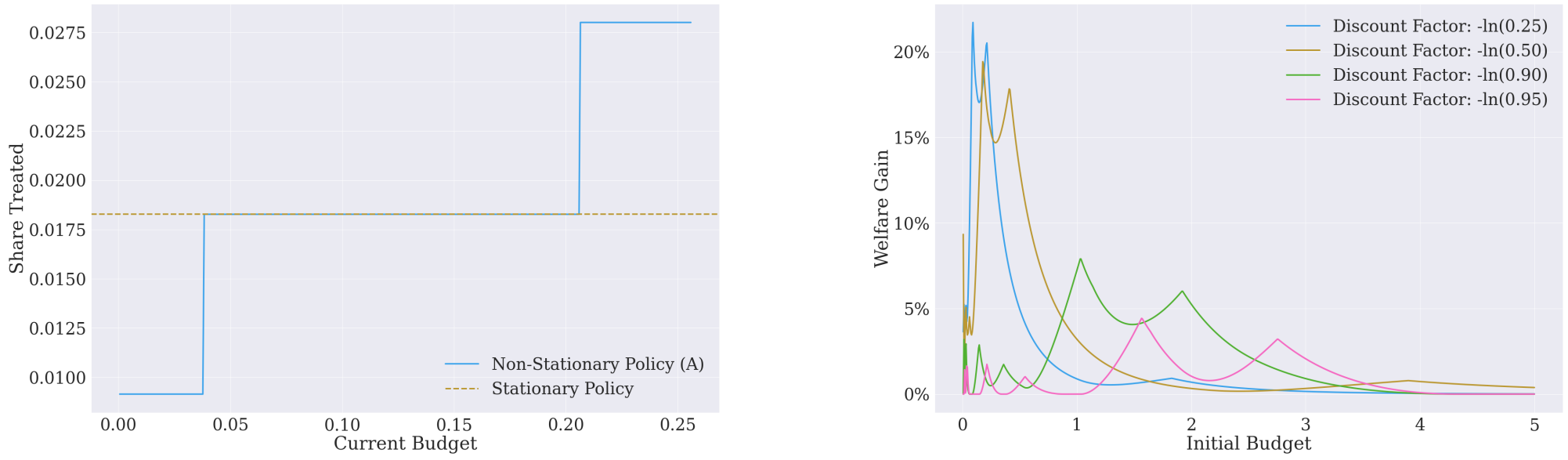

Figure A6: Effects of the Initial Budget

A budget of 1 corresponds to the expected cost of treating all arrivals of one year. The share treated remains constant between an initial budget of 5 and 10. If not indicated otherwise, we use a discount factor of $-\log(0.9)$.

arrivals per week, $z_0 = 1$ indicates the availability of funds to treat 520 candidates.

Since operating costs are a fraction of the intial budget $z_0$, the exact value of $z_0$ now influences the optimal non-stationary policy. The left panel of Figure A6 shows how the policies change with current budget, under a discount factor of $-ln(0.9)$ and with $z_0 = 0.25$. The right panel replicates the analysis from Figure 1 (right) in the main text, but applied to the current context of social media-detox. As before, the welfare advantage of the non-stationary policy over the stationary one is highly sensitive to both the initial budget and the discount factor. Notably, the equivalent of Figure 1 (right) in this application resembles the findings in the main text (not displayed).

# E Alternative Boundary Conditions

Our methodology can be extended to allow for other types of boundary conditions, beyond the Dirichlet boundary condition considered in the main text. Different examples and constraints require different boundary conditions. We consider the following possibilities, which arguably cover the majority of relevant applications:

**Periodic boundary condition.** When the program continues indefinitely, $t$ is a relevant state variable only as it relates to some periodic quantity, e.g., seasonality. So, in this setting, $\mathscr{U} \equiv \mathbb{R} \times \mathbb{R}$, and we impose the periodic boundary condition:

$$h_\theta(z,t) = h_\theta(z,t+T_p) \ \forall \ (z,t) \in \mathbb{R} \times \mathbb{R}. \tag{45}$$

Here, $T_p$ denotes the period length. This boundary condition can only be valid if $\lambda(t), \bar{G}_\theta(z,t), \bar{r}_\theta(z,t)$ are periodic in $t$ with period length $T_p$. This implies that $\pi_\theta$ should also be periodic.

**Neumann boundary condition.** The Neumann boundary condition is useful for imposing a no-borrowing constraint. Specifically, suppose that instead of a fixed budget, the policy maker receives a flow of funds at the rate $\sigma(z,t)$ with respect to time, and the rate of interest is $b$. Also assume that the policy maker faces a no-borrowing constraint at $z = \underline{z}$, i.e., the budget is not allowed to go below this level. Then, at $z = \underline{z}$, the flow rate of change of $z$ is $\sigma(\underline{z},t) + b\underline{z}$ and flow rewards are 0 (since no individual can be treated). Hence, the evolution of $h_\theta(z,t)$ at $z = \underline{z}$ is governed by

$$\beta h_\theta(z,t) - \{\sigma(z,t) + bz\}\partial_z h_\theta(z,t) - \partial_t h_\theta(z,t) = 0, \quad \text{on } \{\underline{z}\} \times [t_0, T). \tag{46}$$

Intuitively, one can imagine that the coefficients $\bar{G}_\theta(z,t), \bar{r}_\theta(z,t)$ of PDE (6) are discontinuous at $z = \underline{z}$ and this discontinuity gives rise to the boundary condition (46). Thus, boundary conditions of this form allow the dynamics at the boundary to be different from those in the interior. This can also be useful in examples with queues or capacity constraints (e.g., when the queue length is 0, or the capacity is full). The following generalization of (46) accommodates all these examples: set

$\mathscr{U} \equiv (\underline{z}, \infty) \times [t_0, T)$ and the boundary condition to be

$$\beta h_\theta(z,t) - \bar{\sigma}_\theta(z,t)\partial_z h_\theta(z,t) - \partial_t h_\theta(z,t) - \bar{\eta}_\theta(z,t) = 0, \quad \text{on } \{\underline{z}\} \times [t_0, T), \tag{47}$$
$$h_\theta(z,T) = 0, \quad \text{on } (\underline{z}, \infty) \times \{T\}.$$

Here, $\bar{\sigma}_\theta(\underline{z},t)$ and $\bar{\eta}_\theta(\underline{z},t)$ are known functions. We require $\bar{\sigma}_\theta(\underline{z},t) > \delta > 0$ for all $t$, to ensure $z$ cannot decrease after hitting the boundary.

## E.1 Formal Results for the Alternative Boundary Conditions

Lemma E.1 below extends Lemma 1 to the other boundary conditions. First, we need the following assumption.

**Assumption 5.** *For the Neumann boundary condition, $\bar{\sigma}_\theta(\underline{z},t), \bar{\eta}_\theta(\underline{z},t)$ are bounded and Lipschitz continuous in $t$ uniformly over $\theta \in \Theta$. Furthermore, $\bar{\sigma}_\theta(\underline{z},t)$ is uniformly bounded away from* 0*, i.e., $\bar{\sigma}_\theta(\underline{z},t) \geq \delta > 0$.*

**Lemma E.1.** *Suppose that Assumption 1 holds, along with $\beta \geq 0$ for the periodic boundary condition and Assumption 5 for the Neumann boundary condition. Then for each $\theta$, there exists a unique viscosity solution $h_\theta(z,t)$ to PDE (6) under the periodic and Neumann boundary conditions, (45) and (47).*

Unfortunately, extending Theorem 2 to the other boundary conditions is not straightforward. Also, Theorem 2 does not allow for $\beta = 0$ in the Dirichlet setting as well. For these remaining cases, we offer alternative results bounding the regret from using $\hat{\theta}$, obtained as

$$\hat{\theta} = \arg\max_\theta \hat{h}_\theta(z_0, t_0), \tag{48}$$

where $\hat{h}_\theta(z,t)$ is the solution to the empirical PDE (12). While estimation of $\hat{\theta}$ is not always feasible (though PDE methods can be employed in some simple cases), the bounds we obtain are useful as a baseline for the regret when there is no numerical error.

Existence of $\hat{h}_\theta(z,t)$ does not follow from Lemma 1. We need a comparison theorem (Crandall *et al* 1992), which will guarantee existence and uniqueness. A sufficient condition for this is: $G_a(x,z,t), \pi_\theta(x,z,t)$ are uniformly continuous in $(z,t)$ for each $(x,\theta)$. We will therefore assume this below. While certainly onerous - it precludes deterministic policies that vary with $(z,t)$ in the soft-max class (though any $\sigma > 0$ is fine) - we believe more powerful comparison theorems can be devised that eliminate this requirement and leave this as an avenue for future research.

**Theorem E.1.** *Suppose that Assumptions 1-4 hold, along with Assumption 5 for the Neumann boundary condition. Also suppose that $G_a(x,z,t), \pi_\theta(x,z,t)$ are uniformly continuous in $(z,t)$ for each $(x,\theta)$. Then, with probability approaching one,*

$$\sup_{(z,t)\in\bar{\mathscr{U}},\theta\in\Theta} \left|\hat{h}_\theta(z,t) - h_\theta(z,t)\right| \leq C\sqrt{\frac{\nu}{n}},$$

*for the boundary conditions (8) , (47) and (45), the latter requiring $\beta \geq \beta_0$.*

The theorem allows $\beta$ to be arbitrary, even negative, under Dirichlet and Neumann boundary conditions.

## E.2 Proof of Lemma E.1 and Theorem E.1

We start by formally defining the viscosity solution concept under the Neumann boundary condition. Specifically, consider a Cauchy problem with a nonlinear Neu-mann boundary condition (in what follows, let $\mathscr{Z}$ denote the domain of $z$):

$$\begin{aligned} F\left(y,u(y),D(y)\right) &= 0 \text{ on } \mathscr{Z}\times(0,\bar{T}]; \\ B\left(y,u(y),Du(y)\right) &= 0 \text{ on } \partial\mathscr{Z}\times(0,\bar{T}]; \\ u(y) &= 0 \text{ on } \mathscr{Z}\times\{0\}. \end{aligned} \tag{49}$$

Here, $B(\cdot)=0$ is the nonlinear boundary condition. In general, this boundary condition may be over-determined and may not hold everywhere. We thus need some weaker notion of the boundary condition as well. This is provided in the definition below, due to Barles and Lions (1991); see also Crandall *et al* (1992).

**Definition 2.** *A bounded continuous $u(\cdot)$ is a viscosity sub-solution to (49) if:*

*(i) $u(z,0)\leq 0$ for all $z\in\mathscr{Z}$, and*

*(ii) for each $\phi\in\mathscr{C}^2(\bar{\mathscr{Z}}\times[0,\bar{T}])$, if $u-\phi$ has a local maximum at $y\in\bar{\mathscr{Z}}\times(0,\bar{T}]$,*

$$\begin{aligned} F\left(y,u(y),D\phi(y)\right) &\leq 0 \text{ if } y\in\mathscr{Z}\times(0,\bar{T}]; \\ \min\left\{F\left(y,u(y),D\phi(y)\right),B\left(y,u(y),D\phi(y)\right)\right\} &\leq 0 \text{ if } y\in\partial\mathscr{Z}\times(0,\bar{T}]. \end{aligned}$$

*Similarly, a bounded continuous $u(\cdot)$ is a viscosity super-solution to (49) if:*

*(i) $u(z,0)\geq 0$ for all $z\in\mathscr{Z}$, and*

*(ii) for each $\phi\in\mathscr{C}^2(\bar{\mathscr{Z}}\times[0,\bar{T}])$, if $u-\phi$ has a local minimum at $y\in\bar{\mathscr{Z}}\times(0,\bar{T}]$,*

$$\begin{aligned} F\left(y,u(y),D\phi(y)\right) &\geq 0 \text{ if } y\in\mathscr{Z}\times(0,\bar{T}]; \\ \max\left\{F\left(y,u(y),D\phi(y)\right),B\left(y,u(y),D\phi(y)\right)\right\} &\geq 0 \text{ if } y\in\partial\mathscr{Z}\times(0,\bar{T}]. \end{aligned}$$

*Finally, $u(\cdot)$ is a viscosity solution to (49) if it is both a sub- and a super-solution.*

As in Appendix A, we shall suppose that PDE (49) can be written in Hamiltonian form, i.e., $F(y,u,Du)=\partial_t u+H(y,u,\partial_z u)$ for some $H_\theta(\cdot)$. Following Bar-les and Lions (1991), we require $H_\theta(\cdot)$ to satisfy (R1)-(R3) in Appendix A (with $\mathscr{X} \times (0, \bar{T}]$ replacing $\mathscr{U}$ in (R2)) for existence of a unique viscosity solution to (49). In addition, we also need some regularity conditions on $B(\cdot)$:

(R4) $B(y,u,q)$ is uniformly continuous in all its arguments.

(R5) There exists a modulus of continuity $\omega(\cdot)$ such that $\forall\ (y_1, y_2) \in \partial\mathscr{X} \times (0, \bar{T}]$,

$$|B(y_1,u,q) - B(y_2,u,q)| \leq \omega\left(\|y_1 - y_2\|\,|1 + \|q\||\right).$$

(R6) $B(y,u,q)$ is non-decreasing in $u$ for all $(y,q)$.

(R7) Let $n(y)$ denote the outward normal to $\Gamma$ at $y$. There exists $\nu > 0$ such that $B(y,u,q+\lambda n(y)) - B(y,u,q+\mu n(y)) \geq \nu(\lambda - \mu)$ for all $\lambda \geq \mu$.

#### E.2.1 Proof of Lemma E.1

*Periodic boundary condition.* We construct the solution to the periodic boundary condition as the long run limit of a Cauchy problem. In particular, we employ a change of variables $\tau(t) = t^* - t$, where $t^*$ is arbitrary, and let $v_\theta(\cdot)$ denote a solution to the Cauchy problem

$$\begin{aligned} \partial_\tau v_\theta(z,\tau) + \bar{H}_\theta\left(z,\tau,v_\theta(z,\tau),\partial_z v_\theta(z,\tau)\right) &= 0 \quad \text{on } \mathbb{R} \times (0,\infty); \\ v_\theta(z,\tau) &= v_0 \quad \text{on } \mathbb{R} \times \{0\}, \end{aligned}$$

where

$$\bar{H}_\theta(z,\tau,u,p) := \beta u - \lambda(\tau)\bar{G}_\theta(z,\tau)p - \lambda(\tau)\bar{r}_\theta(z,\tau),$$

and $v_0$ is some arbitrary Lipschitz continuous function, e.g., $v_0 = 0$. We then claim that if $\bar{H}_\theta(\cdot)$ is periodic in $\tau$ (which is guaranteed by the fact $\lambda(\cdot), \bar{G}_\theta(z,\cdot), \bar{r}_\theta(z,\cdot)$ are $T_p$-periodic, see Section E), a unique periodic viscosity solution, $h_\theta$, satisfying (45) can be identified as $h_\theta(z,t^*-\tau) = \lim_{m\to\infty} v_\theta(z, mT_p+\tau)$ for all $\tau \in [0,T_p]$. We show this claim by following the arguments of Bostan and Namah (2007, Proposition 5). First, we note that existence of a solution $v_\theta$ to the Cauchy problem is assured by the regularity conditions (R1)-(R3), which are clearly satisfied under Assumption 1 when $\beta \geq 0$. Now, define $v_\theta^+(z,\tau) = v_\theta(z,\tau+T_p)$. By periodicity of $\bar{H}_\theta(\cdot)$, $v_\theta^+(z,\tau)$ is also a viscosity solution to $\partial_\tau v_\theta + \bar{H}_\theta\left(z,\tau,v_\theta,\partial_z v_\theta\right) = 0$ on $\mathbb{R}\times(0,\infty)$. By Lemma G.2, $|v_\theta| \leq M < \infty$ for some $M<\infty$. Combined with the Comparison Theorem for Cauchy problems (Lemma G.4), we obtain

$$\sup_{(z,w)\in\mathbb{R}\times[0,\infty)} |v_\theta^+(z,w) - v_\theta(z,w)| \leq e^{-\beta w} \sup_{z\in\mathbb{R}} |v_\theta^+(z,0) - v_\theta(z,0)| \leq 2e^{-\beta w} M.$$

Setting $w = \tau + mT_p$ in the above equation, and denoting $h_{m,\theta}(z,t^*-\tau) := v_\theta(z, mT_p + \tau)$, we have

$$\sup_{(z,\tau)\in\mathbb{R}\times[0,T_p]} |h_{m+1,\theta}(z,t^*-\tau) - h_{m,\theta}(z,t^*-\tau)| \leq 2e^{-\beta m T_p} M.$$

Thus, when $\beta > 0$, there exists a limit, $h_\theta(z,t^*-\tau)$, to the sequence $\{h_{m,\theta}(z,t^*-\tau)\}_{m=1}^{\infty}$ on the domain $(z,\tau)\in\mathbb{R}\times[0,T_p]$. This limit is periodic in $T_p$, as can be seen from the fact

$$|h_{m,\theta}(z,t^*-T_p-\tau) - h_{m,\theta}(z,t^*-\tau)| := |h_{m+1,\theta}(z,t^*-\tau) - h_{m,\theta}(z,t^*-\tau)| \to 0$$

uniformly over all $\tau \in [0,T_p]$. Additionally, since $h_{m,\theta}(z,t^*-\tau)$ is a viscosity solution to $\partial_\tau v_\theta + H_\theta\left(z,\tau,v_\theta,\partial_\tau v_\theta\right) = 0$ on $\mathbb{R}\times(0,\infty)$ for each $m$, the stability property of viscosity solutions (see, Crandall and Lions, 1983) implies that $h_\theta(z,t^*-\tau)$ is a viscosity solution for this PDE as well. We have thus shown that there ex-ists a periodic viscosity solution, $h_\theta(z,t^*-\tau)$, to $\partial_\tau v_\theta + \bar{H}_\theta(z,\tau,v_\theta,\partial_\tau v_\theta) = 0$ on $\mathbb{R}\times\mathbb{R}$. That it is also unique follows from the Comparison Theorem G.2 for periodic boundary condition problems. But $\partial_\tau v_\theta + H_\theta(z,\tau,v_\theta,\partial_\tau v_\theta) = 0$ is just the time-reversed version of the population PDE (6). Since $t^*$ was arbitrary, this implies $h_\theta(z,t)$ is the unique periodic viscosity solution to PDE (6).

*Neumann boundary condition.* As in the proof of the Dirichlet setting, we start by considering the transformed PDE problem

$$\begin{aligned} \partial_\tau u_\theta + H_\theta(z,\tau,\partial_z u_\theta) &= 0 \quad \text{on } (\underline{z},\infty)\times(0,T]; \\ B_\theta(z,\tau,Du_\theta) &= 0 \quad \text{on } \{\underline{z}\}\times(0,T]; \\ u_\theta(z,\tau) &= 0 \quad \text{on } [\underline{z},\infty)\times\{0\}, \end{aligned} \tag{50}$$

where $H_\theta(\cdot)$ is defined in (19), and (here $q := (q_1,q_2)^\intercal$)

$$B_\theta(z,\tau,q) := -e^{\beta\tau}\bar{\eta}_\theta(z,\tau) - \lambda(\tau)\bar{\sigma}_\theta(z,\tau)q_1 + q_2. \tag{51}$$

As in the Dirichlet setting, it suffices to show existence of a unique solution, $u_\theta$, to (50) since this is related to $h_\theta$ by $h_\theta(z,t) = e^{-\beta(T-t)}u_\theta(z,T-t)$. By Barles and Lions (1991, Theorem 3), a unique solution to (50) exists as long as $H_\theta(\cdot)$ and $B_\theta(\cdot)$ satisfy the regularity conditions (R1)-(R8). It is straightforward to verify (R1)-(R7) under Assumptions 1 and 5 (note that the outward normal to the plane $\{\underline{z}\}\times(0,T]$ is $n = (-1,0)^\intercal$, so (R7) holds as long as $\bar{\sigma}_\theta(z,\tau) > 0$, as assured by Assumption 1(iv)). For (R8), a set of sub- and super-solutions to (50) is given by $-L\tau$ and $L\tau$, where $L > \sup_{\theta,z,\tau}\max\{|\lambda(\tau)\bar{G}_\theta(z,\tau)|,|\bar{\eta}_\theta(z,\tau)|\}$ and Assumptions 1 and 5 guarantee such an $L<\infty$ exists.

### E.2.2 Proof of Theorem E.1

We prove the theorem seperately for the various boundary conditions.

### E.2.3 Dirichlet Boundary Condition

We distinguish two sub-cases, depending on whether $T < \infty$ or $T = \infty$. For our proof we choose the case of $T < \infty$. We later sketch how the proof can be modified to deal with the other, arguably simpler, case where $T = \infty$ (but $\underline{z} > -\infty$).

Without loss of generality, we may set $t_0 = 0$. As in the proof of Lemma 1, we make a change of variable $\tau(t) := T - t$, and employ the transformation $u_\theta(z,\tau) := e^{\beta\tau} h_\theta(z, T-\tau)$. Note that $u_\theta$ is the viscosity solution to (18). Similarly, we also define $\hat{u}_\theta(z,\tau) := e^{\beta\tau}\hat{h}_\theta(z, T-\tau)$, and note that $\hat{u}_\theta$ is the viscosity solution to

$$\partial_\tau \hat{u}_\theta + \hat{H}_\theta(z,\tau,\partial_z \hat{u}_\theta) = 0 \text{ on } \Upsilon, \;\; \hat{u}_\theta = 0 \text{ on } \mathscr{B}; \qquad \text{where} \tag{52}$$

$$\hat{H}_\theta(z,\tau,p) := -e^{\beta\tau}\lambda(\tau)\hat{r}_\theta(z,\tau) - \lambda(\tau)\hat{G}_\theta(z,\tau)p, \tag{53}$$

and $\Upsilon$, $\mathscr{B}$ are defined in (18). Here, existence and uniqueness of $\hat{u}_\theta$, and by extension, of $\hat{h}_\theta$, follows by similar arguments as in the proof of Lemma 1. Indeed, under the conditions for Theorem E.1, $\hat{H}_\theta(\cdot)$ satisfies the regularity properties (R1)-(R3) for all $\theta \in \Theta$ (in particular, note that uniform continuity of $G_a(x,z,t), \pi_\theta(x,z,t)$ implies $\hat{G}_\theta(z,t), \hat{r}_\theta(z,t)$ are also uniformly continuous).

We claim that for each $\theta \in \Theta$, $u_\theta(z,\tau) + \tau C\sqrt{v/n}$ is a viscosity super-solution to (52) on $\Upsilon$, for some appropriate choice of $C$. We show this using the definition of a viscosity super-solution. First, note that $u_\theta(z,\tau) + \tau C\sqrt{v/n}$ is bounded and continuous on $\bar{\Upsilon}$ since so is $u_\theta$ (see, Lemmas G.2 and G.3). Now, take any arbitrary point $(z^*,\tau^*) \in \Upsilon$, and let $\phi(z,\tau) \in C^2(\Upsilon)$ be any function such that $u_\theta(z,\tau) + \tau C\sqrt{v/n} - \phi(z,\tau)$ attains a local minimum at $(z^*,\tau^*)$. This implies $u_\theta(z,\tau) - \varphi(z,\tau)$ attains a local minimum at $(z^*,\tau^*)$, where $\varphi(z,\tau) := -\tau C\sqrt{v/n}+\phi(z,\tau)$. Since $u_\theta(z,\tau)$ is a viscosity solution to (18), it follows

$$\partial_\tau\varphi(z^*,\tau^*)+H_\theta\left(z^*,\tau^*,\partial_z\varphi(z^*,\tau^*)\right)\geq 0.$$

The above implies, after some algebra, that

$$\begin{aligned}
&\partial_\tau\phi(z^*,\tau^*)-e^{\beta\tau^*}\lambda(\tau^*)\hat{r}_\theta(z^*,\tau^*)-\lambda(\tau^*)\hat{G}_\theta(z^*,\tau^*)\partial_z\phi(z^*,\tau^*) \qquad (54)\\
&\geq C\sqrt{\frac{v}{n}}-e^{\beta\tau^*}\bar{\lambda}\left|\hat{r}_\theta(z^*,\tau^*)-\bar{r}_\theta(z^*,\tau^*)\right|-\bar{\lambda}\left|\hat{G}_\theta(z^*,\tau^*)-\bar{G}_\theta(z^*,\tau^*)\right|\left|\partial_z\phi(z^*,\tau^*)\right|,
\end{aligned}$$

where $\bar{\lambda} := \sup_\tau \lambda(\tau)<\infty$ by Assumption 1(ii). We will now show that under some $C<\infty$, the right hand side of (54) is non-negative for all $(\theta,z^*,\tau^*)$. To this end, we first note that Lemma G.3 assures $u_\theta(\cdot,\tau)$ is Lipschitz continuous in its first argument, with a Lipschitz constant $L_1<\infty$ independent of $z,\tau,\theta$. Consequently, for $u_\theta(z,\tau)-\varphi(z,\tau)$ to attain a local minimum at $(z^*,\tau^*)$, it has to be the case $|\partial_z\varphi(z^*,\tau^*)|\leq L_1$. This in turn implies

$$|\partial_z\phi(z^*,\tau^*)|\leq L_1. \qquad (55)$$

In view of (54) - (55) and (20), we can set $C>C_0\bar{\lambda}(e^{\beta T}+L_1)$, under which the right hand side of (54) is bounded away from 0 wpa1. Hence, wpa1, $u_\theta(z,\tau)+\tau C\sqrt{v/n}$ is a viscosity super-solution to (52) on $\Upsilon$. Since $C<\infty$ is independent of $\theta,z,\tau$, this result holds true for all $\theta\in\Theta$.

The function $\hat{u}_\theta$ is a viscosity solution, and therefore, a sub-solution to (52) on $\Upsilon$. At the same time, $u_\theta(z,\tau)+\tau C\sqrt{v/n}\geq 0=\hat{u}_\theta(z,\tau)$ on $\mathscr{B}$ and we have already shown that $u_\theta(z,\tau)+\tau C\sqrt{v/n}$ is a viscosity super-solution to (52) on $\Upsilon$. Furthermore, as noted earlier, $\hat{H}_\theta(\cdot)$ satisfies the regularity conditions (R1)-(R3) for all $\theta \in \Theta$. Consequently, we can apply the Comparison Theorem G.1 to conclude

$$\hat{u}_\theta(z,\tau) - u_\theta(z,\tau) \leq \tau C \sqrt{\frac{\nu}{n}} \quad \forall\,(z,\tau,\theta) \in \bar{\Upsilon} \times \Theta, \quad \text{wpa1}.$$

A symmetric argument involving $u_\theta(z,\tau) - \tau C\sqrt{\nu/n}$ as a sub-solution to (52) implies that an equivalent bound also holds for $u_\theta(z,\tau) - \hat{u}_\theta(z,\tau)$. Converting the above results back to $h_\theta$ and $\hat{h}_\theta$, we obtain

$$\left|\hat{h}_\theta(z,t) - h_\theta(z,t)\right| \leq C(T-t)e^{-\beta(T-t)}\sqrt{\frac{\nu}{n}} \quad \forall\,(z,t,\theta) \in \bar{\mathscr{U}} \times \Theta, \quad \text{wpa1}.$$

This completes the proof of Theorem E.1 for the Dirichlet setting with $T < \infty$.

We now briefly sketch how the proof can be modified in the setting with $T = \infty$, but $\underline{z} > -\infty$. We make the transformation $u_\theta(z,t) = e^{-\beta t}h_\theta(z,t)$, and write the PDE for $u_\theta(z,t)$ in the form

$$\partial_z u_\theta + H_\theta^{(1)}(t,z,\partial_t u_\theta) = 0 \text{ on } \mathscr{U}; \quad u_\theta = 0 \text{ on } \Gamma, \tag{56}$$

where now

$$H_\theta^{(1)}(t,z,p) := e^{-\beta t}\frac{\bar{r}_\theta(z,t)}{\bar{G}_\theta(z,t)} + \frac{p}{\lambda(t)\bar{G}_\theta(z,t)}.$$

Note that $\bar{G}_\theta(z,t) < 0$ by Assumption 2(ii). The rest of the proof can then proceed as before with straightforward modifications, after reversing the roles of $z$ and $t$.

### Periodic Boundary Condition

Choose some arbitrary $t^* > T_p$. Denote $u_\theta(z,\tau) = e^{\beta\tau}h_\theta(z,t^*-\tau)$ and $\hat{u}_\theta(z,\tau) = e^{\beta\tau}\hat{h}_\theta(z,t^*-\tau)$. Existence of $\hat{u}_\theta, \hat{h}_\theta$ follows by similar reasoning as in the proof of Lemma 1. Set $v_0 := u_\theta(z,0)$ and $\hat{v}_0 := \hat{u}_\theta(z,0)$. By Lemma G.1, $u_\theta$ is the viscosity solution to (note that the boundary condition is satisfied by definition of $v_0$)

$$\partial_\tau f + H_\theta(z,\tau,\partial_z f) = 0 \text{ on } \Upsilon \equiv \mathbb{R}\times(0,\infty); \quad f(\cdot,0) = v_0, \tag{57}$$

where $H_\theta(\cdot)$ is defined in (19). Similarly, $\hat{u}_\theta(z,\tau)$ is the viscosity solution to

$$\partial_\tau f + \hat{H}_\theta(z,\tau,\partial_z f) = 0 \text{ on } \Upsilon; \quad f(\cdot,0) = \hat{v}_0, \tag{58}$$

where $\hat{H}_\theta(\cdot)$ is defined in (19). Finally, we also define $\tilde{u}_\theta(z,\tau)$ as the viscosity solution to the Cauchy problem

$$\partial_\tau f + \hat{H}_\theta(z,\tau,\partial_z f) = 0 \text{ on } \Upsilon; \quad f(\cdot,0) = v_0. \tag{59}$$

Note that $\tilde{u}_\theta$ exists and is unique, by the same reasoning as in the proof of Lemma E.1. Also, let $\tilde{h}_\theta(z,t) := e^{-\beta t}\tilde{u}_\theta(z,t^*-t)$.

Observe that $u_\theta$ and $\tilde{u}_\theta$ share the same boundary condition in (57) and (59). Furthermore, Lemma G.5 assures $u_\theta(\cdot,\tau)$ is Lipschitz continuous in its first argument, with a Lipschitz constant $L_1<\infty$ independent of $z,\tau,t,\theta$. Consequently, we can employ the same arguments as those used in the Dirichlet setting to show

$$|\tilde{u}_\theta(z,\tau) - u_\theta(z,\tau)| \le C_1\tau\sqrt{\frac{v}{n}}, \quad \text{wpa1},$$

for some constant $C_1<\infty$ independent of $\theta,z,\tau,t^*$. In terms of $\tilde{h}_\theta$ and $h_\theta$, this is equivalent to

$$\left|\tilde{h}_\theta(z,t^*-\tau) - h_\theta(z,t^*-\tau)\right| \le C_1\tau e^{-\beta\tau}\sqrt{\frac{v}{n}}, \quad \text{wpa1}.$$

Setting $\tau = T_p$ in the above expression, and noting that $h_\theta$ is $T_p$-periodic, we obtain

$$\left|\tilde{h}_\theta(z, t^* - T_p) - h_\theta(z, t^*)\right| \le C_1 T_p e^{-\beta T_p} \sqrt{\frac{v}{n}}, \quad \text{wpa1}. \tag{60}$$

Now, we can also compare $\tilde{u}_\theta$ and $\hat{u}_\theta$ on $\Upsilon$, using the Comparison Theorem G.1 (it is straightforward to note that the regularity conditions are satisfied under the statement of Theorem E.1). This gives us (henceforth, $(f)_+ := \max\{f, 0\}$)

$$\left(\tilde{u}_\theta(z, T_p) - \hat{u}_\theta(z, T_p)\right)_+ \le \left(\tilde{u}_\theta(z, 0) - \hat{u}_\theta(z, 0)\right)_+, \quad \text{wpa1}.$$

Recall that $\tilde{u}_\theta(z, 0) = v_0 = u_\theta(z, 0)$, by definition. Hence,

$$\left(\tilde{u}_\theta(z, T_p) - \hat{u}_\theta(z, T_p)\right)_+ \le \left(u_\theta(z, 0) - \hat{u}_\theta(z, 0)\right)_+, \quad \text{wpa1}.$$

Rewriting the above in terms of $\tilde{h}_\theta, \hat{h}_\theta$ and $h_\theta$, and noting $\hat{h}_\theta$ is $T_p$-periodic, we get

$$e^{\beta T_p}\left(\tilde{h}_\theta(z, t^* - T_p) - \hat{h}_\theta(z, t^*)\right)_+ \le \left(h_\theta(z, t^*) - \hat{h}_\theta(z, t^*)\right)_+, \quad \text{wpa1}. \tag{61}$$

In view of (60) and (61), wpa1,

$$\begin{aligned}
\left(h_\theta(z, t^*) - \hat{h}_\theta(z, t^*)\right)_+ &\le \left(\tilde{h}_\theta(z, t^* - T_p) - \hat{h}_\theta(z, t^*)\right)_+ + C_1 T_p e^{-\beta T_p} \sqrt{\frac{v}{n}} \\
&\le e^{-\beta T_p}\left(h_\theta(z, t^*) - \hat{h}_\theta(z, t^*)\right)_+ + C_1 T_p e^{-\beta T_p} \sqrt{\frac{v}{n}}.
\end{aligned}$$

Rearranging the above expression gives

$$\left(h_\theta(z, t^*) - \hat{h}_\theta(z, t^*)\right)_+ \le C_1 \frac{T_p e^{-\beta T_p}}{1 - e^{-\beta T_p}} \sqrt{\frac{v}{n}}, \quad \text{wpa1}.$$

We can analogously obtain an equivalent upper bound for $\left(\hat{h}_\theta(z, t^*) - h_\theta(z, t^*)\right)_+$.

Since $t^*$ was arbitrary, this concludes the proof of Theorem E.1 for the periodic setting.

**Neumann Boundary Condition**

As before, denote $u_\theta(z,\tau) := e^{\beta\tau} h_\theta(z, T-\tau)$ and $\hat{u}_\theta(z,\tau) := e^{\beta\tau}\hat{h}_\theta(z, T-\tau)$. Existence of $\hat{u}_\theta, \hat{h}_\theta$ follows by a similar reasoning as in the proof of Lemma E.1. Now, $u_\theta(z,\tau)$ is the viscosity solution to (see, Lemma G.1)

$$\begin{aligned} \partial_\tau u_\theta + H_\theta(z,\tau,\partial_z u_\theta) &= 0 \quad \text{on } (\underline{z},\infty)\times(0,T]; \\ B_\theta(z,\tau,\partial_z u_\theta,\partial_\tau u_\theta) &= 0 \quad \text{on } \{\underline{z}\}\times(0,T]; \\ u_\theta(\cdot,0) &= 0, \end{aligned} \tag{62}$$

where $H_\theta(\cdot)$ and $B_\theta(\cdot)$ have been defined earlier in (19) and (51). Similarly, $\hat{u}_\theta$ is the viscosity solution to

$$\begin{aligned} \partial_\tau u_\theta + \hat{H}_\theta(z,\tau,\partial_z \hat{u}_\theta) &= 0 \quad \text{on } (\underline{z},\infty)\times(0,T]; \\ B_\theta(z,\tau,\partial_z \hat{u}_\theta,\partial_\tau \hat{u}_\theta) &= 0 \quad \text{on } \{\underline{z}\}\times(0,T]; \\ \hat{u}_\theta(\cdot,0) &= 0, \end{aligned} \tag{63}$$

where $\hat{H}_\theta(\cdot)$ is defined in (19). As before, the proof strategy is to show that $u_\theta(z,\tau)+\tau C\sqrt{\nu/n}$ and $u_\theta(z,\tau)-\tau C\sqrt{\nu/n}$ are viscosity super- and sub-solutions to (63) for some $C<\infty$.

Denote $w_\theta(z,\tau) := u_\theta(z,\tau)+\tau C\sqrt{\nu/n}$. Clearly, $w_\theta(z,0) = 0 = \hat{u}_\theta(z,0)$. Furthermore, by Lemma G.6, $u_\theta$ is Lipschitz continuous uniformly over $\theta\in\Theta$.[12] Hence, we can recycle the arguments from the Dirichlet setting to show that in a viscosity sense,

$$\partial_\tau w_\theta + \hat{H}_\theta(z,\tau,\partial_z w_\theta) \geq 0 \quad \text{on } (\underline{z},\infty)\times(0,T], \quad \text{wpa1},$$

for some suitable choice of $C$. Thus, to verify that $w_\theta(z,\tau)$ is a super-solution to (63), it remains to show that in a viscosity sense and wpa1,

$$\max\left\{\partial_\tau w_\theta + \hat{H}_\theta(z,\tau,\partial_z w_\theta), B_\theta(z,\tau,\partial_z w_\theta,\partial_\tau w_\theta)\right\} \geq 0 \text{ on } \{\underline{z}\}\times(0,T]. \tag{64}$$

[12] It is straightforward to verify that under Assumption 1, the functions $H_\theta(\cdot)$ and $B_\theta(\cdot)$ satisfy conditions (R1)-(R7) and (R9)-(R10) uniformly over all $\theta\in\Theta$.

Take an arbitrary point $(\underline{z},\tau^*) \in \{\underline{z}\}\times(0,T]$, and let $\phi(z,\tau) \in C^2([\underline{z},\infty)\times(0,T])$ be any function such that $w_\theta(z,\tau)-\phi(z,\tau)$ attains a local minimum at $(\underline{z},\tau^*)$. We then show below that wpa1,

$$\max\left\{\partial_\tau \phi + \hat{H}_\theta(\underline{z},\tau^*,\partial_z \phi), B_\theta(\underline{z},\tau^*,\partial_z \phi,\partial_\tau \phi)\right\} \geq 0, \tag{65}$$

which proves (64).

Observe that if $w_\theta(z,\tau)-\phi(z,\tau)$ attains a local minimum at $(\underline{z},\tau^*)$, then $u_\theta(z,\tau)-\varphi(z,\tau)$ attains a local minimum at $(\underline{z},\tau^*)$, where $\varphi(z,\tau) := -\tau C\sqrt{\nu/n}+\phi(z,\tau)$. Lemma G.6 assures $u_\theta$ is Lipschitz continuous with Lipschitz constant $L_1$. Hence, for $(\underline{z},\tau^*)$ to be a local minimum of $u_\theta(z,\tau)-\varphi(z,\tau)$ relative to the domain $[\underline{z},\infty)\times[0,T]$, it must be the case[13]

$$|\partial_\tau \varphi(\underline{z},\tau^*)| \leq L_1, \text{ and } \partial_z \varphi(\underline{z},\tau^*) \leq L_1. \tag{66}$$

Now, by the fact $u_\theta(z,\tau)$ is a viscosity solution of (62), we have

$$\max\{\partial_\tau \varphi + H_\theta(\underline{z},\tau^*,\partial_z \varphi), B_\theta(\underline{z},\tau^*,\partial_z \varphi,\partial_\tau \varphi)\} \geq 0.$$

[13] It is possible that $\partial_z \varphi(\underline{z},\tau^*) < -L_1$ since $(\underline{z},\tau^*)$ lies on the boundary and we only define maxima or minima relative to the domain $[\underline{z},\infty)\times[0,T]$.

Suppose $B_\theta(\underline{z},\tau^*,\partial_z\varphi,\partial_\tau\varphi)\geq 0$. Then by $\partial_z\varphi=\partial_z\phi$ and $\partial_\tau\varphi=\partial_\tau\phi-C\sqrt{v/n}$, it is easy to verify $B_\theta(\underline{z},\tau^*,\partial_z\phi,\partial_\tau\phi)\geq C\sqrt{v/n}\geq 0$, which proves (65). So let us suppose instead that $B_\theta(\underline{z},\tau^*,\partial_z\varphi,\partial_\tau\varphi)<0$. We will use this to obtain a lower bound on $\partial_z\varphi(\underline{z},\tau^*)$. Indeed, $B_\theta(\underline{z},\tau^*,\partial_z\varphi,\partial_\tau\varphi)<0$ implies

$$\bar{\sigma}_\theta(\underline{z},\tau^*)\partial_z\varphi(\underline{z},\tau^*)>-e^{\beta\tau}\bar{\eta}_\theta(\underline{z},\tau^*)+\partial_\tau\varphi(\underline{z},\tau^*)\geq -C_\eta e^{\beta T}-L_1,$$

where the last inequality follows from Assumption 1(iv) - which ensures $\bar{\eta}_\theta(\underline{z},\tau)$ is bounded above by some constant, say, $C_\eta$ - and (66). But Assumption 1(iv) also assures that $\bar{\sigma}_\theta(\underline{z},\cdot)$ is uniformly bounded away from 0. Hence we conclude $\partial_z\varphi(\underline{z},\tau^*)\geq -L_2$ if $B_\theta(\underline{z},\tau^*,\partial_z\varphi,\partial_\tau\varphi)<0$, where $L_2<\infty$ is independent of $\theta,\tau^*$. Combined with (66), this implies

$$|\partial_z\varphi(\underline{z},\tau^*)|\leq\max\{L_1,L_2\},\quad\text{if}\quad B_\theta(\underline{z},\tau^*,\partial_z\varphi,\partial_\tau\varphi)<0. \tag{67}$$

Now, if $B_\theta(\underline{z},\tau^*,\partial_z\varphi,\partial_\tau\varphi)<0$ as we supposed, it must be the case $\partial_\tau\varphi+H_\theta(\underline{z},\tau^*,\partial_z\varphi,\partial_\tau\varphi)\geq 0$ to satisfy the requirement for the viscosity boundary condition. Then by similar arguments as in the Dirichlet case, we obtain via (67) and (20) that[14]

$$\partial_\tau\phi+\hat{H}_\theta(\underline{z},\tau^*,\partial_z\phi)\geq 0,\quad\text{wpa1},$$

as long as $C>C_0(\exp(\beta T)+\bar{\lambda}\max\{L_1,L_2\})$. We have thereby shown (65).

Returning to the main argument, we have shown by the above that $u_\theta(z,\tau)+\tau C\sqrt{v/n}$ is a super-solution to (63), wpa1. At the same time, $\hat{u}_\theta(z,t)$ is the solution to (63). Furthermore, in view of the assumptions made for Theorem E.1, it is straightforward to verify that $\hat{H}_\theta(\cdot), B_\theta(\cdot)$ satisfy the regularity conditions (R1)-(R7) for all $\theta \in \Theta$. Hence, we can apply the Comparison Theorem G.3 for the Neumann setting to conclude

$$\hat{u}_\theta(z,\tau) - u_\theta(z,\tau) \leq \tau C\sqrt{\frac{v}{n}} \quad \forall\,(z,\tau,\theta) \in [\underline{z},\infty)\times[0,T]\times\Theta, \quad \text{wpa1}.$$

[14] In terms of the notation in (20), the domain $\Upsilon$ should be replaced with $\bar{\Upsilon}\equiv[\underline{z},\infty)\times[0,T]$ here. So, to get the rates in (20), we use the fact that Assumption 2(iii) continuously extends $\pi_\theta(1|s)$ and $G_a(s)$ to the boundary, see Footnote 18 in the main text.

A symmetric argument involving $u_\theta(z,\tau) - \tau C\sqrt{v/n}$ as a sub-solution to (63), implies that an equivalent bound also applies for $u_\theta(z,\tau) - \hat{u}_\theta(z,\tau)$. Rewriting these bounds in terms of $h_\theta$ and $\hat{h}_\theta$, we have thus shown

$$\sup_{z\in[\underline{z},\infty);\theta\in\Theta} \left|\hat{h}_\theta(z,t) - h_\theta(z,t)\right| \leq (T-t)e^{-\beta(T-t)}C\sqrt{\frac{v}{n}}.$$

This concludes our proof of Theorem E.1 for the Neumann boundary condition.

# F Extensions and Additional Details

## F.1 Time Varying Distribution of Covariates

In realistic settings, the distribution of the covariates may change with time. Let $F_t$ denote the joint distribution of covariates and potential outcomes at time $t$. We assume that the conditional distribution of the potential outcomes given $x$ is time-invariant, so the variation in $F_t$ is driven solely by variation in the covariate distribution. The distribution $F_t$ is also in general different from $F$, the distribution from which the data is drawn.

Assuming that the support of $F_t(\cdot)$ lies within that of $F(\cdot)$ for all $t$, we can write $F_t(x) = \int_{\tilde{x}\leq x} w_t(\tilde{x})dF(\tilde{x})$, for some weight function $w_t(\cdot)$.[15] Let $\lambda_x(t)$ denote the covariate specific arrival rates. Then,

$$w_t(x) = \frac{\lambda_x(t)}{\int \lambda_{\tilde{x}}(t) dF(\tilde{x})}.$$

[15] As before, we use the same notation, $F$, for the marginal and joint distributions of $\{x, Y(1), Y(0)\}$.

Our previous results amounted to assuming $\lambda_x(t) \equiv \lambda(t)$ for all $x$. But now the net arrival rate (i.e., averaging across all covariates) is given by $\lambda(t) := \int \lambda_{\tilde{x}}(t) dF(\tilde{x})$.

With the above in mind, the PDE for the evolution of $h_\theta(z,t)$ is the same as (6), but with $F_t$ replacing $F$ in the definitions of $\bar{r}_\theta(z,t), \bar{G}_\theta(z,t)$. If $w_t(x)$, or equivalently, $\lambda_x(t)$, is known or forecast, we can estimate $F_t$ using $F_{n,t} := n^{-1} \sum_i w_t(x_i) \delta(x_i)$, where $\delta(\cdot)$ denotes the Dirac delta function. Based on this, we can construct our sample dynamic environment by replacing $F_n$ with $F_{n,t}$ in Section 4 (e.g., for the PPO algorithm we would draw observations at random from $F_{n,t}$ instead of $F_n$). With known weights, an extension of the methods of Athey and Wager (2021) shows that Theorems E.1 and 2 continue to hold.

More realistically, however, $\lambda_x(\cdot)$ can often only be estimated or forecast at the level of finite bins, with $\lambda_x(t) \equiv \lambda_j(t)$ for each $x$ in bin $j$. This is the case, for instance, in our empirical example (Setting 2), where we estimate the arrival distributions and arrival rates separately for each calendar month (they are assumed to be constant within the month). In such cases, we would approximate $w_t(\cdot)$ with a piece-wise constant function $\hat{w}_t(\cdot)$ given by $\hat{w}_t(j) = \lambda_j(t) / \sum_j \lambda_j(t)$ for each bin $j$.

## F.2 Regret Bounds when the Outcomes are affected by $z, t$

In some applications $E[Y(a,z,t)|s] = \mu_a(s)$ may depend on all of $s$. Suppose that consistent estimation of $\mu_a(s)$ is possible. Then, we can estimate the rewards as $\hat{r}(s,1) = \hat{\mu}_1(s) - \hat{\mu}_0(s)$, and use this along with $F_n$ to obtain the sample PDE (12).

Suppose there exists a sequence $\psi_n$ such that, for $a \in \{0,1\}$,

$$\sup_{x,(z,t)\in\bar{\mathscr{U}}} |\hat{\mu}_a(x,z,t) - \mu_a(x,z,t)| = O_p(\psi_n^{-1}). \tag{68}$$

Primitive conditions for the above can be obtained on a case-by-case basis. Also, letting $\mathrm{VC}(\cdot)$ denote the VC dimension, suppose that for $a \in \{0,1\}$,

$$\mathrm{VC}\left(\bar{\mathscr{I}}_a\right) < \infty; \text{ where } \bar{\mathscr{I}}_a := \left\{\mu_a(\cdot,z,t)\pi_\theta(1|\cdot,z,t) : (z,t) \in \bar{\mathscr{U}}, \theta \in \Theta\right\}. \tag{69}$$

Under these assumptions, we can follow Kitagawa and Tetenov (2018) to show[16]

$$\sup_{(z,t)\in\bar{\mathscr{U}},\theta\in\Theta} |\hat{r}_\theta(z,t) - \bar{r}_\theta(z,t)| = O_p(\psi_n^{-1}).$$

We then have the following counterpart to Theorem 2 (a similar counterpart to Theorem E.1 also exists), the proof of which follows the same reasoning and is thus omitted:

**Theorem F.1.** *Suppose that Assumptions 1-3 hold, along with (68) & (69), and* $\beta > 0$. *Then, with probability approaching one, there exists* $C < \infty$ *such that*

$$\sup_{(z,t)\in\bar{\mathscr{U}},\theta\in\Theta} \left|\tilde{h}_\theta(z,t) - h_\theta(z,t)\right| \le C\left\{\psi_n^{-1} + b_n^{-1/2}\right\}$$

*under the Dirichlet boundary condition (8). Furthermore, there exists* $\beta_0 > 0$ *such that the above also holds under the periodic boundary condition (45) as long as* $\beta \ge \beta_0$.

[16] On the other hand, the rate for $\left|\hat{G}_\theta(z,t) - \bar{G}_\theta(z,t)\right|$ is still $\sqrt{v/n}$.

## F.3 Updating the Data and Online Learning

If the relevant treatment rewards, $Y$, are observable outside of a formal study (e.g., without an RCT) but subject to delays, the data can be regularly updated to re-estimate the coefficients with each new data set. This approach also allows for periodic recalibration of candidate arrival rate forecasts. If rewards are observed instantly (though this is not true in our setting), the PPO algorithm used to compute our optimal policy can be applied in an online manner. However, this approach is not welfare-efficient, as it does not leverage known dynamics, such as the budget's law of motion or the fact that the distribution $F$ is independent of $z$. Instead, we propose a more efficient hybrid approach that combines a general Actor-Critic (AC) algorithm, such as PPO, with Gradient Bandit algorithms, leading to what we term 'AC with *decision-time estimation* of value functions'. Specifically, at each state $(x,z,t)$, and before administering an action, $h_\theta$ is re-estimated.

To describe the procedure, let $\{Y_i, X_i, Z_i, T_i, A_i\}_{i=1}^n$ denote the sequence of $n$ observations before the state $(X_{n+1}, Z_{n+1}, T_{n+1})$ corresponding to the $n+1$-th arriving individual. Here, $A_i \sim \text{Bernoulli}(\pi_{\theta_i}(1|S_i))$ are the actions of the policy maker, with $\theta_i$ denoting the policy parameter at observation $i$, and $S_i := (X_i, Z_i, T_i)$. Based on these observations, we estimate the rewards as

$$\hat{r}^{(n)}(X_i,1) = \hat{\mu}(X_i,1) - \hat{\mu}(X_i,0) + (2A_i - 1)\frac{Y_i - \hat{\mu}(X_i,A_i)}{A_i\pi_{\theta_i}(1|S_i) + (1-A_i)(1-\pi_{\theta_i}(1|S_i))}$$

for $i = 1,\dots,N$, where $\hat{\mu}(x,w)$ is estimated non-parametrically. Let $F_n$ denote the empirical distribution implied by $\{Y_i, A_i, X_i\}_{i=1}^n$. Then, using $\hat{F}_n$ and $\hat{r}^{(n)}$, along with the current forecast $\lambda(\cdot)$, we compute an estimate $\hat{h}_{\theta_n}(z,t) := \nu_n^\intercal \phi_{z,t}$ of $h_{\theta_n}$, where $h_{\theta_n}$ is the integrated value function under the current policy parameter $\theta_n$. In particular, the value weight $\nu_n$ is computed using TD learning (see, Sutton and Barto, 2018, Chapter 6) by simulating multiple episodes using the sample dynamics generated by $\hat{F}_n, \hat{r}^{(n)}(\cdot)$. We suggest initializing the TD-learning step with the previous weights $v_{n-1}$ to ensure fast convergence to the new $v_n$ (this can be further speeded up with parallel updates). Based on the value of $\hat{h}_{\theta_n}$, we update $\theta$ as

$$\theta_{n+1} = \theta_n + \alpha_{n,\theta} e^{-\beta(T_n - T_0)} \delta_n(S_n, S'_{n+1}, A_n) \nabla_\theta \ln \pi_{\theta_n}(A_n|S_n),$$

for some learning rate $\alpha_{n,\theta}$, where

$$\begin{aligned}&\delta_n(S_n, S'_{n+1}, A_n)\\ &:= \hat{r}^{(n)}(S_n, A_n) + \mathbb{I}\{(Z_{n+1}, T_{n+1}) \in \mathscr{U}\}\, e^{-\beta(T_{n+1} - T_n)} \hat{h}_{\theta_n}(Z_{n+1}, T_{n+1}) - \hat{h}_{\theta_n}(Z_n, T_n).\end{aligned}$$

Following this, we administer an action $A_{n+1} \sim \text{Bernoulli}(\pi_{\theta_{n+1}}(1|S_n))$. This results in an instantaneous outcome $Y_{n+1}$, and an evolution to a new state $S_{n+2}$. We then repeat the above steps with the new state, and continue in this fashion indefinitely.

Note that in contrast to the estimation of the value function $\hat{h}_\theta$, the policy function is only updated online, once at each state (i.e., unlike the estimation of $\hat{h}_\theta$, we do not update it with simulated data). This strategy is similar to Gradient Bandit algorithms (see, Sutton and Barto, 2018, Chapter 2).

A key tuning parameter in this approach is the learning rate, $\alpha_{n,\theta}$, which must be carefully selected to achieve an optimal balance between exploration and exploitation. The theoretical requirements for the learning rates are similar to those needed for convergence in stochastic gradient descent, namely, $\sum_n \alpha_{n,\theta} = \infty$ and $\sum_n \alpha_{n,\theta}^2 < \infty$ (these conditions also apply to Gradient Bandit algorithms). While $\alpha_{n,\theta} = 1/n$ technically meets these conditions, it may be too slow in practical applications. Determining the optimal choice of $\alpha_{n,\theta}$ is, however, beyond the scope of this paper.

Under this approach, the estimation error for $\hat{h}_\theta$ decreases with the number of individuals observed, regardless of the degree of exploration over the $(z,t)$ space. By Theorems 2 and E.1, if there are $n$ observations before reaching state $s$, then $\sup_{\theta,z,t} |\hat{h}_\theta(z,t) - h_\theta(z,t)| \lesssim \sqrt{v/n}$. This property is beneficial because, in many of our examples, we only occasionally revisit the vicinity of any given state (e.g., if the policy duration spans a year, we may encounter similar values of $(z,t)$ only annually).

## F.4 Alternative Welfare Criteria

In the main text, we treat the arrival rates $\lambda(\cdot)$ as forecasts and measure welfare based on their "forecasted" values. Alternatively, rather than relying on point forecasts, we could use density forecasts.

### F.4.1 Density Forecasts

Denote each forecast for the arrival rates by $\lambda(t;\xi)$, where $\xi$ indexes the forecasts. For example, in consensus or ensemble forecasts, each $\xi$ may represent a different estimate or model. For each $\xi$, we can obtain the integrated value function $h_\theta(z,t;\xi)$ by replacing $\lambda(t)$ in (6) with $\lambda(t;\xi)$. Let $P(\xi)$ denote some - possibly subjective - probability distribution that the policy maker places over the forecasts. We take this distribution as given. Then we define the 'forecast' welfare as

$$W_\theta(z,t) = \int h_\theta(z,t;\xi)dP(\xi).$$

The policy maker's problem is to then choose $\theta^*$ such that

$$\theta^* = \arg\max_{\theta\in\Theta} W_\theta(z_0,t_0).$$

It is straightforward to extend our AC algorithm to multiple forecasts: we simply draw a value of $\xi$ from $P(\xi)$ at the start of every new episode. In consensus or ensemble forecasts, this involves drawing a model at random based on the weights given to each of them.

### F.4.2 Estimation of $\lambda(t)$

One could alternatively calculate the welfare based on an unknown but true value of $\lambda(t)$, denoted by $\lambda_0(\cdot)$. Denote the integrated value function under $\lambda_0(\cdot)$ by $h_\theta(z,t;\lambda_0)$. Under this alternative welfare criterion, the optimal choice of $\theta$ is

$$\theta_0^* = \arg\max_{\theta\in\Theta} h_\theta(z_0,t_0;\lambda_0).$$

To simplify matters, assume that we only have access to a single point forecast or estimate of $\lambda_0(\cdot)$, denoted by $\hat{\lambda}(\cdot)$. The extension to density estimates is straightforward, so we do not consider it here. We can use the observational data and the estimate $\hat{\lambda}(\cdot)$ to obtain an empirical counterpart, $\hat{h}_\theta(z,t;\hat{\lambda})$, of $h_\theta(z,t;\lambda_0)$, where $\hat{h}_\theta(\cdot;\hat{\lambda})$ is the solution to PDE (12) in the main text, with $\lambda(\cdot)$ replaced by $\hat{\lambda}(\cdot)$. This suggests the following maximization problem for estimating the optimal policy:

$$\hat{\theta} = \arg\max_{\theta\in\Theta} \hat{h}_\theta(z_0,t_0;\hat{\lambda}).$$

Note that the definition of $\hat{\theta}$ above is similar to that in the main text (cf. equation 48), except for employing $\hat{\lambda}(\cdot)$ in place of $\lambda(\cdot)$.

In terms of the statistical properties, we now have to take into account the statistical uncertainty between $\hat{\lambda}(\cdot)$ and $\lambda_0(\cdot)$ while calculating the regret. Typically, estimation of $\lambda_0(\cdot)$ is orthogonal to estimation of the treatment effects (which are used for estimating $\bar{r}_\theta(z,t)$). Indeed, $\hat{\lambda}(\cdot)$ may be obtained from a completely different and much bigger dataset, e.g., for estimating unemployment rates, we can potentially make use of large survey data, whereas the observational dataset for estimating the rewards is typically much smaller.

We can decompose the regret into two parts: the first dealing with estimation of the treatment effects, and the other with the estimation of $\lambda_0(\cdot)$. Formally, letting $\mathscr{R}_0(\hat{\theta})$ denote the regret under the present welfare criterion, we have

$$\begin{aligned}
\mathscr{R}_0(\hat{\theta}) &:= h_{\hat{\theta}}(z_0,t_0;\lambda_0) - h_{\theta_0^*}(z_0,t_0;\lambda_0) \\
&= \left\{h_{\hat{\theta}}(z_0,t_0;\hat{\lambda}) - h_{\theta_0^*}(z_0,t_0;\hat{\lambda})\right\} + \dots \\
&\quad \dots + \left\{h_{\hat{\theta}}(z_0,t_0;\lambda_0) - h_{\hat{\theta}}(z_0,t_0;\hat{\lambda}) + h_{\theta_0^*}(z_0,t_0;\lambda_0) - h_{\theta_0^*}(z_0,t_0;\hat{\lambda})\right\} \\
&\leq \left\{h_{\hat{\theta}}(z_0,t_0;\hat{\lambda}) - h_{\theta_0^*}(z_0,t_0;\hat{\lambda})\right\} + 2\sup_{\theta\in\Theta}\left|h_{\theta}(z_0,t_0;\lambda_0) - h_{\theta}(z_0,t_0;\hat{\lambda})\right| \\
&:= \mathscr{R}_0^{(I)} + \mathscr{R}_0^{(II)}.
\end{aligned}$$

The first term, $\mathscr{R}_0^{(I)}$, can be analyzed using the techniques developed so far. Indeed,[17]

$$\mathscr{R}_0^{(I)} \leq 2\sup_{\theta\in\Theta}\left|\hat{h}_{\theta}(z_0,t_0;\hat{\lambda}) - h_{\theta}(z_0,t_0;\hat{\lambda})\right| \leq 2C\sqrt{\frac{v}{n}} \quad \text{wpa1}.$$

As for the second term, we can analyze it using the same PDE techniques as that used in the proof of Theorem E.1. This gives us

$$\mathscr{R}_0^{(II)} \leq 2C_1 \sup_{t\in[t_0,\infty)}\left|\lambda_0(t) - \hat{\lambda}(t)\right|,$$

where the constant $C_1$ depends only on (1) the upper bound $M$ for $|\bar{G}_\theta(z,t)|$ and $|\bar{r}_\theta(z,t)|$, and (2) the uniform Lipschitz constants for $\bar{G}_\theta(z,t)$ and $\bar{r}_\theta(z,t)$.[18] In

[17] We require $\hat{\lambda}(\cdot)$ to be uniformly upper bounded and bounded away from 0. This is clearly satisfied wpa1 if $\hat{\lambda}(\cdot) - \lambda_0(\cdot) = o_p(1)$ and $\lambda_0(\cdot)$ is upper bounded and bounded away from 0.

[18] Assumption 1 assures that all these quantities are indeed finite.

particular, we emphasize that $\mathscr{R}_0^{(II)}$ is independent of the complexity $v$ of the policy space. It may even be independent of $n$ if, e.g., $\hat{\lambda}(\cdot)$ is constructed using a different dataset.

Combining the above, we have thus shown

$$\mathscr{R}_0(\hat{\theta}) \leq 2C\sqrt{\frac{v}{n}} + 2C_1 \sup_{t\in[t_0,\infty)} \left|\lambda_0(t) - \hat{\lambda}(t)\right|.$$

Thus, the regret rate is exactly the same as that derived in the main text, except for an additional term dealing with estimation of $\lambda_0(\cdot)$.

## F.5 Discussion of Assumption 1

We provide some primitive conditions under which the soft-max policy class (4) satisfies Assumption 1(i). Recall that the soft-max class of policy functions is of the form

$$\pi_\theta(1|s) = \frac{\exp(\theta^\intercal f(s))}{1+\exp(\theta^\intercal f(s))},$$

where $f(\cdot)$ denotes a vector of basis functions over $s$. Let $\mathbb{S}^{k-1} \equiv \{\tilde{\theta} \in \mathbb{R}^k : \|\tilde{\theta}\| = 1\}$. The following conditions are sufficient to show Assumption 1(i):

**Assumption R.** *(i) $G_a(s)$ and $r(s,1)$ are uniformly bounded. Furthermore, there exists $C < \infty$ such that $E_{x\sim F}[|\nabla_{(z,t)}G_a(s)|] < C$ and $E_{x\sim F}[|\nabla_{(z,t)}r(s,1)|] < C$ uniformly over all $(z,t) \in \mathscr{U}$.*

*(ii) There exists $M < \infty$ independent of $(x,z,t)$ such that $|\nabla_{(z,t)}f(s)| \leq M$. This can be relaxed to $E_{x\sim F}[|\nabla_{(z,t)}f(s)|] \leq M$ if $\|\theta\|$ is bounded away from $\infty$.*

*(iii) Either $\|\theta\|$ is bounded away from $\infty$, or, there exists $\delta > 0$ such that the probability density function of $\tilde{\theta}^\intercal f(s)$ in the interval $[-\delta,\delta]$ is bounded for each $\tilde{\theta} \in \mathbb{S}^{k-1}$ and $(z,t) \in \mathscr{U}$.*

Assumption R(i) imposes some regularity conditions on $G_a(s)$ and $r(s,1)$. In our empirical example, these quantities do not even depend on $(z,t)$, so the assumption is trivially satisfied there. Assumption R(ii) ensures that $f(s)$ varies smoothly with $(z,t)$. Assumption R(iii) provides two possibilities. If $1\|\theta\|$ is compactly supported, it easy to see that the derivatives of $\pi_\theta(\cdot|s)$ with respect to $(z,t)$ are bounded, but this constrains the ability of the policy class to approximate deterministic policies. As an alternative, we can require that the probability density function of $\tilde{\theta}^\intercal f(s)$ around 0 is bounded for any given $(z,t)$ and $\tilde{\theta} \in \mathbb{S}^{k-1}$. It is easy to verify that this alternative condition holds as long there exists at least one continuous covariate, the coefficient of $\tilde{\theta}$ corresponding to that covariate is non-zero, and the conditional density of that covariate given the others is bounded away from $\infty$. The case of discrete covariates with $\|\theta\| \to \infty$ presents some difficulties and is discussed in the next sub-section. Under Assumptions R, it is straightforward to show the following result:

**Proposition 1.** *Suppose that Assumptions R(i)-R(iii) hold. Then $\bar{G}_\theta(z,t)$ and $\bar{r}_\theta(z,t)$ are Lipschitz continuous uniformly over $\theta$.*

### F.5.1 Discrete Covariates with arbitrary $\|\theta\|$

With purely discrete covariates and arbitrary $\|\theta\|$, $\bar{G}_\theta(z,t)$ and $\bar{r}_\theta(z,t)$ are generically discontinuous, except when the policy is independent of $(z,t)$. Nevertheless, depending on the boundary condition, we can allow for some discontinuities and still end up with a Lipschitz continuous solution. For instance, the results of Ishii (1985) imply a comparison theorem (akin to Theorem G.1 in Section G) can be derived under the following alternative to Assumption 1(i):

**Assumption 1a.** *Suppose that the boundary condition is either a periodic one, or of the Cauchy form $h_\theta(z,T) = 0\ \forall\ z$. We can then replace Assumption 1(i) with the following: $\bar{G}_\theta(z,t)$ and $\bar{r}_\theta(z,t)$ are integrable in t for any $(z,\theta)$, and Lipschitz* *continuous in $z$ uniformly over $(t,\theta)$. A similar condition also holds, with the roles of $z,t$ reversed, if the boundary condition is in the form $h_\theta(\underline{z},t)=0\ \forall\, t$.*

The above condition is also sufficient for proving (uniform) Lipschitz continuity of $h_\theta(z,t)$. This enables us to verify our main Theorems 2 and E.1 under the weaker Assumption 1a.

The general case, when $\bar{G}_\theta(z,t)$ and $\bar{r}_\theta(z,t)$ may be discontinuous in both arguments, is more difficult, but we offer the following comments: Suppose that there are $K$ distinct covariate groups in the population. Then we can create $2^K$ strata, each corresponding to regions of $(z,t)$ where the (deterministic) policy function takes the value 1 for exactly one particular subgroup from the $K$ groups. In this way, we can divide the space $\mathscr{U}$ into discrete regions, also called stratified domains, within which $\bar{G}_\theta(z,t)$ and $\bar{r}_\theta(z,t)$ are constant. Discontinuities occur at the boundaries between the strata. Under some regularity conditions, Barles and Chasseigne (2014) demonstrate existence and uniqueness of a solution in this context, and also prove a comparison theorem. It is unknown (to us), however, whether this solution is Lispchitz continuous.

# G Properties of Viscosity Solutions

This section collects various properties of viscosity solutions for the proofs of Theorems 2 and E.1. Our first lemma concerns the relationship between the population PDE (6) from the main text, and the 'transformed' PDE (18) introduced in Appendix A. The population PDE (6) is given by

$$F_\theta(z,t,f,Df)=0 \text{ on } \mathscr{U};\quad \text{where,} \tag{70}$$

$$F_\theta(z,t,u,q_1,q_2):=\beta u-\lambda(t)\bar{G}_\theta(z,t)q_1-q_2-\lambda(t)\bar{r}_\theta(z,t),$$

and $\mathscr{U}$ is some open set. The transformed PDE is given by

$$\partial_\tau f + H_\theta(z,\tau,\partial_z f) = 0 \text{ on } \Upsilon; \quad \text{where,} \tag{71}$$
$$H_\theta(z,\tau,p) := -e^{\beta\tau}\lambda(\tau)\bar{r}_\theta(z,\tau) - \lambda(\tau)\bar{G}_\theta(z,\tau)p,$$

and $\Upsilon := \{(z, T-t) : (z,t) \in \mathscr{U}\}$. The following lemma shows that there is a one-to-one relationship between the viscosity solutions to these PDEs:

**Lemma G.1.** *If $u_\theta$ is a viscosity solution to (71) on $\Upsilon$, then $e^{-\beta(T-t)}u_\theta(z,T-t)$ is a viscosity solution to (70) on $\mathscr{U}$. Similarly, if $h_\theta$ is a viscosity solution to (70) on $\mathscr{U}$, then $e^{\beta\tau}h_\theta(z,T-\tau)$ is a viscosity solution to (71) on $\Upsilon$.*

*Proof.* We only prove the first claim, as the proof of the other claim is analogous. Suppose that $u_\theta$ is a viscosity solution to (71). We will show that $\tilde{h}_\theta(z,t) := e^{-\beta(T-t)}u_\theta(z,T-t)$ is a viscosity solution to (70) on $\mathscr{U}$. To this end, consider any $\phi \in \mathscr{C}^2(\mathscr{U})$ such that $\tilde{h}_\theta(z,t) - \phi(z,t)$ attains a local maximum at some $(z^*,t^*) \in \mathscr{U}$. It is without loss of generality to suppose that $\tilde{h}_\theta(z^*,t^*) - \phi(z^*,t^*) = 0$, as the requirements for a viscosity solution only involve the derivatives of $\phi$ and we can therefore always add or subtract a constant to $\phi$. We then have $e^{\beta(T-t)}\{h_\theta(z,t) - \phi(z,t)\} \leq 0$ for all $(z,t)$ in a neighborhood of $(z^*,t^*)$, i.e., $e^{\beta(T-\tau)}\{h_\theta(z,t) - \phi(z,t)\}$ also attains a local maximum at $(z^*,t^*)$. This implies $e^{\beta\tau}\{h_\theta(z,T-\tau) - \phi(z,T-\tau)\}$, as a function of $(z,\tau)$, attains a local maximum at $(z^*,T-t^*) \in \Upsilon$, or, equivalently, $u_\theta(z,\tau) - \tilde{\phi}(z,\tau)$ attains a local maximum at $(z^*,T-t^*) \in \Upsilon$, where $\tilde{\phi}(z,\tau) := e^{\beta\tau}\phi(z,T-\tau)$. Since $u_\theta$ is a viscosity solution, we thus have

$$\partial_\tau\tilde{\phi}(z^*,T-t^*) + H_\theta(z^*,T-t^*,\partial_z\tilde{\phi}(z^*,T-t^*)) \leq 0,$$

and therefore, after getting rid of a positive multiplicative constant, $e^{\beta(T-t^*)}$, we get

$$\beta\tilde{h}_\theta(z^*,t^*)-\lambda(t^*)\bar{G}_\theta(z^*,t^*)\partial_z\phi(z^*,t^*)-\partial_t\phi(z^*,t^*)-\lambda(t^*)\bar{r}_\theta(z^*,t^*)\leq 0,$$

where we have made use of the definitions of $H_\theta(\cdot)$ and $\tilde{\phi}(\cdot)$, along with the fact $\tilde{h}_\theta(z^*,t^*)=\phi(z^*,t^*)$. The above implies that $\tilde{h}_\theta(z,t)$ is a viscosity sub-solution to PDE (70) on $\mathscr{U}$. By an analogous argument, we can similarly show $\tilde{h}_\theta(z,t)$ is a viscosity super-solution to PDE (70) on $\mathscr{U}$. Hence, $\tilde{h}_\theta(z,t):=e^{-\beta(T-t)}u_\theta(z,T-t)$ is a viscosity solution to PDE (70) on $\mathscr{U}$. □

While Lemma G.1 is only stated for the interior domain $\mathscr{U}$, it is straightforward to extend it to the boundary. For the Dirichlet boundary condition, it is easy to verify that if $u_\theta=0$ on $\mathscr{B}\equiv\{(z,T-t):(z,t)\in\Gamma\}$, then $\tilde{h}_\theta(z,t):=e^{-\beta(T-t)}u_\theta(z,T-t)=0$ on $\Gamma$ (an analogous statement also holds for $h_\theta$). One can prove similar claims for the Neumann boundary conditions as well, using the same arguments. Hence, the relationship between the viscosity solutions $h_\theta(z,t)$ and $u_\theta(z,t)$ holds under all the boundary conditions considered in this paper. These results imply that $h_\theta$ exists and is unique if and only if $u_\theta$ exists and is unique as well.

In the remainder of this section, we collect properties of viscosity solutions broken down into separate cases for each of the boundary conditions:

## G.1 Dirichlet Boundary Condition

Consider the Dirichlet problem for PDEs in Hamiltonian form:

$$\partial_t f+H(z,t,f,\partial_z f)=0 \text{ on } \mathscr{U}; \quad u=0 \text{ on } \Gamma. \tag{72}$$

The following comparison theorem is due to Crandall and Lions (1986, Theorem 1). Recall the notation $(f)_+:=\max\{f,0\}$.

**Theorem G.1.** ***(Comparison Theorem - Dirichlet form)*** *Suppose that the function $H(\cdot)$ satisfies conditions (R1)-(R3) from Appendix A. Let $u,v$ be respectively, a viscosity sub- and super-solution to*

$$\partial_t f + H(z,t,f,\partial_z f) = 0 \text{ on } \mathscr{U},$$

*where $\mathscr{U}$ is an open set. Then*

$$\sup_{\bar{\mathscr{U}}}(u-v)_+ \leq \sup_{\partial\mathscr{U}}(u-v)_+. \tag{73}$$

*If, alternatively, $\mathscr{U}$ is the of the form $\mathscr{Z}\times(0,T]$, where $\mathscr{Z}$ is any open set, we can replace $\partial\mathscr{U}$ in the statement with $\Gamma \equiv \{\partial\mathscr{Z}\times[0,T]\}\cup\{\mathscr{Z}\times\{0\}\}$.*

The above theorem can be applied on any open set $\mathscr{U}$; we do not need to specify the actual boundary condition.

**Lemma G.2.** *Suppose that Assumptions 1-4 hold for PDE (6) under the Dirichlet boundary condition (8). Then there exists $L_0 < \infty$ independent of $\theta, z, t$ such that $|h_\theta(z,t)| \leq L_0$. In addition, for the setting with $T < \infty$, there exists $K < \infty$ such that $|h_\theta(z,t)| \leq K|T-t|$. In a similar vein, for the setting with $\underline{z} > -\infty$, there exists $K_1 < \infty$ such that $|h_\theta(z,t)| \leq K_1|z-\underline{z}|$.*

*Proof.* Consider the Dirichlet problem with $T<\infty$. Define $u_\theta(z,\tau) := e^{\beta\tau}h_\theta(z,T-\tau)$. Then $u_\theta(\cdot)$ is the viscosity solution to (18). Now, it is straightforward to verify that $\phi(z,\tau) := K\tau$ is a super-solution to (18) as long as $K \geq \sup_{z,\tau}|\lambda(\tau)\bar{r}_\theta(z,\tau)|$ (note that $|\lambda(\tau)\bar{r}_\theta(z,\tau)|$ is uniformly bounded by virtue of Assumption 2(i)). Hence, by the Comparison Theorem G.1, it follows $u_\theta \leq \phi$ on $\bar{\mathscr{U}}$.[19] This implies $u_\theta \leq KT$ everywhere, and furthermore, since $h_\theta(z,t) = e^{-\beta(T-t)}u_\theta(z,T-t)$, that $|h_\theta(z,t)| \leq$

[19] It is straightforward to verify that all relevant regularity conditions for the application of Theorem G.1 are satisfied under Assumption 1.

$K|T-t|$. A similar argument, after switching the roles of $z,\tau$, also proves $|h_\theta(z,t)| \leq K_1|z-\underline{z}|$. □

**Lemma G.3.** *Suppose that Assumptions 1-4 hold for PDE (6) under the Dirichlet boundary condition (8). Then there exists $L_1 < \infty$ independent of $\theta, z, t$ such that $h_\theta(z,t)$ is locally Lipschitz continuous in both arguments with Lipschitz constant $L_1$.*

*Proof.* We consider three cases: (i) $\underline{z} = -\infty$, (ii) $T = \infty$, and (iii) $\underline{z} > -\infty$ and $T < \infty$. For cases (i) and (ii), the claim follows by similar arguments as in Souganidis (1985, Proposition 1.5). For case (iii), the claim follows by similar arguments as in Crandall and Lions (1986, Theorem 4). We refer the reader to the supplementary material for a detailed proof. □

## G.2 Periodic Boundary Condition

Consider time-periodic PDEs of the form

$$\partial_t f + H(z,t,f,\partial_z f) = 0 \text{ on } \mathscr{U}; \tag{74}$$
$$f(z,t) = f(z,t+T_p) \ \forall \ (z,t) \in \mathscr{U}.$$

We first present a stronger version of the Comparison Theorem for Cauchy problems, due to Crandall and Lions (1983). This turns out to be useful to prove a comparison theorem for the periodic problem (74).

**Lemma G.4.** *Suppose $\beta \geq 0$, and the function $H(\cdot)$ satisfies conditions (R1)-(R3) from Appendix A. Let $u, v$ be, respectively, viscosity sub- and super-solutions to*

$$\partial_t f + H(z,t,f,\partial_z f) = 0 \text{ on } \mathbb{R} \times (t_0, \infty).$$

*Then for all $t \in [t_0, \infty)$,*

$$e^{\beta(t-t_0)} \sup_{z\in\mathbb{R}} (u(z,t) - v(z,t))_+ \le \sup_{z\in\mathbb{R}} (u(z,t_0) - v(z,t_0))_+ .$$

**Theorem G.2.** ***(Comparison Theorem - Periodic form)*** *Suppose that the function $H(\cdot)$ satisfies conditions (R1)-(R3) from Appendix A, and that it is $T_p$-periodic in $t$. Also suppose that $\beta \ge 0$. Let $u, v$ be respectively, $T_p$-periodic viscosity sub- and super-solutions to (74) on $\mathscr{U}$. Then $u(x,t) \le v(x,t)$ on $\mathbb{R} \times \mathbb{R}$.*

*Proof.* By Lemma G.4, for any $t_0 \in \mathbb{R}$,

$$e^{\beta T_p} \sup_{z\in\mathbb{R}} (u(z, T_p + t_0) - v(z, T_p + t_0))_+ \le \sup_{z\in\mathbb{R}} (u(z,t_0) - v(z,t_0))_+ .$$

But by periodicity, $u(z, T_p + t_0) - v(z, T_p + t_0) = u(z,t_0) - v(z,t_0)$. Hence, $\sup_{z\in\mathbb{R}} (u(z,t_0) - v(z,t_0))_+ = 0$. But the choice of $t_0$ was arbitrary; therefore, $u(z,t) \le v(z,t)$ on $\mathbb{R} \times \mathbb{R}$. $\square$

**Lemma G.5.** *Suppose that Assumptions 1-4 hold for the periodic boundary condition (45), and the discount factor $\beta$ is sufficiently large. Then there exists $L_1 < \infty$ independent of $\theta, z, t$ such that $h_\theta$ is locally Lipschitz continuous with Lipschitz constant $L_1$.*

*Proof.* That $h_\theta(z,t)$ is Lipschitz continuous in $z$ follows by similar arguments as in Crandall and Lions (1986, Theorem 3). For the Lipschitz continuity of $h_\theta(z,t)$ in $t$, we can argue as in Souganidis (1985, Proposition 3.4). We refer the reader to the supplementary material for a detailed proof. $\square$

## G.3 Neumann Boundary Condition

For results on the Neumann boundary condition, we impose additional regularity conditions on $H(\cdot)$ and $B(\cdot)$, i.e., in addition to (R1)-(R8) in Appendices A and E.2.

These are needed to prove Lipschitz continuity of solutions, and are given by (as before, we use the notation $y := (z,t)$):

(R9) There exist $C_1, C_2 < \infty$ such that

$$|H(y_1,u,p_1) - H(y_2,u,p_2)| \le C_1 (\|y_1 - y_2\| + \|p_1 - p_2\|), \quad \text{and}$$
$$|H(y_1,u,p) - H(y_2,u,p)| \le C_2 \|p\| \|y_1 - y_2\|.$$

(R10) There exist $C_3, C_4 < \infty$ such that

$$|B(y_1,u,p_1) - B(y_2,u,p_2)| \le C_3 (\|y_1 - y_2\| + \|p_1 - p_2\|), \quad \text{and}$$
$$|B(y_1,u,p) - B(y_2,u,p)| \le C_4 \|p\| \|y_1 - y_2\|.$$

It is straightforward to verify that under Assumptions 1-4, the regularity conditions (R1)-(R7) and (R9)-R(10) are satisfied by $H_\theta(\cdot)$ and $B_\theta(\cdot)$ in PDE (50), with constants $C_1, C_2, C_3 C_4$ independent of $\theta$ (this is due to uniform boundedness and Lipschitz continuity of $\lambda(t)$, $\bar{G}_\theta(z,t)$ and $\bar{r}_\theta(z,t)$ imposed in Assumption 1). The condition (R8) is not required for the results below (it is only used to show existence of a solution). The conditions (R1)-(R7) are also satisfied by $\hat{H}_\theta(\cdot)$ in the sample PDE (63) under the additional assumption - made is the statement of Theorem E.1 - that $G_a(x,z,t), \pi_\theta(x,z,t)$ are uniformly continuous in $(z,t)$. (R9) and (R10) may not be satisfied by $\hat{H}_\theta(\cdot)$. But they are not needed to prove a comparison theorem, being used only to show Lipschitz continuity of solutions, which we do not require for the sample PDE (63).

The following results are taken from Barles and Lions (1991), but see also Crandall *et al* (1992, Theorem 7.12). We refer to those papers for the proofs.

**Theorem G.3.** ***(Comparison Theorem - Neumann form)*** *Suppose that the func-*

*tions $H(\cdot)$ and $B(\cdot)$ satisfy conditions (R1)-(R7) in Appendices A and E.2. Let $u, v$ be respectively, a viscosity sub- and super-solutions to (49). Then $u(x,t) \le v(x,t)$ on $\bar{\mathscr{Z}} \times [0, \bar{T}]$.*

**Lemma G.6.** *Suppose that the functions $H(\cdot)$ and $B(\cdot)$ satisfy conditions (R1)-(R7) in Appendices A and E.2, along with (R9)-(R10). Then the unique viscosity solution, u, of PDE (49) is Lipschitz continuous on $\bar{\mathscr{Z}} \times [0, \bar{T}]$, where the Lipschitz constant depends only on the values of $C_1$-$C_4$ in (R9)-(R10).*

# H Semi-Convexity, Sup-Convolution etc.

In this section, we collect various properties of semi-convex/concave functions, and sup/inf-convolutions used in the proof of Theorem 2. Since the results are well-known in the literature, we will only state them here. The interested reader is referred to the supplementary material for the proofs.

## H.1 Semi-Convexity and Concavity

In what follows, we take $y$ to be a vector in $\mathbb{R}^d$. Moreover, for any vector $y$, $|y|$ denotes its Euclidean norm.

**Definition 3.** *A function u on $\mathbb{R}^d$ is said to be semi-convex with the coefficient c if $u(y) + \frac{c}{2}|y|^2$ is a convex function. Similarly, u is said to be semi-concave with the coefficient c if $u(y) - \frac{c}{2}|y|^2$ is concave.*

**Lemma H.1.** *Suppose that u is semi-convex. Then u is twice differentiable almost everywhere. Furthermore, at every point y for which Du exists,*

$$u(y+h) \ge u(y) + h^{\intercal} Du(y) - \frac{c}{2}|h|^2 \quad \forall\, h \in \mathbb{R}^d.$$

An analogous property also holds for semi-concave functions. We can also extend the scope of the theorem to points where $Du$ does not exist by considering one-sided derivatives, which can be shown to exist everywhere for semi-convex functions.

## H.2 Sup and Inf Convolutions

Let $u(y)$ denote a continuous function on some open set $\mathscr{Y}$. Let $\partial\mathscr{Y}$ denote the boundary of $\mathscr{Y}$, and $\bar{\mathscr{Y}}$ its closure. Also, $\|Du\|$ denotes the Lipschitz constant for $u$, with the convention that it is $\infty$ if $u$ is not Lipschitz continuous.

**Definition 4.** *The function $u^{\varepsilon}$ is said to be the sup-convolution of $u$ if*

$$u^{\varepsilon}(y) = \sup_{\tilde{y}\in\mathscr{Y}} \left\{ u(\tilde{y}) - \frac{1}{2\varepsilon}|\tilde{y}-y|^2 \right\}.$$

*Similarly, $u_{\varepsilon}$ is said to be the inf-convolution of $u$ if*

$$u_{\varepsilon}(y) = \inf_{\tilde{y}\in\mathscr{Y}} \left\{ u(\tilde{y}) + \frac{1}{2\varepsilon}|\tilde{y}-y|^2 \right\}.$$

The following lemma characterizes the properties of sup-convolutions (similar results apply for inf-convolutions):

**Lemma H.2.** *Suppose that $u$ is continuous on $\bar{\mathscr{Y}}$. Then,*

*(i) $u^{\varepsilon}$ is semi-convex with coefficient $1/\varepsilon$.*

*(ii) For all $y \in \bar{\mathscr{Y}}$, $|u^{\varepsilon}(y) - u(y)| \leq 4\|Du\|^2\varepsilon$.*

*(iii) $\|Du^{\varepsilon}\| \leq 4\|Du\|$.*

Our next lemma concerns PDEs of the form

$$F(y, u(y), Du(y)) = 0 \text{ on } \mathscr{Y}.$$

We assume that $F(\cdot)$ satisfies the following property:

$$|F(y_1,q_1,p)-F(y_2,q_2,p)| \leq C|p|\{|q_1-q_2|+|y_1-y_2|\}, \tag{75}$$

where $C<\infty$ is some constant. Define $\mathscr{Y}_\varepsilon$ as the set of all points in $\mathscr{Y}$ that are at least $2\|Du\|\varepsilon$ distance away from $\partial\mathscr{Y}$, i.e.,

$$\mathscr{Y}_\varepsilon := \{y \in \mathscr{Y} : |y-w| > 2\|Du\|\varepsilon \ \forall\, w \in \partial\mathscr{Y}\}.$$

**Lemma H.3.** *Suppose that $u$ is a viscosity solution to $F(y,u,Du)=0$ on $\mathscr{Y}$, and $\|Du\| \leq m < \infty$. Suppose also that $F(\cdot)$ satisfies (75). Then, there exists some $c$ depending on only $C$ (from 75) and $m$ such that $F(y,u^\varepsilon,Du^\varepsilon) \leq c\varepsilon$ in the viscosity sense for all $y \in \mathscr{Y}_\varepsilon$.*